\documentclass[11pt,aps,showpacs,notitlepage,nofootinbib]{revtex4-1}
\usepackage{amssymb,amsmath}
\usepackage{bm}
\usepackage[pdftex]{graphicx}
\usepackage{subfig}
\usepackage{color}
\usepackage{amsthm}
\usepackage[update,prepend]{epstopdf}

\usepackage{footnote}

\newcommand{\beq}{\begin{equation}}
\newcommand{\beqn}{\begin{eqnarray}}
\newcommand{\eeq}{\end{equation}}
\newcommand{\eeqn}{\end{eqnarray}}

\begin{document}

\title{Universality and scaling in multi-field $\alpha$-attractor preheating
}
\author{
Oksana~Iarygina$^{1}$, Evangelos~I.~Sfakianakis$^{1,2}$, Dong-Gang~Wang$^{1,3}$, Ana~Ach\'ucarro$^{1,4}$
}
\email{\baselineskip 11pt Email addresses: achucar@lorentz.leidenuniv.nl;  iarygina@lorentz.leidenuniv.nl; e.sfakianakis@nikhef.nl; wdgang@strw.leidenuniv.nl}
\affiliation{
$^1$Institute Lorentz of Theoretical Physics, Leiden University, 2333 CA Leiden, The Netherlands
\\
$^2$ Nikhef, Science Park 105, 1098 XG Amsterdam, The Netherlands
\\
$^3$ Leiden Observatory, Leiden University, 2300 RA Leiden, The Netherlands
\\
$^4$ Department of Theoretical Physics, University of the Basque Country, UPV-EHU 48080 Bilbao, Spain
}
\date{\today}
\begin{abstract}
We explore preheating in multi-field models of inflation in which the field-space metric is a highly curved hyperbolic manifold. One broad family of such models is called $\alpha$-attractors, whose single-field regimes have been extensively studied in the context of inflation and supergravity.
We focus on a simple two-field generalization of the $T$-model, which has received renewed attention in the literature.
Krajewski et al. concluded, using lattice simulations, that multi-field effects can dramatically speed-up preheating.
 We recover their results and further demonstrate that significant analytical progress can be made for preheating in these models using the WKB approximation and Floquet analysis.
We find a simple scaling behavior of the Floquet exponents for large values of the field-space curvature, that enables  a quick estimation of the $T$-model reheating efficiency for any large value of the field-space curvature.
In this regime we further observe and explain universal preheating features that arise for different values of the potential steepness.  In general preheating is faster for larger negative values of the field-space curvature and steeper potentials. For very highly curved field-space manifolds preheating is essentially instantaneous.
\\
\end{abstract}
\pacs{Preprint Nikhef 2018-050}
\maketitle

\tableofcontents

\section{Introduction}

Inflation remains the leading paradigm for the very early universe, providing an elegant solution to the horizon and flatness problems of big bang cosmology \cite{Guth:1980zm, Linde:1981mu}. However, the biggest success of inflation is undoubtedly that it provides a framework for computing the primordial density fluctuations that can be observed as temperature variations of the Cosmic Microwave Background Radiation (CMB) and that provide the seeds for structure formation.

The recent results from the {\it Planck }  satellite \cite{Akrami:2018odb} are the latest in a long line of experiments, starting in 1989 with COBE, trying to constrain the characteristics of the primordial power spectrum through measuring the spectral index of scalar fluctuations ($n_s$). Attempts to measure the running of the spectral index $\alpha_s$ and  the tensor to scalar ratio $r$ have resulted so far only in placing upper bounds on both. While large-field models of inflation are tightly constrained and the simplest ones, like quadratic inflation, are practically ruled out, large families of models are still compatible with the data, providing predictions that match those of the Starobinsky model \cite{Starobinsky:1980te}
\beq
n_s = 1-{2\over N_*} \, , \quad r = {{12} \alpha \over N_*^2}
\label{eq:nsr}
\eeq
{where $N_*$ is the time in $e$-folds where the CMB modes exit the horizon during inflation.}
The two main families of models that provide the observables of Eq.~\eqref{eq:nsr} are models with non-minimal coupling to gravity \cite{KS,nonm1} (sometimes called $\xi$-attractors\footnote{It is worth noting that in the Palatini formulation of gravity the behavior and predictions of $\xi$-attractor models change significantly, as is discussed  for example in Refs.~\cite{Jarv:2017azx, Carrilho:2018ffi}.}) and models with hyperbolic field-space geometry, also called $\alpha$-attractors \cite{Kallosh:2013yoa, Lindealpha, Achucarro:2017ing, Carrasco}. Higgs inflation \cite{Bezrukov:2007ep, GKS} is an example of the former.
For the Starobinsky model and $\xi$-attractors, $\alpha=1$ in Eq.~\eqref{eq:nsr}, hence the prediction for the tensor mode amplitude is fixed. For $\alpha$-attractors, the parameter $\alpha$ corresponds to the curvature of the field-space, as we will see, hence the tensor power is suppressed for highly curved field-space manifolds \cite{Kallosh:2013yoa, Lindealpha, Carrasco}.
At some level, the unifying feature of all these approaches can be attributed to a singularity in the kinetic sector \cite{Galante:2014ifa}.
We will focus only on $\alpha$-attractors, drawing similarities and differences with the other observationally related models when necessary.

While a lot of theoretical and phenomenological work on inflation has focused on single-field scenarios, realistic models of high-energy particle physics typically include many distinct scalar fields at high energies \cite{LythRiotto,WandsReview,MazumdarRocher,VenninWands,GongReview}.
Furthermore, multiple fields with a curved field-space manifold (see e.g.~\cite{Gordon:2000hvETAL, Peterson:2011ytETAL, heavy1, heavy2, KMS, SSK, Achucarro:2016fby, Achucarro:2015bra, Cremonini:2010ua, Lalak:2007vi}) can display  a variety of effects, including non-gaussianities, isocurvature modes, imprints from heavy fields during turns in field space, curvature fluctuations from ultra-light entropy modes,  as well as geometric destabilization of the inflationary trajectory \cite{Renaux-Petel:2015mga, Renaux-Petel:2017dia, Garcia-Saenz:2018ifx, Cicoli:2018ccr}.
Several models that lead to the predictions of Eq.~\eqref{eq:nsr} display strong single field attractors \cite{KS, GKS, Achucarro:2017ing} that persist during and after inflation.
In particular, the multi-field analysis of $\alpha$-attractors has become an interesting topic recently \cite{Achucarro:2017ing, Christodoulidis:2018qdw, Linde:2018hmx, Dias:2018pgj, Bartolo:2018hjc, Maeda:2018sje}.

{During inflation, the inflaton field dominates the total energy density budget. However, the universe must be in a radiation dominated stage before Big Bang Nucleosynthesis (BBN), in order  to produce the observed abundance of light elements \cite{Steigman,FieldsBBN,Cyburt} (see e.g. Refs.~\cite{BTW,Allahverdi,Frolov,AHKK,AminBaumann} for recent reviews). } The period during which the energy density locked in the inflaton condensate is transferred to radiation modes is called reheating. While inflation  is tightly constrained by measurements of the CMB and Large Scale Structure \cite{GuthKaiser,BTW,LythLiddle,Baumann,MartinRingeval,GKN,LindePlanck,MartinRev}, the period after inflation and before Big Bang Nucleosynthesis (BBN),
 provides far fewer observational handles, due to the very short length-scales involved.  {This is due to the fact that most dynamics during reheating takes place at sub-horizon scales, following causality arguments, hence it does not leave an imprint on larger scales, like the CMB\footnote{{While this is true for most models, there are well motivated cases where reheating can excite super-horizon modes and thus affect CMB observables. This does not occur for the $\alpha$-attractor models that we are examining and we will not be discussing it further.}}. Furthermore, the thermalization processes that have to occur before BBN wash out many of the ``fingerprints" of reheating.}
Despite its inherent complexity, knowledge of the reheating era is essential, in order to relate inflationary predictions to present-day observations. The evolution history of the universe determines the relation between the times of horizon-crossing and re-entry of primordial fluctuations \cite{AdsheadEasther,Dai,Creminelli,MartinReheat,GongLeungPi,CaiGuoWang,Cook,Heisig,Liddle:2003as}. Furthemore, preheating in multi-field models of inflation can alter the evolution of cosmological observables \cite{Bassett:2005xm, Taruya, FinelliBrandenberger, ShinjiBassett, ChambersRajantie, BondFrolov, Leung:2012ve}.

The reheating era can proceed either through perturbative decay of the inflaton, or through non-perturbative processes, such as  parametric and tachyonic resonance, also  called preheating (see e.g. \cite{Traschen:1990sw, Kofman:1997yn, Greene:1997fu}  and Ref.~\cite{Amin:2014eta} for a review).
A recent paper \cite{Krajewski:2018moi} used lattice simulations to compute the preheating behavior of a specific two-field realization of the T-model, a  member of the $\alpha$-attractor family \cite{Tmodel}. In this paper we use linear analysis to recover and interpret the results of Ref.~\cite{Krajewski:2018moi} and examine their dependence on the potential steepness and field-space curvature.
We find that the Floquet charts for a specific value of the potential steepness collapse into a single ``master diagram" for small values of $\alpha$ when plotted against axes properly rescaled by the field-space curvature. Even for different potential parameters, the scaling behavior of the Floquet charts persists, albeit in an approximate rather than exact form. Overall we find slightly faster preheating for steeper potentials and for models with stronger field-space curvature. An important conclusion is that, in the limit of  highly curved manifolds, preheating occurs almost instantaneously regardless of the exact form of the T-model potential. This is important for connecting the predictions of $\alpha$-attractors to CMB observations.

The structure of the paper is as follows. In Section \ref{sec:Model} we describe the model and study its background evolution, both during and after inflation.
In Section \ref{sec:tachyonicresonance} we review the formalism for computing fluctuations in multi-field models with non-trivial field-space metric. We also specify the form of the potential and analyze the resulting particle production using semi-analytic arguments, the WKB approximation and Floquet theory.
Section \ref{sec:potentials} generalizes our results to different potentials.
We conclude in Section  \ref{sec:summary}.


\section{Model}
\label{sec:Model}

We consider a model consisting of two interacting scalar fields on a hyperbolic manifold of constant negative curvature. The specific  Lagrangian corresponds to a two-field extension of the well-known T-model, as described in detail in Appendix A and Ref.~\cite{Krajewski:2018moi}, and can be written as
\beq
{\cal L} = -{1\over 2} \left ( \partial_\mu \chi \partial^\mu \chi + e^{2b(\chi)} \partial_\mu \phi \partial^\mu \phi \right ) - V(\phi,\chi)  \, ,
\label{eq:Lphichi}
\eeq
where $ b(\chi) = \log\left(\cosh(\beta \chi)\right) $. The corresponding two-field
 potential is
\beq
V(\phi,\chi) = M^4 \left ( {\cosh(\beta \phi)\cosh(\beta \chi)-1 \over \cosh(\beta \phi)\cosh(\beta \chi)+1}\right)^n \left( \cosh(\beta \chi) \right )^{2/\beta^2} \, ,
\label{eq:Vphichi}
\eeq
where $\beta = \sqrt{2/3\alpha}$ and $M^4= \alpha \mu^2$. For $\chi=0$ the potential becomes
\beq
V(\phi,0) =M^4 \left ( \left ( \tanh(\beta \phi/2) \right )^2 \right )^n = M^4  \tanh^{2n}(\beta |\phi|/2) .
\eeq
The  background equation of  motion for $\phi(t)$ at $\chi(t)=0$ is
\beq
\ddot \phi + 3 H \dot \phi +
{{2\sqrt{2}M^4 n \over\sqrt{ 3\alpha}} }
\text{csch}\left(
{\sqrt{2\over 3\alpha}}| \phi|
\right)
 \tanh ^{2n}\left(
{\frac{|\phi| }{\sqrt{6 \alpha }}  }   \right)=0 .
\eeq
We rescale the inflaton field $\phi$ and the parameter $\alpha$ by the reduced Planck mass as $\phi = \tilde \phi M_{\rm Pl}$ and $\alpha = \tilde \alpha M_{\rm Pl}^2$. Finally, we rescale time by $\mu$, leading to
\beq
{d^2 {\tilde \phi}\over d(\mu t)^2} + 3 {H\over \mu} {d {\tilde  \phi }\over d(\mu t)}+    {{2\sqrt{2\tilde \alpha} \, n \over\sqrt{ 3}} }
    \text{csch}\left(
{\sqrt{2\over 3\tilde \alpha}} |\tilde \phi |
        \right)
     \tanh ^{2n}\left(
{\frac{|\tilde \phi| }{\sqrt{6\tilde  \alpha }}  }
     \right)=0 ,
\label{eq:eomrescaledfull}
\eeq
where
\beqn
 \left ({H\over \mu} \right )^2 = {1\over 3 } \left [ {1\over 2}\left ({d {\tilde \phi}  \over d(\mu t)} \right )^2 + \tilde \alpha \cdot \tanh^{2n} \left ( {|\tilde\phi|\over \sqrt{6\tilde\alpha}} \right ) \right] .
\eeqn
In Ref.~\cite{Krajewski:2018moi} an alternative rescaling of time was implicitly used, which we describe in Appendix C.

\subsection{Single-field background motion}

\begin{figure}
\centering
\includegraphics[width=0.45\textwidth]{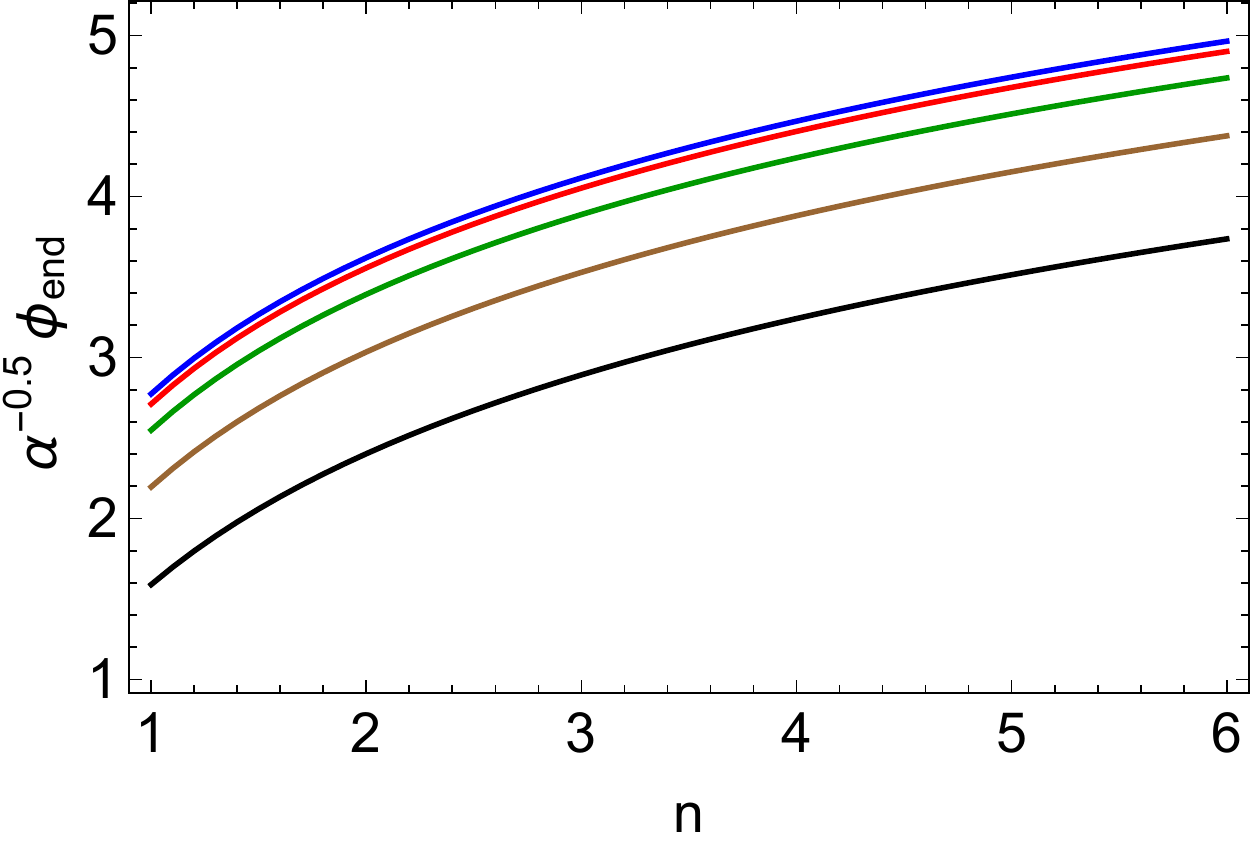}
 \includegraphics[width=0.45\textwidth]{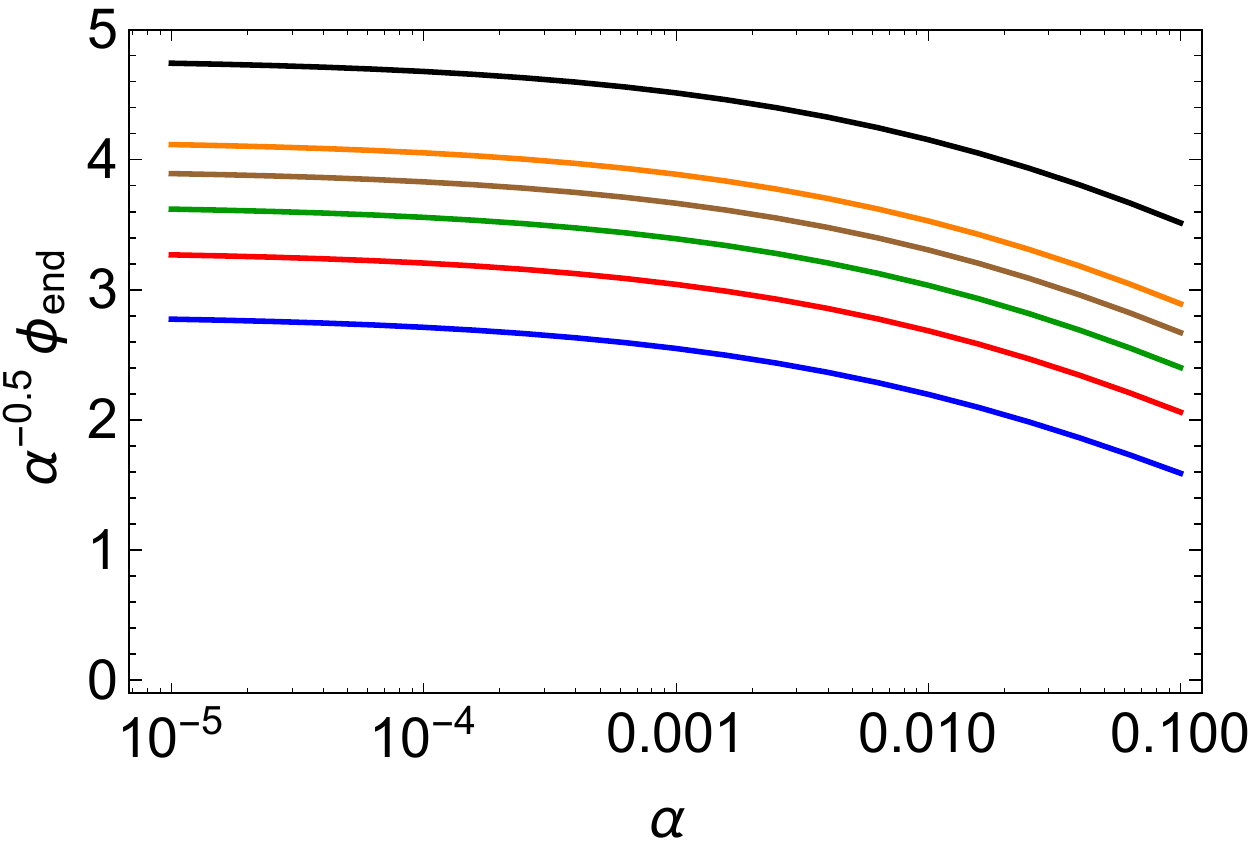}
 \\
 \includegraphics[width=0.45\textwidth]{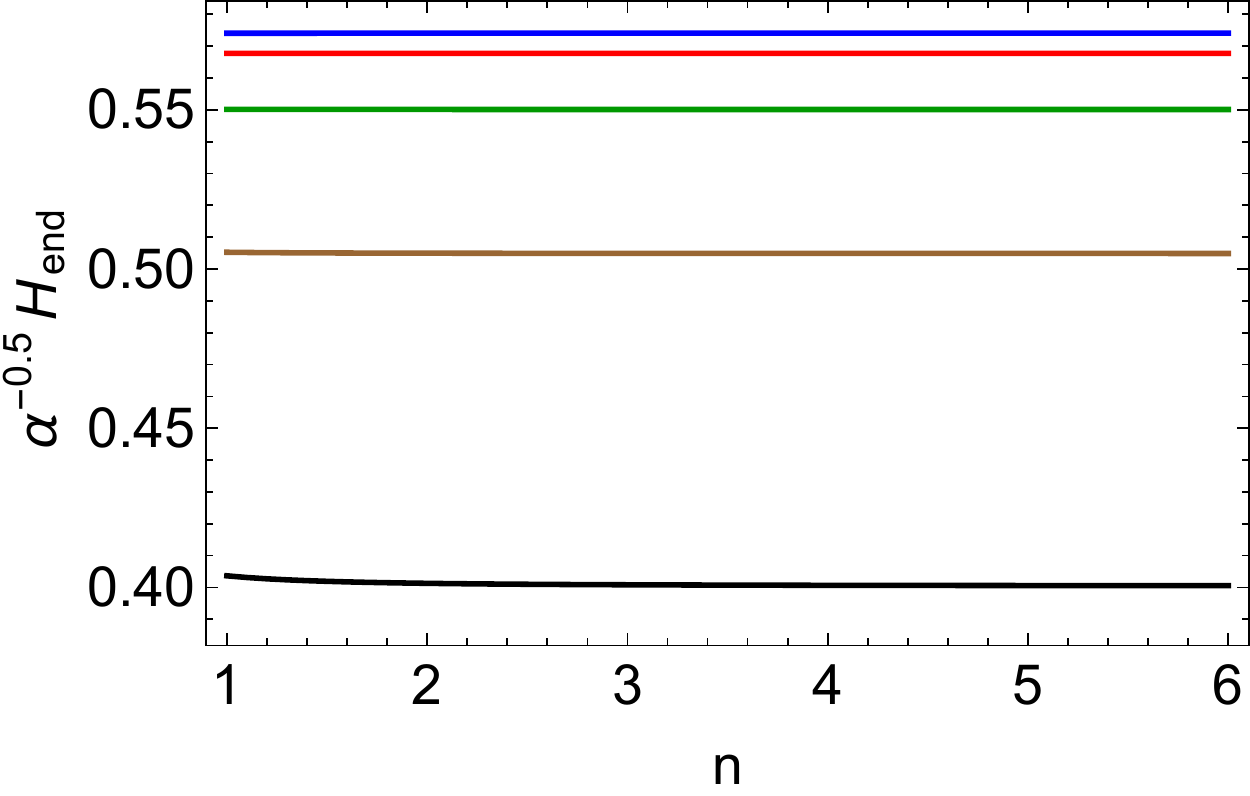}
 \includegraphics[width=0.45\textwidth]{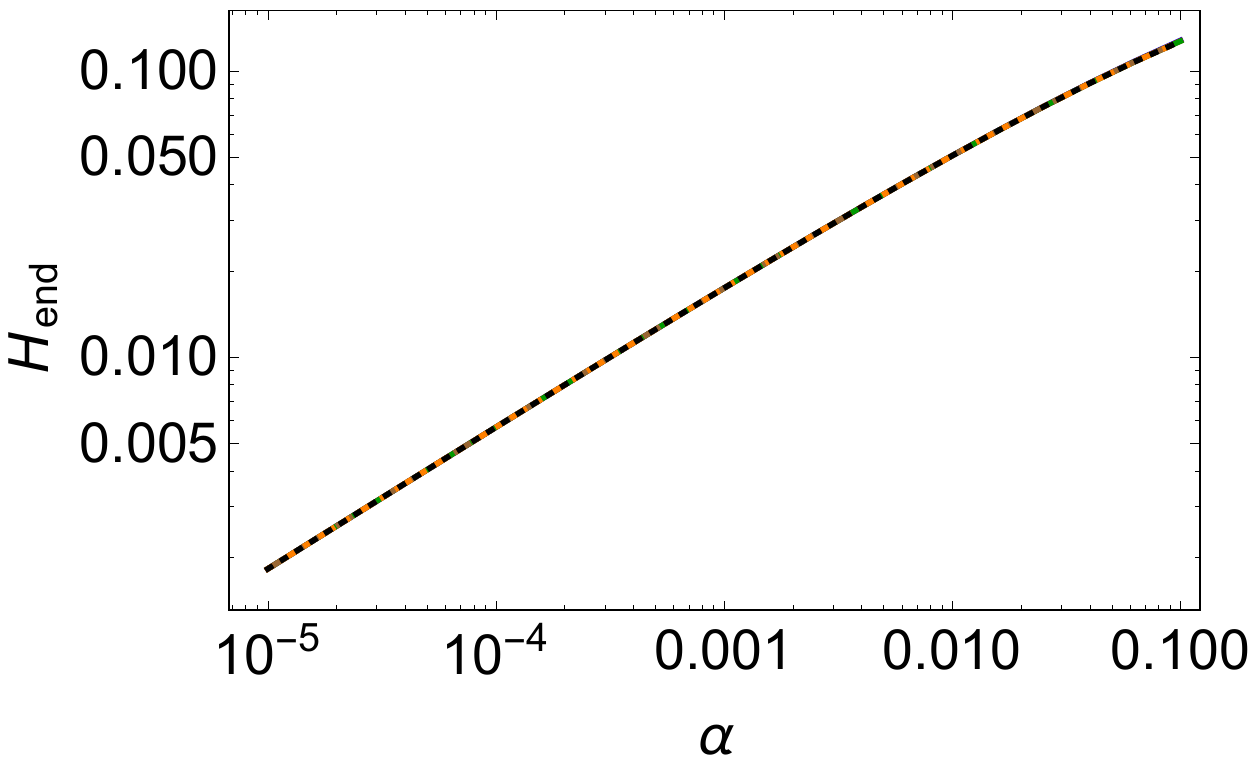}
\caption{
{\it Upper panels:} The rescaled background field at the end of inflation $\phi_{\rm end}/\sqrt{\alpha}$ as a function of $n$ (left) and $\alpha$ (right).
{\it Lower panels:} The rescaled Hubble parameter at the end of inflation $H_{\rm end}/\sqrt{\alpha}$ as a function of $n$ (left) and $\alpha$ (right). The Hubble parameter is measured in units of $\mu$.
 Color coding is as follows:\\
{\it Left:} $\alpha = 10^{-5}, 10^{-4}, 10^{-3}, 10^{-2}, 10^{-1}$ (blue, red, green, brown and black respectively).\\
 {\it Right:} $n = 1, 1.5, 2, 2.5, 3, 5$ (blue, red, green, brown, orange and black respectively).
 }
 \label{fig:phiendvsnvsalpha}
\end{figure}

We start by analyzing the background motion of the $\phi$ and $\chi$ fields, in order to identify the regime of effectively single-field motion and describe CMB constraints on the model parameters.
We initially assume that $\chi(t)=0$ at background level, which is indeed a dynamical attractor, as we will show later. Eq.~\eqref{eq:eomrescaledfull} in the slow-roll approximation and for $\tilde \phi /\sqrt{\tilde\alpha} \gg1$, which holds during inflation, becomes
\beq
3H \dot{ \tilde \phi} + {\frac{4 \left(\sqrt{2 \alpha } n\right)}{\sqrt{3}}
}  e^{ { -\sqrt{2} \tilde \phi / \sqrt{3\tilde\alpha}} }  \simeq 0 ,
\eeq
where $H/\mu \simeq \sqrt{\tilde \alpha / 3}$,
leading to
\beqn
\dot{ \tilde \phi }&=& - {{4\sqrt{2} n \over 3}} e^{ { -\sqrt{2} \tilde \phi / \sqrt{3\tilde\alpha}} } ,
\\
N &=& {{3\tilde \alpha\over 8n } e^{\sqrt{2\over 3\tilde\alpha}\tilde\phi}} .
\eeqn
The slow-roll parameters become
\begin{eqnarray}
\epsilon &\equiv&-{\dot H\over H^2}=
{{16 n^2 \over{3 \tilde \alpha}} e^{-2 \sqrt{2} \tilde\phi / \sqrt{3\tilde\alpha}}
}
\simeq { {3 \tilde \alpha \over 4 N^2}  }
\label{eq:epsilon}
\\
\eta &\equiv& \frac{\dot\epsilon}{\epsilon H}\simeq  \frac{2}{N} .
\label{eq:eta}
\end{eqnarray}
The end of inflation defined as $\epsilon=1$, based on the slow-roll analysis, occurs at
\beq
{{\tilde \phi_{\rm end} \over \sqrt{\tilde \alpha}}={\sqrt{3} \over 2\sqrt{2}} \left(\log {16\over 3 }+ 2\log n  -   \log \tilde \alpha \right) } .
\label{eq:phiend}
\eeq
The last term in Eq.~\eqref{eq:phiend} is subdominant for
{small $\tilde\alpha$}
and can be safely ignored, leading to ${\tilde \phi_{\rm end} / \sqrt{\tilde \alpha}} \simeq {0.6(1.7+2\log n)}$.
Even though the slow-roll approximation fails near the end of inflation, the scaling $\tilde \phi_{\rm end}  / \sqrt{\tilde \alpha} = {\cal O}(1)$ is valid over the whole range of potential parameters $\alpha$ and $n$ that we considered, as shown in Fig.~\ref{fig:phiendvsnvsalpha}.
The Hubble scale at $\epsilon=1$ is
\beq
{H_{\rm end}^2\over \mu^2} = {1\over 2} \tilde \alpha    \cdot \tanh^{2n} \left ( {\tilde\phi_{\rm end}\over \sqrt{6\tilde\alpha}} \right )  \sim {1\over 4} \tilde \alpha ,
\label{eq:Hendslowroll}
\eeq
where the numerical factor in the last equality of Eq.~\eqref{eq:Hendslowroll} is fitted from the bottom right panel of Fig.~\ref{fig:phiendvsnvsalpha}.

The tensor-to-scalar ratio for single-field motion is
\beq
r = 16\epsilon \simeq {12} {\tilde \alpha\over N^2} .
\label{eq:r}
\eeq
In general $r = \alpha\times {\cal O}(10^{-3})$ for modes that exit the horizon at $N\sim 55$ $e$-folds before the end of inflation.
The dimensionless power spectrum of the (scalar) density perturbations is measured to be
\beq
A_s \simeq 2\times 10^{-9} \, .
\eeq
{Using the expression for the scalar power spectrum from single field slow-roll inflation
\beq
A_s = {H^2\over 8\pi^2 M_{\rm Pl}^2 \epsilon} ,
\eeq
and the value of the Hubble scale at the plateau of the potential $H^2 \simeq \tilde \alpha \mu^2 /3$, it is straightforward to see that
 $\mu \sim  10^{-5} M_{\rm Pl}$.
 }
Hence the scale of $\mu$ fixes the amplitude of the scalar power spectrum, independent of $\alpha$ and $n$. By using $\mu$ to re-scale time, it is trivial to connect the preheating calculations performed in the present work to  observational constraints on the potential parameters.

\subsection{Initial condition dependence}

It can be easily seen that the potential of Eq.~\eqref{eq:Vphichi} exhibits a minimum at $\chi=0$ for all values of $\phi$. However, the approach to this potential ``valley" is important and could in principle leave observational signatures, if it occurs close to the time at which the CMB-relevant scales leave the horizon.
\begin{figure}
\centering
\includegraphics[width=0.45\textwidth]{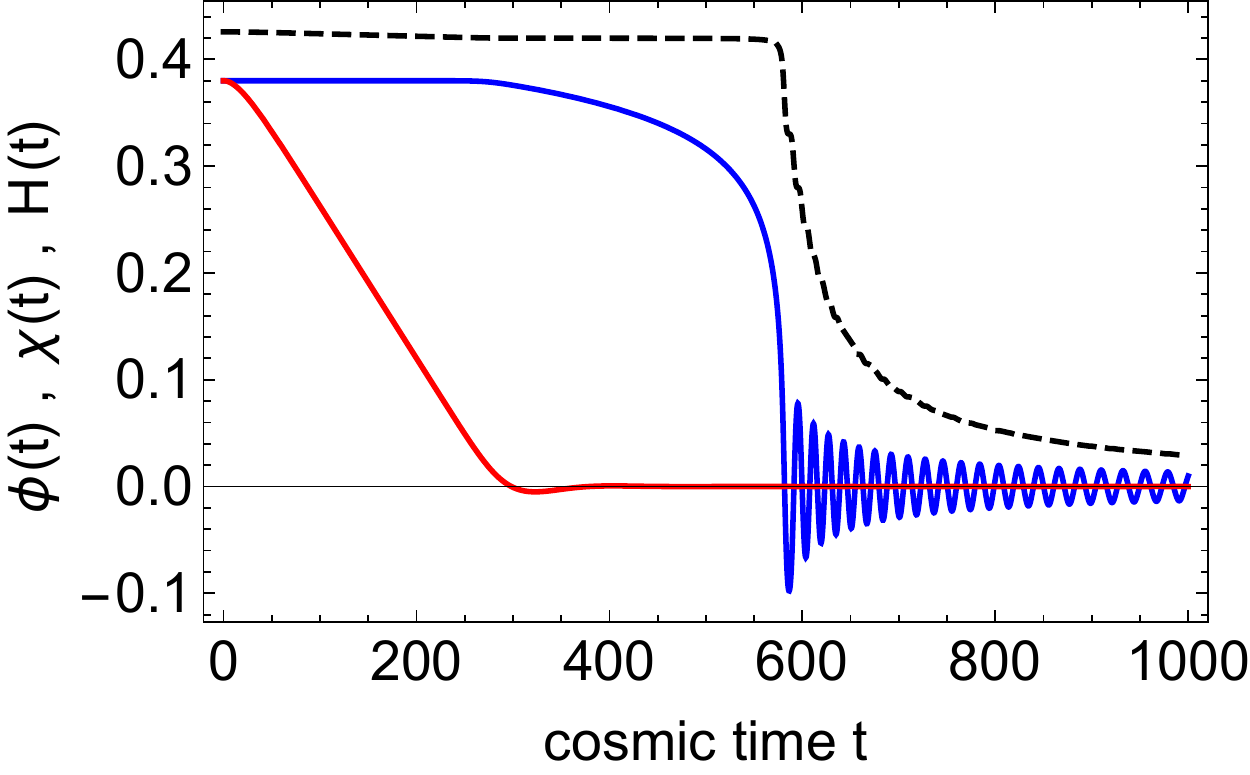}\quad \includegraphics[width=0.45\textwidth]{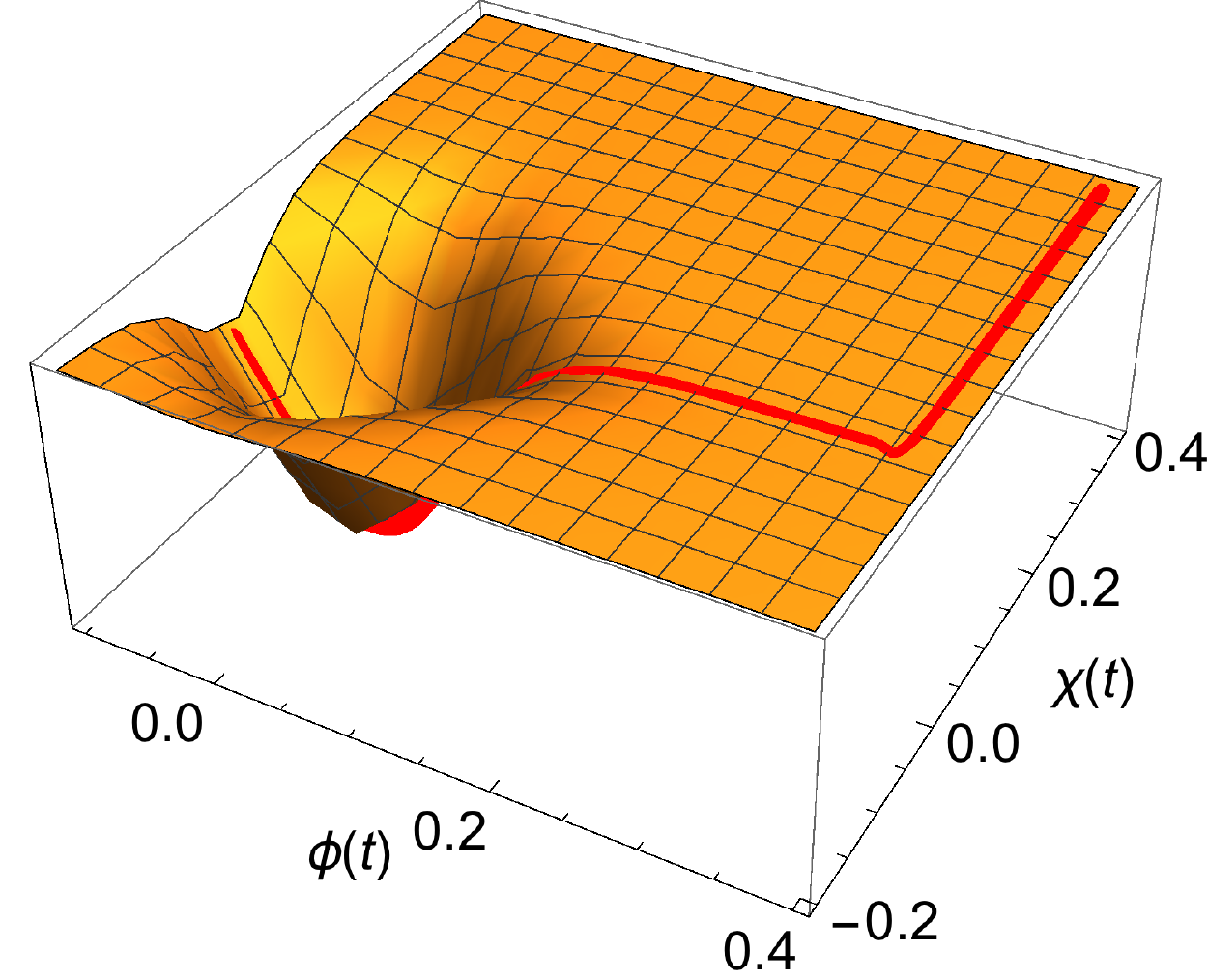}
\caption{
{\it Left:} A characteristic evolution for $\phi$ (blue), $\chi$ (red) and $H$ (black-dashed) for $n=3/2$ and $\tilde \alpha=0.001$, showing the approach to $\chi(t)=0$. The initial conditions are chosen as $\phi_0=\chi_0$ and $\dot \phi_0 = \dot \chi_0=0$.
{\it Right:} The three-dimensional plot of the trajectory on the potential. The two effectively single-field stages are easily visible: $\phi(t) \simeq const.$ followed by $\chi(t)\simeq0$.
 }
 \label{fig:attractorapproach}
\end{figure}

Fig.~\ref{fig:attractorapproach} shows the transition to the single-field trajectory for $n=3/2$ and $\alpha=0.001$. The initial conditions are $\phi_0=\chi_0$, chosen such that there would be $60$ $e$-folds of inflation for $\chi_0=0$. We see two distinct stages of inflation: initially $\phi(t)$ remains almost constant and $\chi(t)$ follows a slow-roll motion until it reaches the minimum $\chi=0$. Then, after a sharp turn in field-space, the field $\phi(t)$ follows a slow-roll motion towards the global minimum of the potential, while $\chi$ stays exponentially close to zero. Hence to a good approximation, the whole inflationary era is separated into  two sequential periods of  distinct single-field motion.

\begin{figure}
\centering
\includegraphics[width=0.45\textwidth]{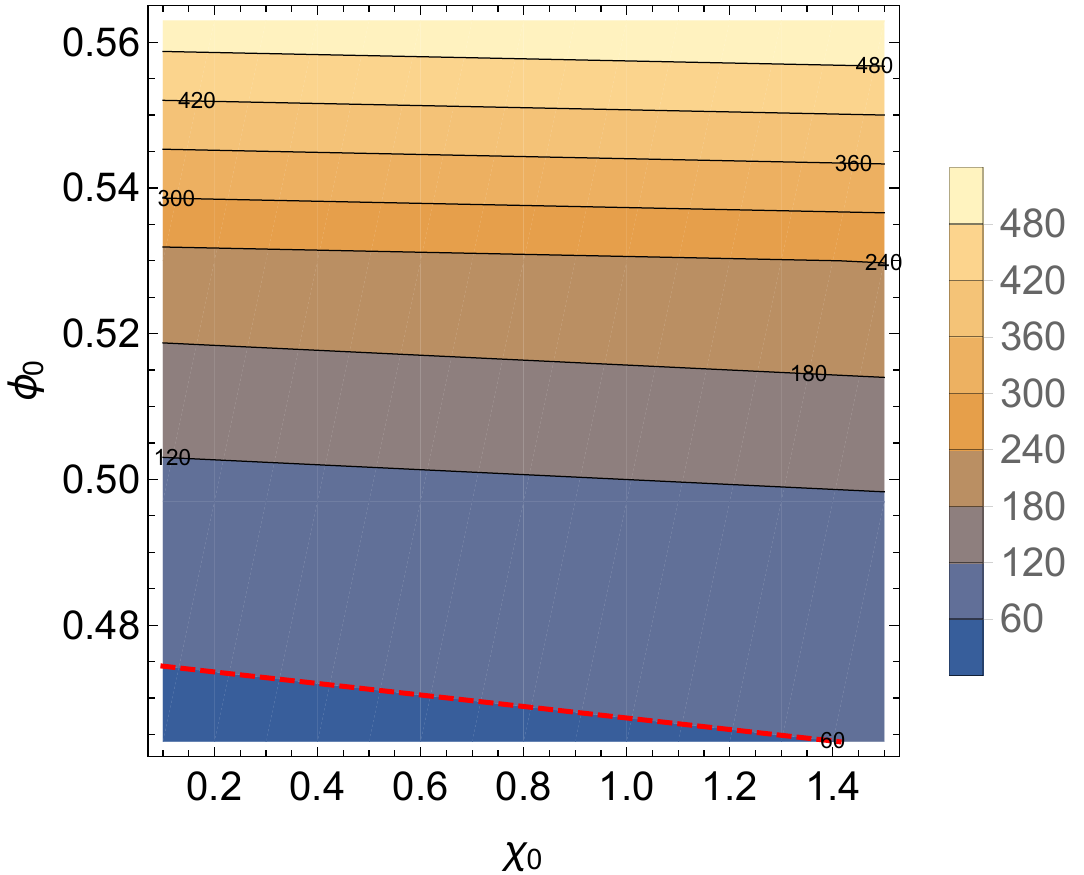}
 \includegraphics[width=0.45\textwidth]{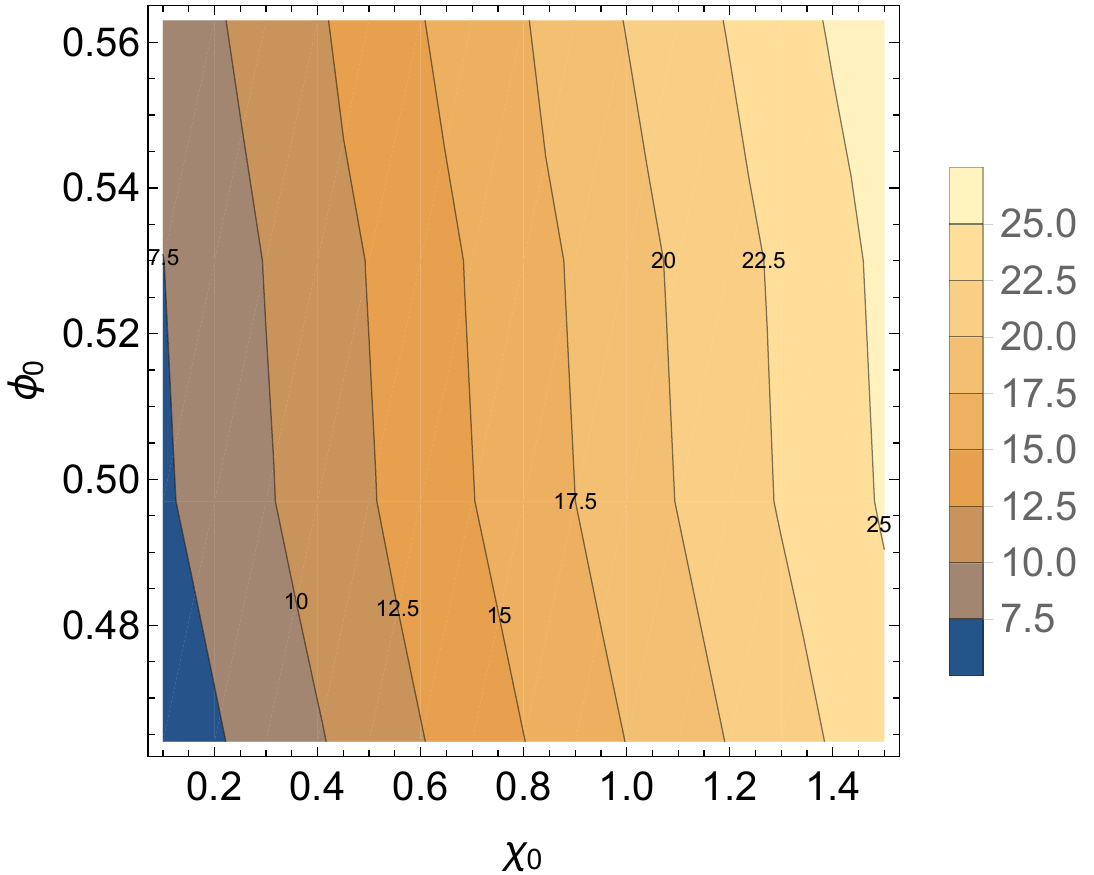}
\caption{
{\it Left:} A contour plot in the $\phi_0\equiv \phi(0)$ and $\chi_0\equiv \chi(0)$ plane for $n=3/2$ and $\tilde\alpha={0.001}$, showing the total number of e-folds of inflation. The initial velocities are chosen as $\dot\phi_0=\dot\chi_0=0$.
{The red-dashed line shows the initial conditions that lead to $60$ $e$-folds of inflation. We see that the total number of $e$-folds are predominately controlled by $\phi_0$.}
{\it Right:} A contour plot in the $\phi_0$ and $\chi_0$ plane for $n=3/2$ and $\tilde\alpha={0.001}$, showing the number of e-folds  from the beginning of inflation until the $\chi=0$ attractor is reached. {As expected, the number of $e$-folds along $\phi \simeq \phi_0$ are mostly determined by $\chi_0$.}
We see that the initial stage of inflation along $\phi\simeq \phi_0$  lasts far less than the second stage of inflation along $\chi=0$, hence it will not leave any observational imprints for non fine-tuned initial conditions.
 }
 \label{fig:contourNef}
\end{figure}

\begin{figure}
\centering
\includegraphics[width=0.45\textwidth]{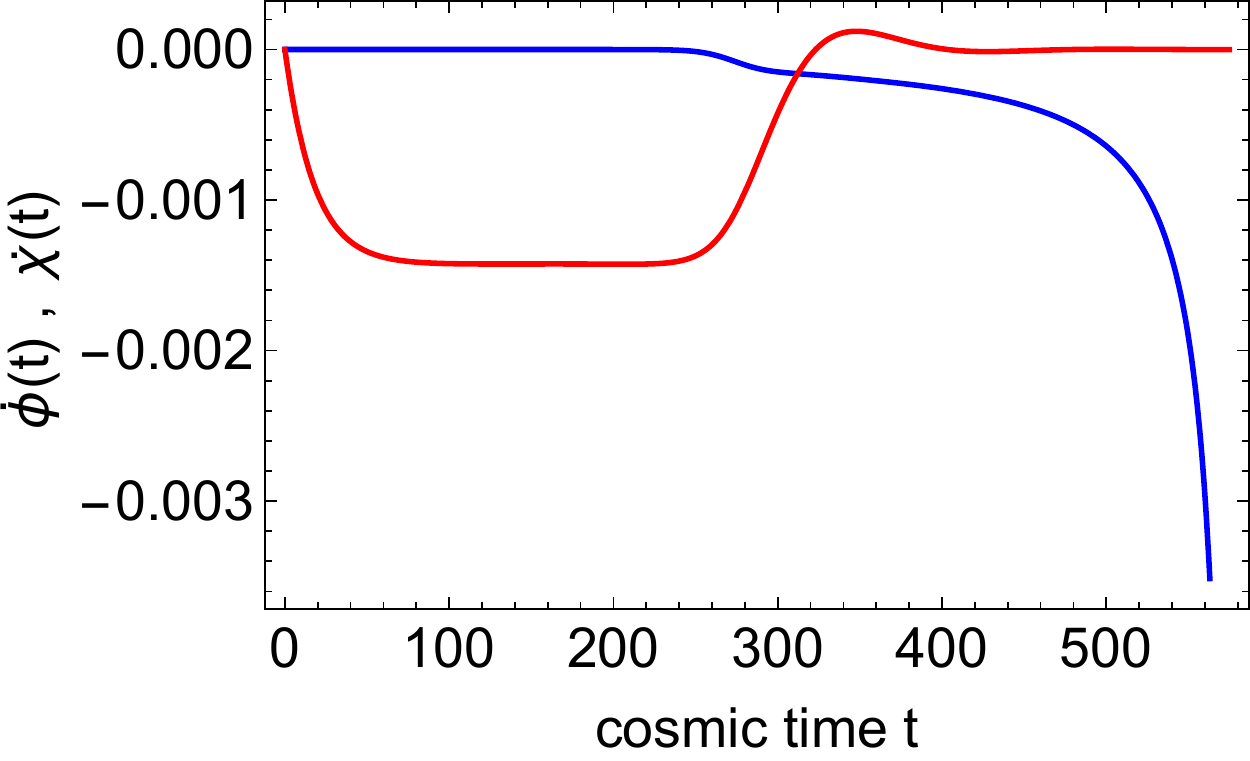}\quad \includegraphics[width=0.45\textwidth]{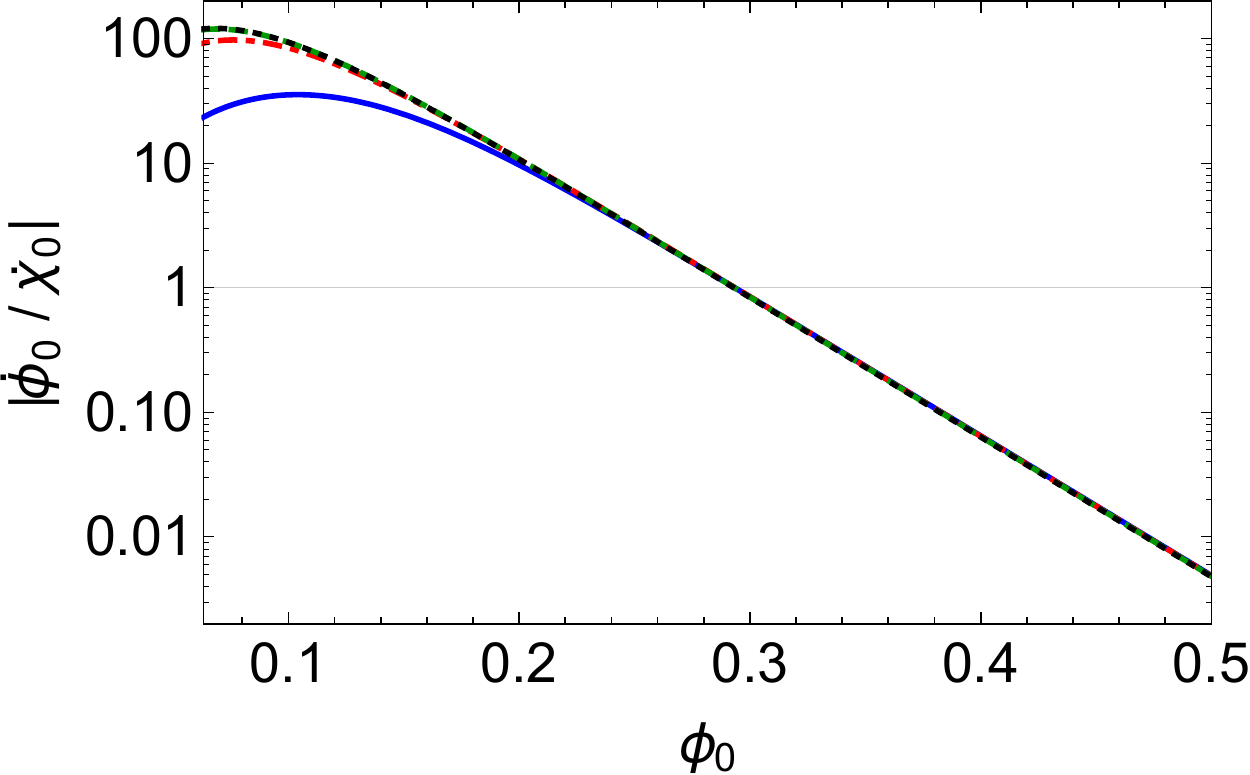}
\caption{
{\it {Left:}}
The field velocities $\dot\phi(t)$ (blue)  and $\dot\chi(t)$ (red) for the example of Fig.~\ref{fig:attractorapproach}. We see that the initial stage of inflation along $\phi=\rm {\rm const.}$
 proceeds with a much larger velocity than the one associated with $\alpha$-attractors, hence it will generate fewer $e$-folds.
{\it {Right:}}
The ratio of the typical velocities $|\dot\phi / \dot \chi|$ as a function of the inflaton field $\phi_0$ for different values of the field amplitude $\chi_0$. We see that for the $\phi$ field values needed  to generate sufficient $e$-folds of inflation the typical $\chi$ velocity is larger than the typical $\phi$ velocity.
 }
 \label{fig:fieldvelocities}
\end{figure}

Starting from a wide range of initial conditions $\phi_0\equiv \phi(0)$ and $\chi_0\equiv \chi(0)$, we see that the system generically follows the two-stage evolution shown in Fig.~\ref{fig:attractorapproach},
proceeding along  $\chi(t)=0$ during the last stage of inflation and during the post-inflationary oscillations.
Figure \ref{fig:contourNef} shows the transition to the single-field motion along $\chi=0$ for broad conditions, constrained to provide more than $60$ $e$-folds of inflation.
Beyond the fact that the single field trajectory along $\chi(t)=0$ is a dynamical attractor for the generalized two-field T-model, its predictions are robust with respect to $\chi_0$. As shown in Fig.~\ref{fig:contourNef}, the number of $e$-folds along the second stage $\chi(t)=0$ is much larger than the number of $e$-folds along the first stage $\phi(t)={\rm const}$. The range of values $\{\phi_0,\chi_0\}$ that place the turn-rate spike (the transition between the two single-field motions) at the observable window $50\lesssim N_* \lesssim 60$ is very narrow, requiring delicate fine-tuning. Hence the generic observational prediction of these models for the CMB is that of usual single-field $\alpha$-attractors. This behavior can be understood analytically. Considering the number of $e$-folds along a single-field trajectory we get
\begin{equation}
N = \int H \, dt = \int {H \over \dot \phi }\, d\phi
\end{equation}
As a quick estimate of the number of $e$-folds we can use $\Delta N_1 \sim (H/ |\dot \chi|) \Delta \chi  \sim (H/ |\dot \chi|)  \chi_0$ during the first stage and $\Delta N_2 \sim (H/ |\dot \phi|) \Delta \phi \sim (H/ |\dot \phi|) \phi_0 $ during the second stage of inflation. Assuming that the Hubble scale does not change much during inflation
\beq
{N_1 \over N_2} \sim  \left |{ {\dot \chi\over \dot \phi} } \right | {\chi_0\over \phi_0}
\eeq
Fig.~\ref{fig:fieldvelocities} shows the ratio $ |\dot \phi / \dot\chi |$ as a function of $\phi$ for several values of $\chi_0$. We see that for large values of $\phi_0$, required to give a sufficient number of $e$-folds of inflation, $|\dot \chi| = {\cal O}(10) | \dot \phi|$, hence $N_1 = {\cal O}(0.1) N_2$ for typical values of $\{\phi_0, \chi_0\}$.
While there is potentially interesting phenomenology from the turning trajectories, it is absent for generically chosen initial conditions.
Since we are only interested in the preheating behavior of the two-field T-model, we will not pursue this subject further here.

\subsection{Geometrical destabilization}

\begin{figure}
\centering
\includegraphics[width=0.7\textwidth]{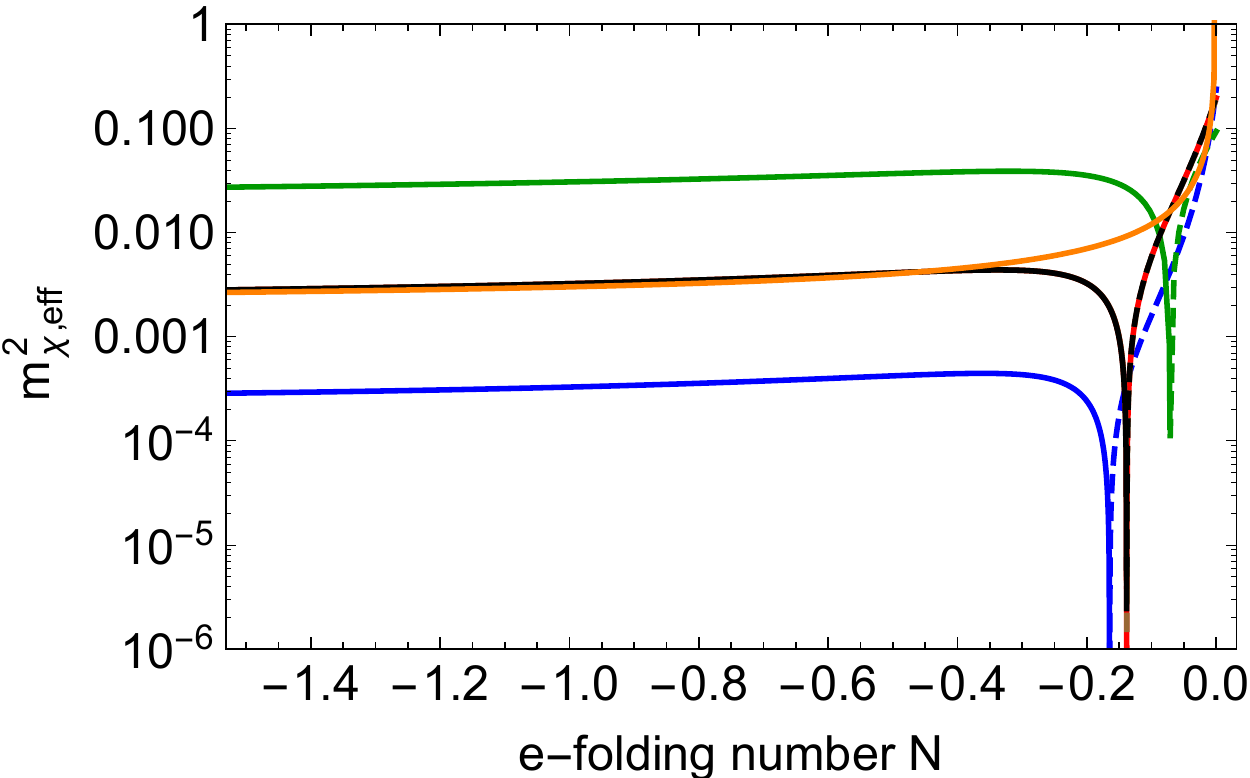}
\caption{
The super-horizon isocurvature effective mass-squared $m^2_{\chi,{\rm eff}}$ given in Eq.~\eqref{eq:meffchi2GD} for several values of $\alpha$ and $n$. In particular $\tilde\alpha=0.0001$ and $n=3/2$ (blue), $\tilde\alpha=0.001$ and $n=3/2$ (red), $\tilde\alpha=0.01$ and $n=3/2$ (green), $\tilde\alpha=0.001$ and $n=1$ (brown),
$\tilde\alpha=0.001$ and $n=10$ (black).
 The dotted parts show the negative part of $m^2_{\chi,{\rm eff}}$. {The three curves that correspond to $\tilde\alpha=0.001$ are visually indistinguishable.}
 The  orange line shows the slow-roll expression of Eq.~\eqref{eq:meffchi2GDSR} for $\tilde\alpha=0.001$.
 We see that the single field trajectory along $\chi=0$ is safe against geometric destabilization effects until close to the end of inflation.
}
 \label{fig:meffGD}
\end{figure}

A novel phenomenon that manifests itself in scalar field systems on a negatively curved manifold is ``geometrical destabilization" \cite{Renaux-Petel:2015mga}, where the presence of a negative field-space Ricci term can turn a stable direction into an unstable one. The study of the effective mass for the $\phi$ and $\chi$ fluctuations will be performed in Section~\ref{sec:tachyonicresonance}. In order to check the stability of the single-field trajectory, it suffices to use the effective super-horizon isocurvature mass
\beqn
\nonumber
m^2_{\chi,{\rm eff}} &=& V_{\chi\chi}(\chi=0) -{1\over 2} {4\over 3\alpha }\dot\phi^2 =
\\ &= &
\frac{  \left(2 \tanh^{2n}\left(\frac{|\phi (t)|}{\sqrt{6} \sqrt{\alpha }}\right) \left(3 \alpha +2 n \coth \left(\frac{\sqrt{\frac{2}{3}} \phi (t)}{\sqrt{\alpha }}\right) \text{csch}\left(\frac{\sqrt{\frac{2}{3}} \phi (t)}{\sqrt{\alpha }}\right)\right)\right)}{3  }-\frac{2 \dot\phi(t)^2 }{3 \alpha }
\label{eq:meffchi2GD}
\eeqn
During inflation and using the slow-roll conditions, we get
\beq
m^2_{\chi,{\rm eff}} =
2\alpha\left ( 1+{1\over 2N}\right)
\label{eq:meffchi2GDSR}
\eeq
which is positive.  However, close to the end of inflation the slow-roll approximation fails and the result cannot be trusted. Hence the model under study is safe against geometrical destabilization effects during inflation. The effective mass of isocurvature fluctuations can become negative after the end of inflation, but this falls under the scope of tachyonic preheating, as will be discussed in  Section~\ref{sec:tachyonicresonance}. Figure~\ref{fig:meffGD} shows the isocurvature effective mass-squared during the last $e$-folds of inflation, showing that it is indeed positive until very close to the end, hence no Geometrical Destabilization will occur\footnote{Recently Ref.~\cite{Christodoulidis:2018qdw} showed the existence of yet another possible evolution for $\alpha$-attractor models, angular inflation, where the background motion proceeds along the boundary of the Poincare disk. We did not see this behavior arise in the context of the two-field T-model studied here, even for highly curved manifolds.}. However Fig.~\ref{fig:meffGD} shows that all computations, either using linear analysis as the ones performed here, or full lattice simulations like in Ref.~\cite{Krajewski:2018moi}, must be initialized more than an $e$-fold before the end of inflation, where the effective isocurvature mass-squared is positive and the connection to the Bunch Davies vacuum is possible.

\subsection{Post-inflationary background oscillations}

\begin{figure}
\centering
\includegraphics[width=0.45\textwidth]{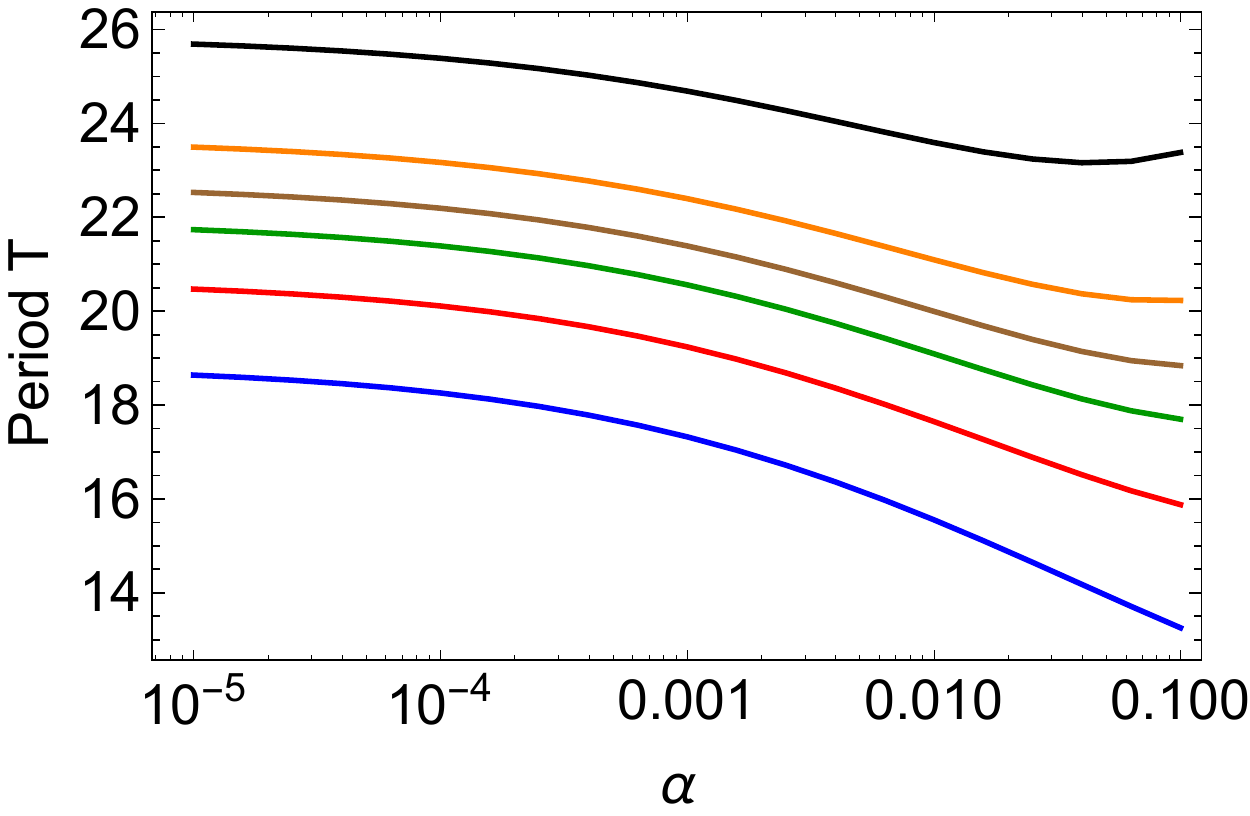} \includegraphics[width=0.45\textwidth]{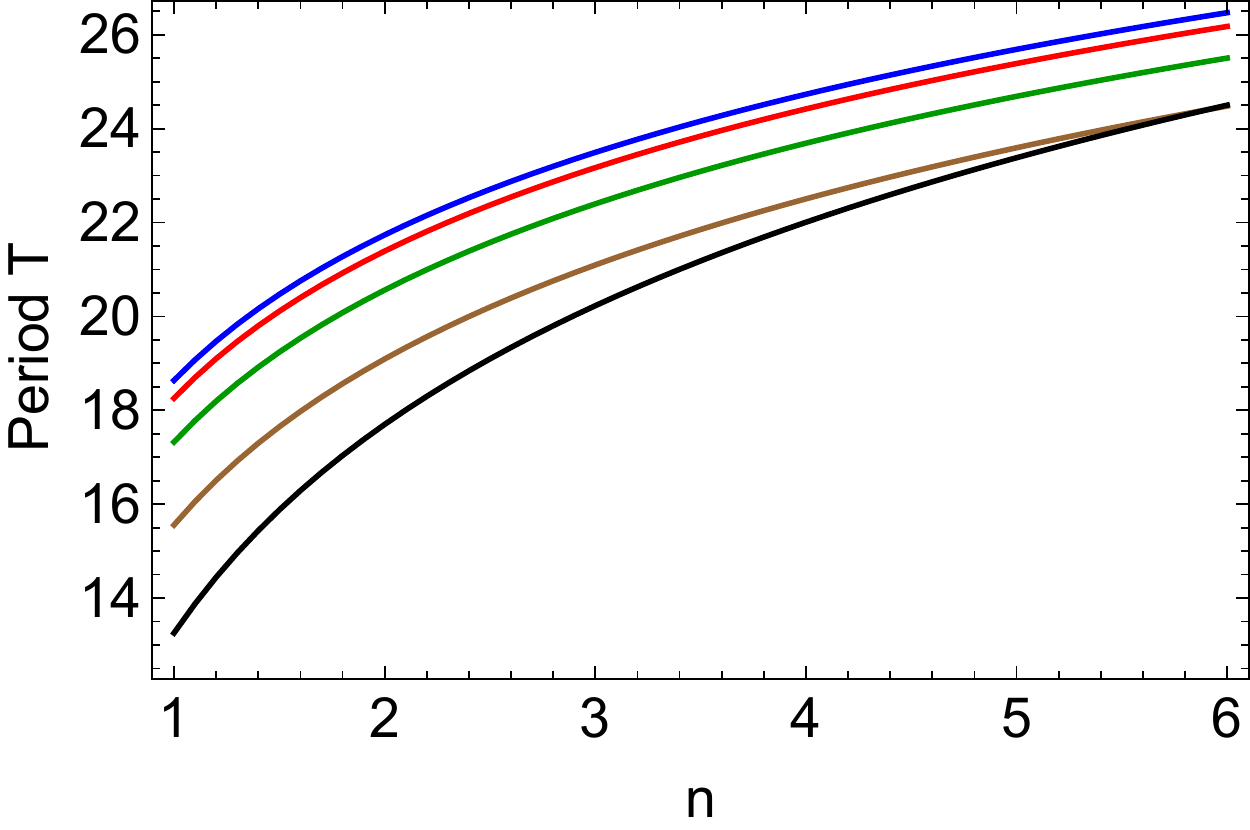}
\includegraphics[width=0.45\textwidth]{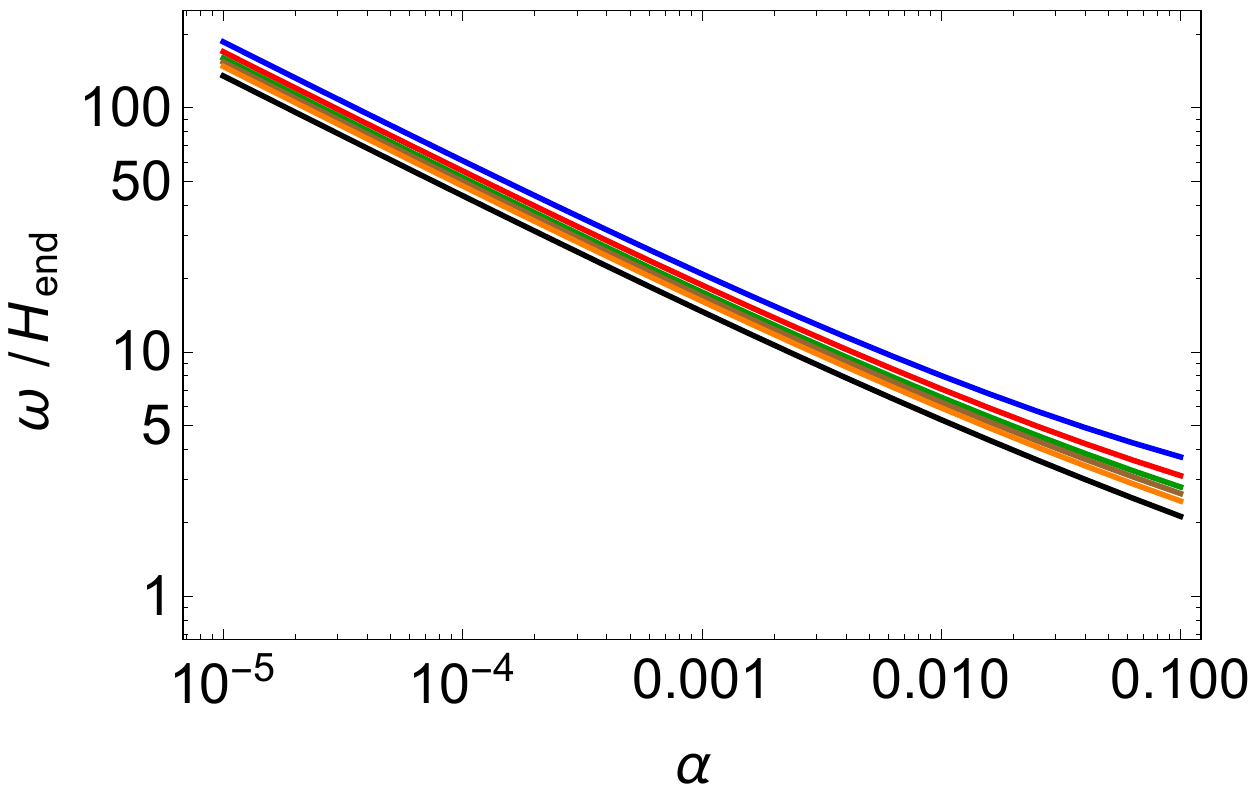} \includegraphics[width=0.45\textwidth]{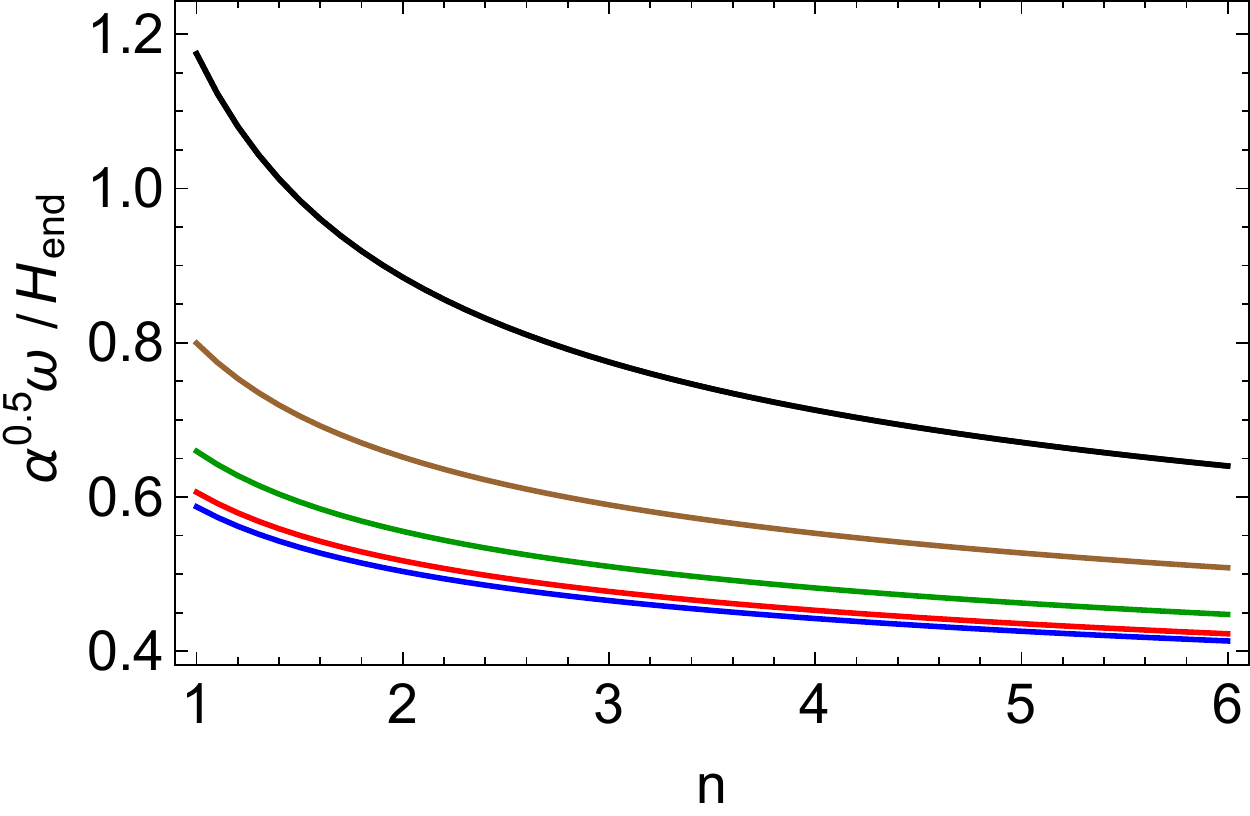}
\caption{
{\it Upper panels:} The background period $T$ as a function of $\alpha$ (left) and $n$ (right).
{\it Lower panels:}  The ratio of the background frequency $\omega = 2\pi /T$ to the Hubble scale at the end of inflation $H_{\rm end}$ as a function of $\alpha$ (left) and $n$ (right). Color-coding follows Fig.~\ref{fig:phiendvsnvsalpha}, specifically:\\
{\it Left:} $\alpha = 10^{-5}, 10^{-4}, 10^{-3}, 10^{-2}, 10^{-1}$ (blue, red, green, brown and black respectively).\\
 {\it Right:} $n = 1, 1.5, 2, 2.5, 3, 5$ (blue, red, green, brown, orange and black respectively)
 }
 \label{fig:periodend}
\end{figure}

In order to study the post-inflationary background evolution of the inflaton field $\phi(t)$, it is convenient to work in terms of  the rescaled field variable ${\delta \equiv \tilde \phi / \sqrt{\tilde \alpha}}$ and re-write the equation of motion for the inflaton field $\phi$ as
\beq
\label{eq:backgroundfield}
\ddot \delta + 3 H \dot {\delta }+   \mu^2 {{2\sqrt{2} \over 3}}
\, n  \cdot
 \text{csch}\left(
{
\sqrt{2\over3}} |\delta|
\right)
 \tanh ^{2n}   \left({1\over \sqrt{6}}
 |\delta| \right)=0
\eeq
where
\beq
\label{eq:backgroundHubble}
\left ({H\over \mu} \right )^2 = {\tilde \alpha \over 3 } \left [ {1\over 2}\left ({d \delta  \over d(\mu t)} \right )^2 +  \tanh^{2n} \left ( {|\delta| \over \sqrt{6}} \right ) \right]
\eeq
The field re-scaling leads to $\delta = {\cal O}(1)$ at the end of inflation and during preheating.
We see that the evolution of $\delta$, if one neglects the Hubble drag term, does  not depend on $\alpha$. This is reminiscent of non-minimally coupled models of inflation, where the background equation of motion approaches one ``master equation", when properly normalized, and thus the background motion is self-similar for large values of the non-minimal coupling $\xi$. In reality the background evolution has a mild dependence on $\alpha$, arising from the (very weak) dependence of $\delta_{\rm end}$  on $\alpha$, which is shown in Fig.~\ref{fig:phiendvsnvsalpha}. Fig.~\ref{fig:periodend} shows the period of background oscillations, if we neglect the Hubble drag and initialize the oscillation at $\delta_{\rm init} = \phi_{\rm end} / \sqrt{ \alpha}$. We see that the period $T\sim 10$. More importantly, there is a significant separation of scales between the background oscillation frequency $\omega = 2\pi/T$ and the Hubble scale. The relation can be roughly fitted as $\omega / H_{\rm end} \sim 1/\sqrt{\tilde \alpha}$. This shows that there are more background oscillations per Hubble time (or per $e$-fold) for smaller values of $\tilde \alpha$, hence for highly curved field-space manifolds.

\section{Tachyonic resonance}
\label{sec:tachyonicresonance}

\subsection{Fluctuations}

The covariant formalism that must be used to study the evolution of fluctuations in models comprised of multiple scalar fields on a curved manifold has been developed and presented in Refs.~\cite{Gong:2011uw, KMS},
described in detail in Ref.~\cite{DeCross:2015uza} and extensively used in Refs.~\cite{DeCross:2016fdz, DeCross:2016cbs, Sfakianakis:2018lzf} for studying preheating in multi-field inflation with non-minimal couplings to gravity. The gauge-invariant perturbations obey
\beq
{\cal D}_t^2 Q^I + 3 H {\cal D}_t Q^I + \left[ \frac{ k^2}{a^2} \delta^I_{\> J} + {\cal M}^I_{\> J} \right] Q^J = 0 ,
\label{eq:eomQ}
\eeq
where the mass-squared matrix is given by
\beq
{\cal M}^I_{\> J} \equiv {\cal G}^{IK} \left( {\cal D}_J {\cal D}_K V \right) - {\cal R}^I_{\> LMJ} \dot{\varphi}^L \dot{\varphi}^M - \frac{1}{M_{\rm pl}^2 a^3} {\cal D}_t \left( \frac{ a^3}{H} \dot{\varphi}^I \dot{\varphi}_J \right)
\label{MIJdef}
\eeq
and ${\cal R}^I_{\> LMJ}$ is the Riemann tensor constructed from ${\cal G}_{IJ} (\varphi^K)$.
For the model at hand, where the background motion is restricted along the $\chi=0$ direction,
the field-space structure simplifies significantly ${\cal G}_{IJ}(\chi=0) = \delta_{IJ}$ and $\Gamma^I_{JK} = 0$, hence all covariant derivatives become partial derivatives and the quantization of the fluctuations proceeds as usual. This is not the case for other parametrizations of the field-space, or other background trajectories, where ${\cal G}_{IJ} \neq \delta_{IJ}$, and one would have to use the field-space vielbeins to properly quantize the system, as done for example in Ref.~\cite{DeCross:2015uza}.

\begin{figure}
\centering
\includegraphics[width=0.3\textwidth]{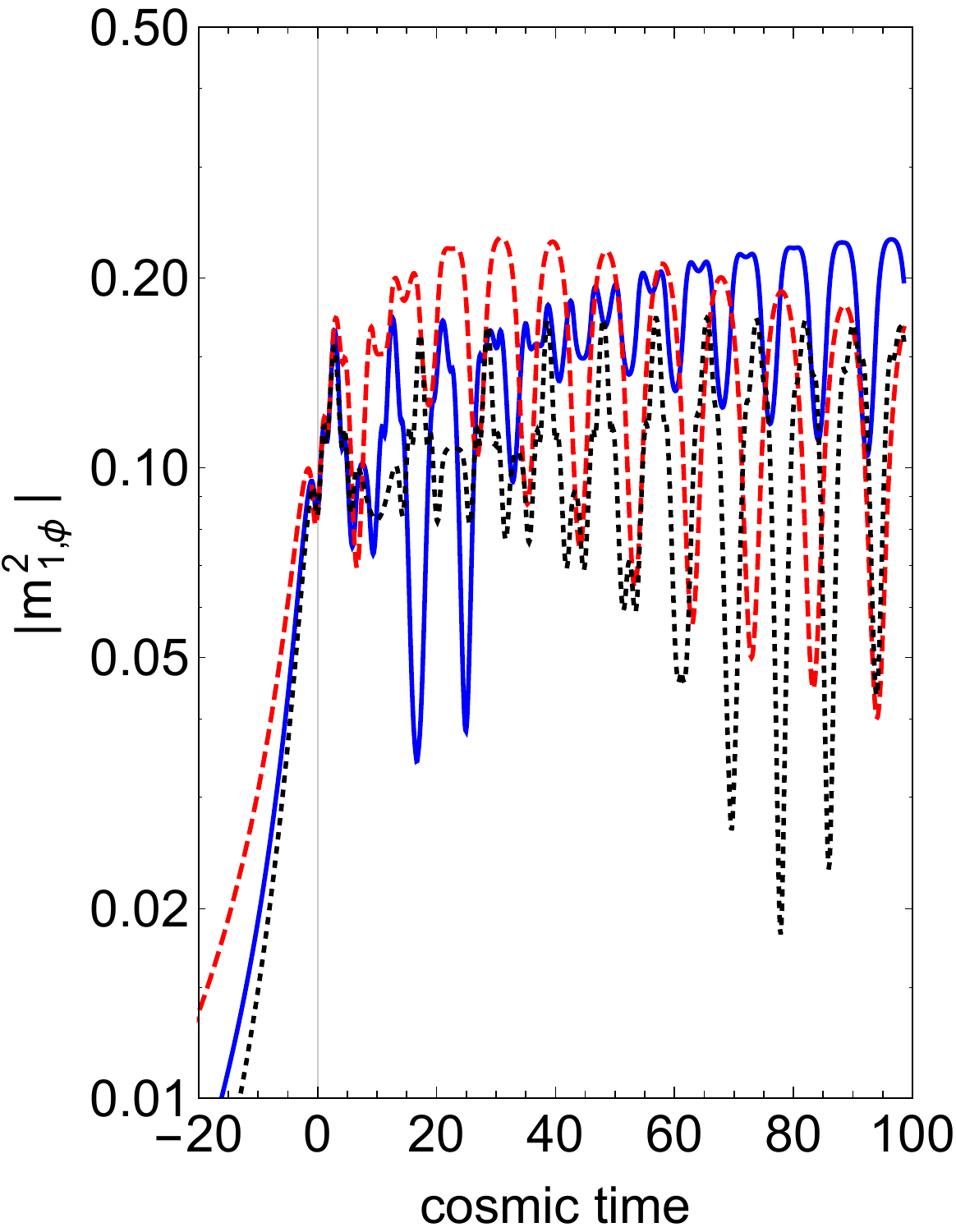} \includegraphics[width=0.3\textwidth]{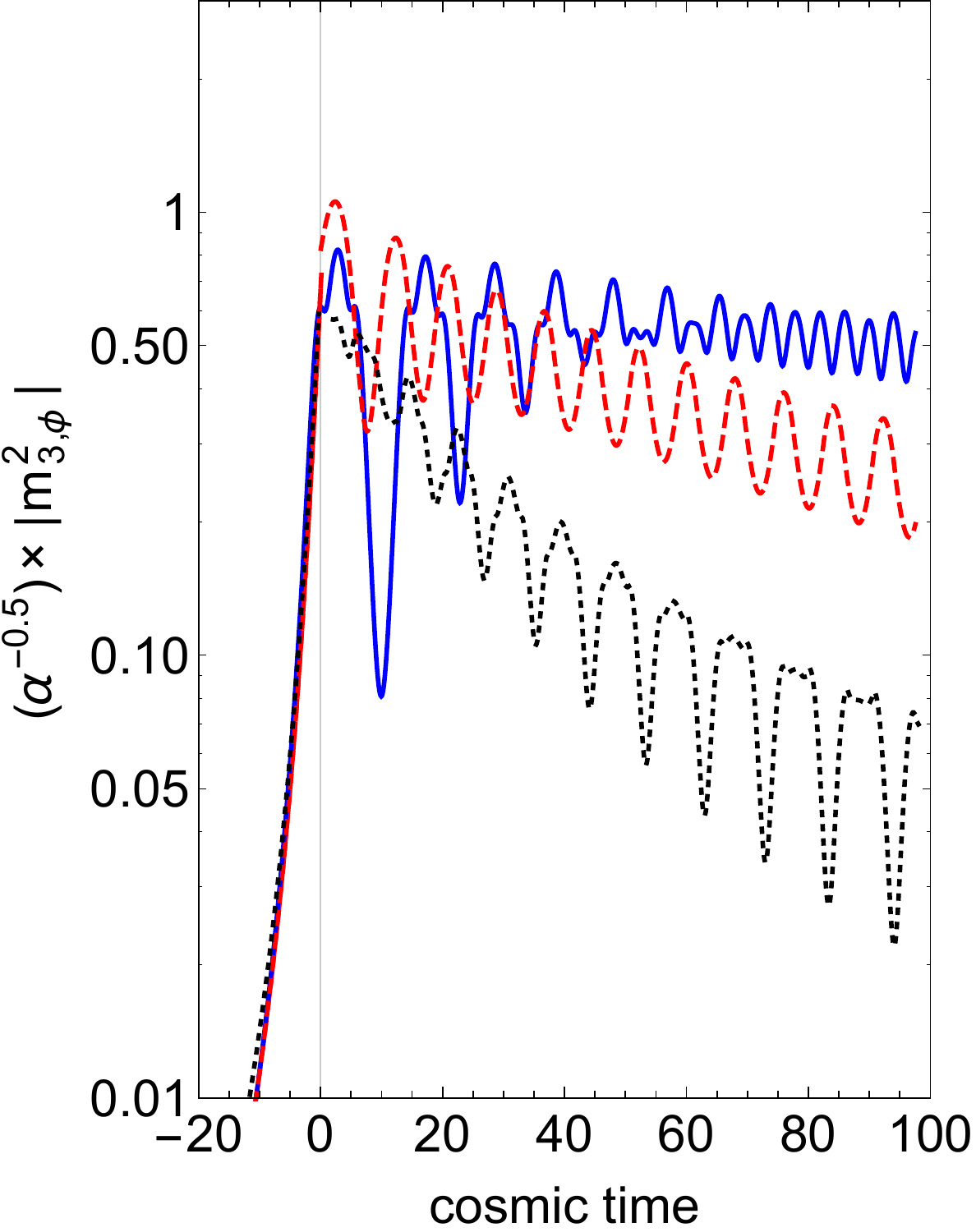}
\includegraphics[width=0.3\textwidth]{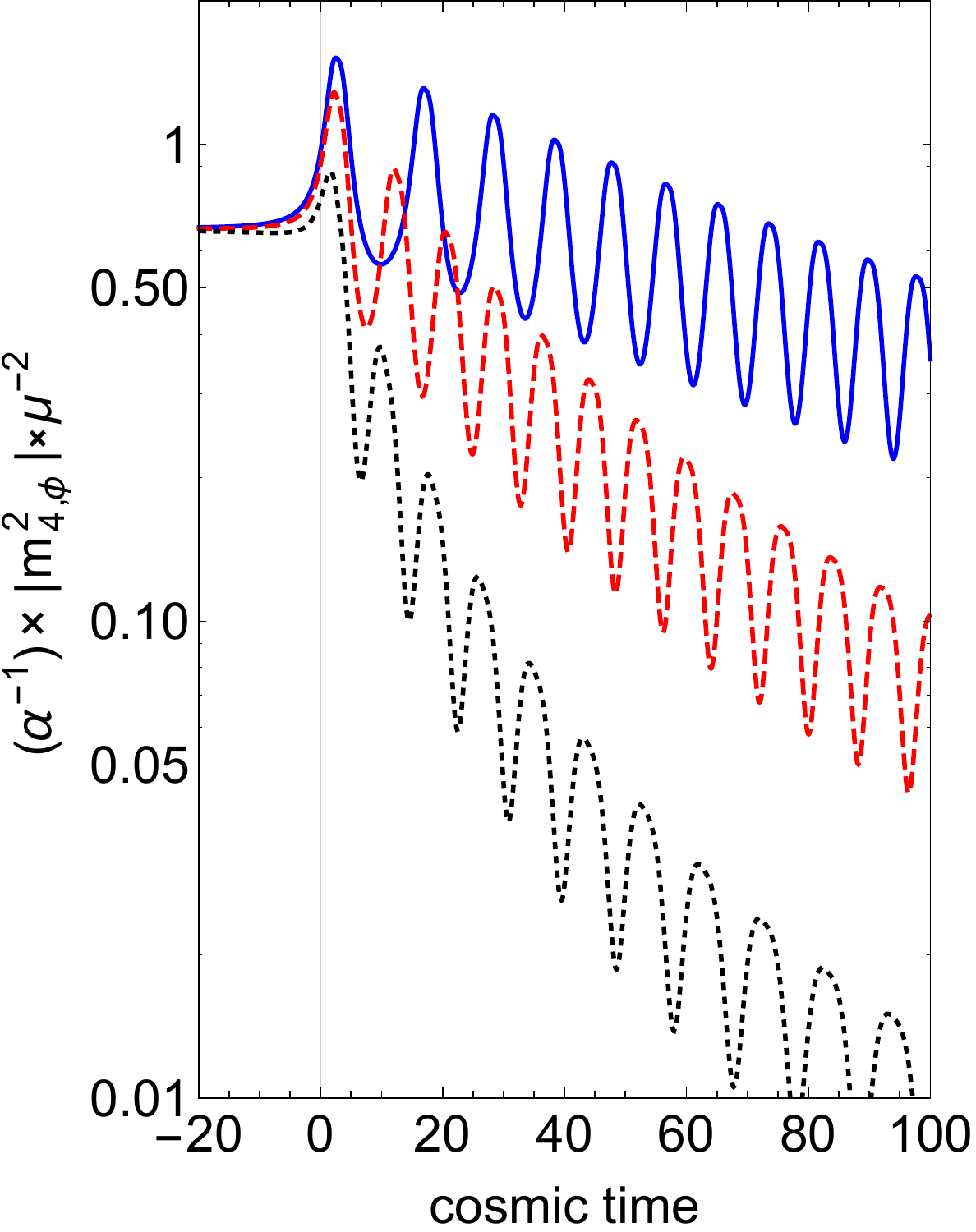}
\caption{
The absolute values of the non-zero components of $m^2_{{\rm eff},\phi}$, (left to right: $m^2_{1,\phi}$, $m^2_{2,\phi} / \sqrt{\alpha}$ and $m^2_{3,\phi} /\alpha$) properly rescaled to showcase the scalings of Eq.~\eqref{eq:mIphiscale}
for $n=3/2$ and $\tilde \alpha=10^{-2},10^{-3},10^{-4}$ (black, red and blue respectively). The fact that all curves within each panel have similar values at the end of inflation is a numerical validation of the scalings shown in Eq.~\eqref{eq:mIphiscale}.
{The curves on the left and middle panels for $t>0$ are generated through using a moving average window on the values of $|m_{\{2,3\},\phi}^2|$. Without this smoothing the curves would exhibit large oscillations and hence would overlap and be very hard to distinguish. The information lost is not important, since at this point we are interested in the scaling properties of the effective mass components, not their exact form.}
 }
 \label{fig:mIphiPLOT}
\end{figure}

We rescale the perturbations as $Q^I (x^\mu) \rightarrow X^I (x^\mu) / a(t)$ and
 work in terms of conformal time, $d \eta = dt / a(t)$. This allows us to write the quadratic action in a form that resembles Minkowski space, which makes their quantization  straightforward.
The quadratic action becomes
\beq
S_2^{(X)} = \int d^3x d\eta \left [ -{1\over 2} \eta^{\mu\nu} \delta_{IJ} \partial_\mu X^I \partial_\nu X^J -{1\over 2}\mathbb{M}_{I J} X^I X^J \right ] \, ,
\eeq
where
\beq
\mathbb{M}_{I J} = a^2 \left ( {\cal M}_{IJ} - {1\over 6} \delta_{IJ} R\right)
\eeq
and $R$ is the space-time Ricci scalar. The energy density of the two fields in momentum-space becomes
\beq
\rho_k^{(X)} = {1\over 2} \delta_{IJ} \partial_\eta X^I_k \partial_\eta X_k^J +{1\over 2} [\omega^2_k(\eta)]_{IJ} X^I_k X^J_k  = {1\over 2} \delta_{IJ} \left [
\partial_\eta X^I_k \partial_\eta X_k^J
- (\partial^2_\eta X^I)X^J \right ]
\eeq
where we used the equation of motion for the second equality and defined the effective frequency-squared as
\beq
[\omega^2_k(\eta)]_{IJ}  = k^2 \delta_{IJ} + \mathbb{M}_{I J}
\eeq
We promote the fields $X^I$ to operators $\hat X^I$
and expand $\hat{X}^\phi$ and $\hat{X}^\chi$ in sets of creation and annihilation operators and associated mode functions
\beq
\hat X^I = \int {d^3k \over (2\pi)^{3/2}} \left [u^I(k,\eta)\hat a e^{i k\cdot x} +  u^{I*}(k,\eta)\hat a^\dagger e^{-i k\cdot x} \right ] \, .
\eeq
and we define $u^\phi \equiv v$ and $u^\chi \equiv z$.
Since the modes decouple on a single-field background with vanishing turn-rate, the equations of motion are
\beqn
\begin{split}
\partial^2_\eta{v}_k & + \omega_\phi^2(k,\eta) v_k \simeq 0 \, , \quad \omega_\phi(k,\eta)^2=k^2 + a^2 m_{\rm eff, \phi}^2 \, ,
 \\
\partial^2_\eta{z}_k & + \omega_\chi^2(k,\eta)z_k \simeq 0 \, , \quad \omega_\chi(k,\eta)^2=k^2 + a^2 m_{\rm eff, \chi}^2 \, .
\end{split}
\label{eq:vzeom}
\eeqn
The effective masses of the two types of fluctuations, along  the background motion and perpendicular to it, consist of four distinct contributions \cite{DeCross:2015uza}:
\beq
m_{ {\rm eff}, I}^2 = m_{1, I}^2 + m_{2, I}^2 + m_{3,I}^2 + m_{4,I}^2 ,
\label{eq:meffgeneral}
\eeq
with
\beqn
\begin{split}
m_{1, I}^2 &\equiv {\cal G}^{I K} \left( {\cal D}_I {\cal D}_K V \right) , \\
m_{2, I}^2 &\equiv - {\cal R}^I_{\> LM I} \dot{\varphi}^L \dot{\varphi}^M , \\
m_{3, I}^2 &\equiv - \frac{1}{ M_{\rm pl}^2 a^3} \delta^I_{\> K} \delta^J _{\> I} \> {\cal D}_t \left( \frac{a^3}{H} \dot{\varphi}^K \dot{\varphi}_J \right) , \\
m_{4, I}^2 &\equiv - \frac{1}{6} R = (\epsilon -2) H^2 \, .
\end{split}
\label{miphi}
\eeqn
The various component of the effective mass-squared arises from a different source:
\begin{itemize}
\item $m_{1,I}^2$ is the usual effective mass term derived from the  curvature of the potential around the minimum.
\item $m_{2,I}^2$ comes from the geometry of field-space and has no analogue in models with a trivial field space.
\item $m_{3,I}^2$ arises due to the presence of coupled metric perturbations by considering linear fluctuations in the metric as well as in the fields. This contribution vanishes in the limit of infinitely rigid space-time.
\item $m_{4,I}^2$ encodes the curvature of space-time.
\end{itemize}
In general $m_{3, \chi}^2 = 0 = m_{2, \phi}^2$, since the coupled metric fluctuations described by $m_{3, I}^2$ only affect the adiabatic modes $\delta\phi$, while the field-space curvature described by $m_{2, I}^2$ only affects the isocurvature modes\footnote{The terms ``adiabatic" and ``isocurvature" refer to fluctuations along and perpendicular to the background trajectory respectively.} $\delta\chi$.
In our case, both $m_{3, I}^2$ and $m_{4, I}^2$ are subdominant for highly curved field spaces $\tilde \alpha \ll 1$, as can be seen from the various scalings of the terms in Eq.~\eqref{eq:meffgeneral}
\beqn
\begin{split}
 m_{1, \phi}^2& \sim \mu^2
\\
 m_{3,\phi}^2 &\sim \mu^2 \sqrt{\tilde \alpha}
\\
m_{4,\phi}^2&=m_{4,\chi}^2 \sim \mu^2 \tilde\alpha
\end{split}
\label{eq:mIphiscale}
\eeqn
The small value of $m_{4,I}^2$ is one further indication that fluctuations behave almost as if they were in flat spacetime.
These scalings agree very well with numerical evaluations for a large range of $\tilde \alpha$, as shown in Fig.~\ref{fig:mIphiPLOT}.
 A closer analysis of scaling relations for $m_{ {\rm eff}, \chi}^2$, will be performed in Section \ref{sec:effectivefrequency}.
 Meanwhile, within the single-field attractor along $\chi = 0$, the energy densities for adiabatic and isocurvature perturbations take the simple form \cite{DeCross:2015uza}
\beqn
\begin{split}
\rho_k^{(\phi)} &= \frac{1}{2} \left[ \vert {v}'_k \vert^2 + \left( k^2+a^2 m_{\rm eff,\phi}^2 \right) \vert v_k \vert^2 \right] , \\
\rho_k^{(\chi)} &= \frac{1}{2} \left[ \vert {z}'_k \vert^2 + \left( k^2+a^2 m_{\rm eff,\chi}^2 \right) \vert z_k \vert^2 \right] , \\
\end{split}
\label{eq:rhodef}
\eeqn
where we thus approximate the two effective masses as
\beqn
m_{{\rm eff},\phi}^2 &\simeq& V_{\phi\phi} (\chi=0)
\\
m_{{\rm eff},\chi}^2  &\simeq& V_{\chi\chi}  (\chi=0) + {1\over 2} {\cal R}\dot\phi^2
\eeqn
where  ${\cal R} = -4/3\alpha$ is the field space Ricci curvature scalar and we dropped the subdominant terms.
We must keep in mind that $Q^\phi \sim v_k / a(t)$ and $Q^\chi \sim z_k / a(t)$. We measure particle production with respect to the instantaneous adiabatic vacuum \cite{AHKK}.
The initial conditions for preheating can be read off from Eq.~\eqref{eq:vzeom}, using the WKB approximation and starting during inflation, when the effective mass is positive
\beqn
v_k^{\rm init} &=& {1\over \sqrt{2\omega_{\phi}(k,\eta)}} e^{-i\int^\eta_{\eta_0} \omega_{\phi}(k, \eta') d\eta'}
\label{eq:WKBv}
\\
z_k^{\rm init} &=& {1\over \sqrt{2\omega_{\chi}(k,\eta)}} e^{-i \int^\eta_{\eta_0} \omega_{\chi}(k, \eta') d\eta'}
\label{eq:WKBz}
\eeqn
In the far past $a(\eta)\to 0$, hence $\{\omega_{\phi}(k,\eta), \omega_{\chi}(k,\eta)\} \to k$, which makes the solutions of Eqs.~\eqref{eq:WKBv} and \eqref{eq:WKBz} match to the Bunch-Davies vacuum during inflation.

Since we will be performing the computations in cosmic time, we write the equations of motion for the two types of fluctuations.
The fluctuation equation for the $\phi$ field (adiabatic direction) is
\beq
\ddot Q_\phi + 3 H \dot Q_\phi + \left [ {k^2\over a^2} + V_{\phi\phi} \right ]Q_\phi =0
\eeq
where we neglected the term arising from the coupled metric fluctuations, that is proportional to $M_{\rm Pl}^{-2}$. We again rescale time by $\mu$ giving us
\beq
{d^2 Q_\phi\over d(\mu t)^2} + 3 {H\over \mu} {d Q_\phi\over d(\mu t)} + \left [ {(k/\mu)^2\over a^2} + {V_{\phi\phi}\over \mu^2} \right ]Q_\phi =0
\eeq
where the potential-dependent term of the effective frequency is
\beq
{V_{\phi\phi}\over \mu^2} =\sqrt{\tilde \alpha} {d^2\over d\tilde\phi^2} \left  [ \tanh^{2n}\left (|\tilde\phi| \over \sqrt{6 \tilde \alpha} \right ) \right]
\eeq
The results for the isocurvature modes $\delta\chi$ or $Q_\chi$ are similar
\beq
{d^2 Q_\chi \over d(\mu t)^2}+ 3 {H\over m} {d Q_\chi\over d(\mu t)} +\left [ {(k/\mu)^2\over a^2} + {V_{\chi\chi}\over \mu^2} + {1\over 2} \left ({d \phi\over d(\mu t)}\right)^2 {\cal R} \right ]Q_\chi =0
\eeq
The last term in the above equation is the Riemann contribution to the effective mass-squared $\omega^2_\chi$
\beq
m_{2,\chi}^2 = -{\cal R}^\chi_{\phi\phi\chi} \dot \phi^2 = -{2 \over 3\alpha}\dot\phi^2 = {1\over 2} {\cal R}\dot\phi^2 \, .
\eeq
Since the self-resonance of $\delta \phi$ modes in these models has been extensively studied (see for example Ref.~\cite{MustafaDuration}), we will focus our attention on  $\delta \chi$ fluctuations, which can undergo tachyonic excitation, which is generally more efficient than parametric amplification. Also the excitation of $\delta\chi$ modes is  a truly multi-field phenomenon that depends crucially on the field-space geometry.

\subsection{Effective frequency}
\label{sec:effectivefrequency}

We examine in detail the effective frequency-squared for the $\delta\chi$ fluctuations, $\omega_\chi^2(k,t)$. For simplicity we will focus on the case of $n=3/2$, which matches the potential used in the lattice simulations presented in Ref.~\cite{ Krajewski:2018moi}. The generalization of our results for other potentials is discussed in Section~\ref{sec:potentials}.

\begin{figure}
\centering
\includegraphics[width=0.45\textwidth]{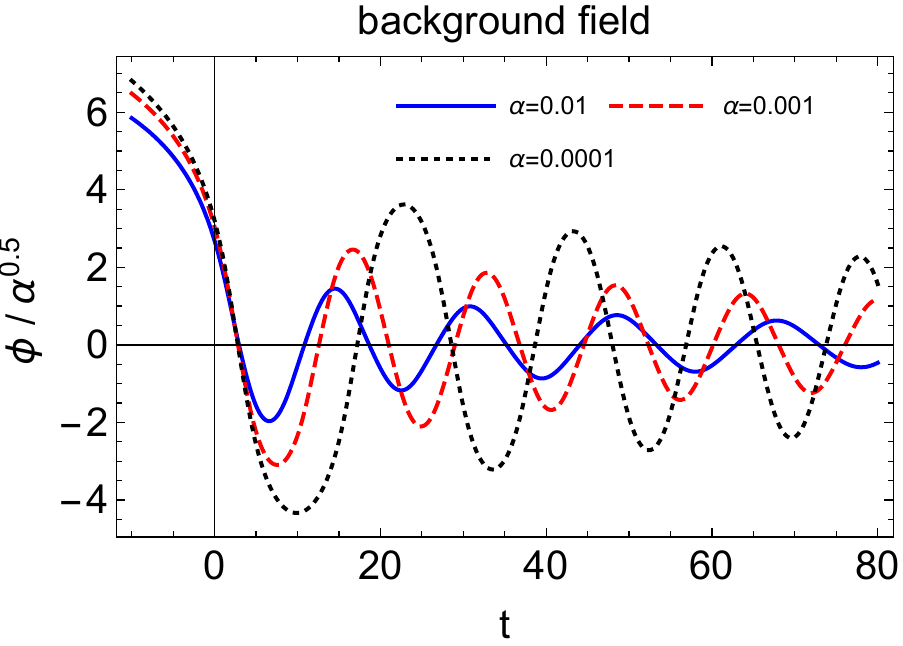} \includegraphics[width=0.45\textwidth]{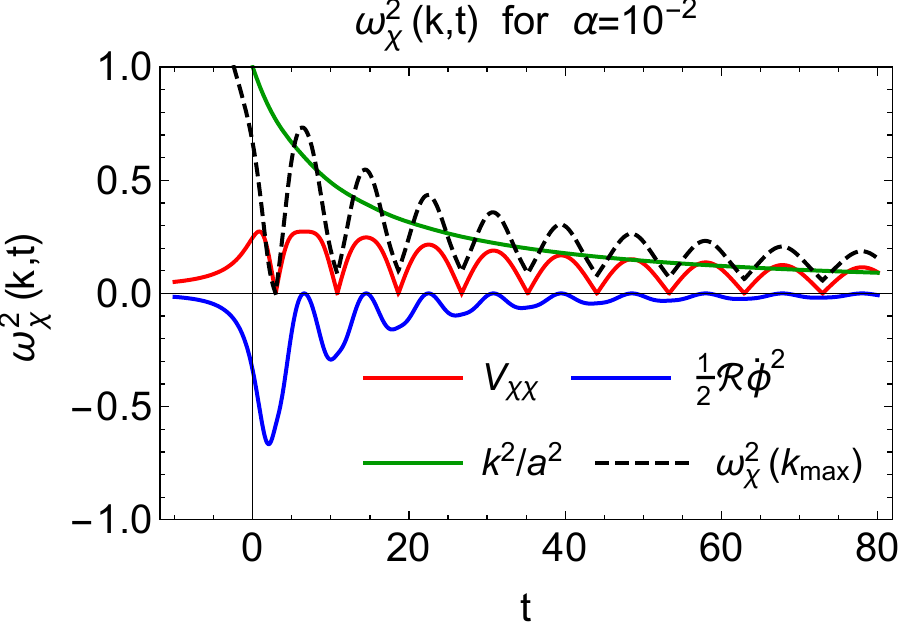}
\includegraphics[width=0.45\textwidth]{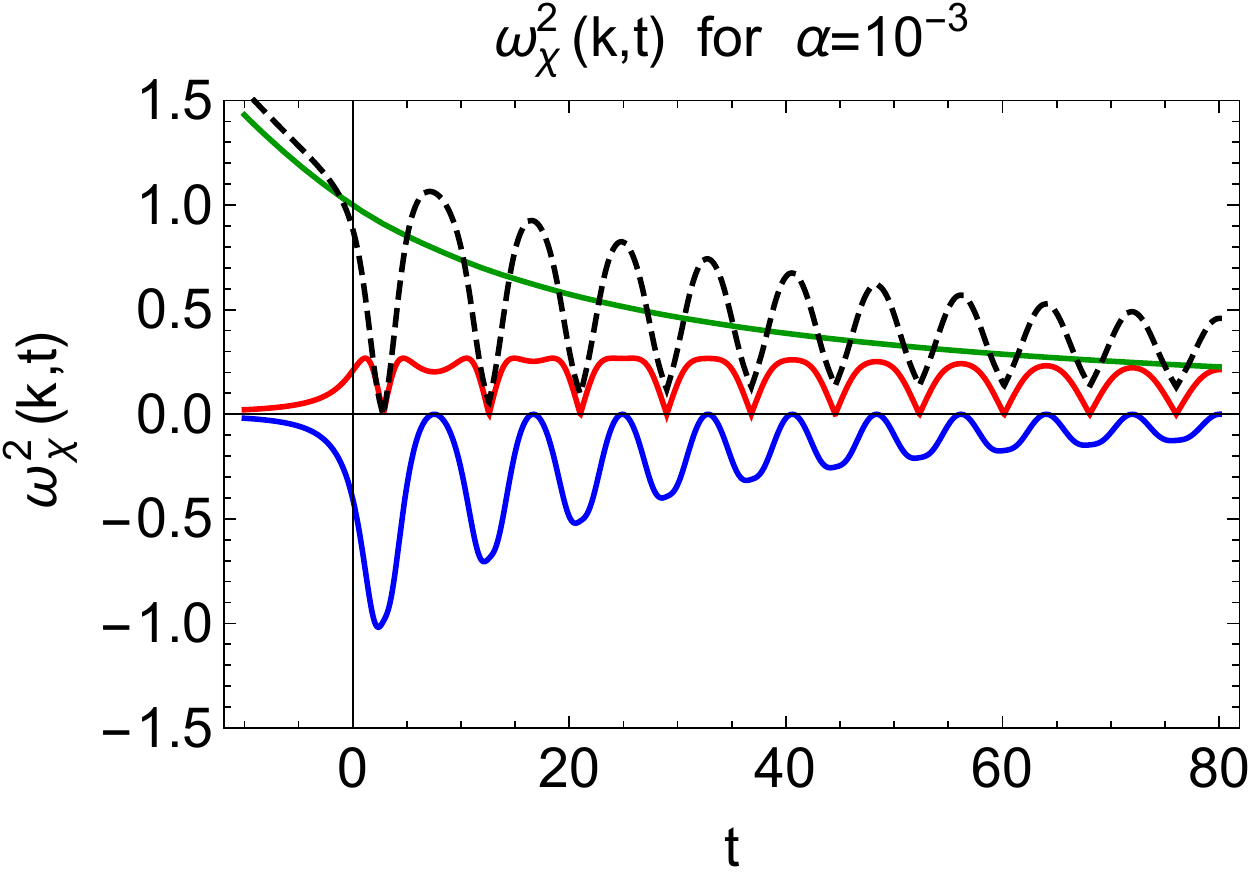} \includegraphics[width=0.45\textwidth]{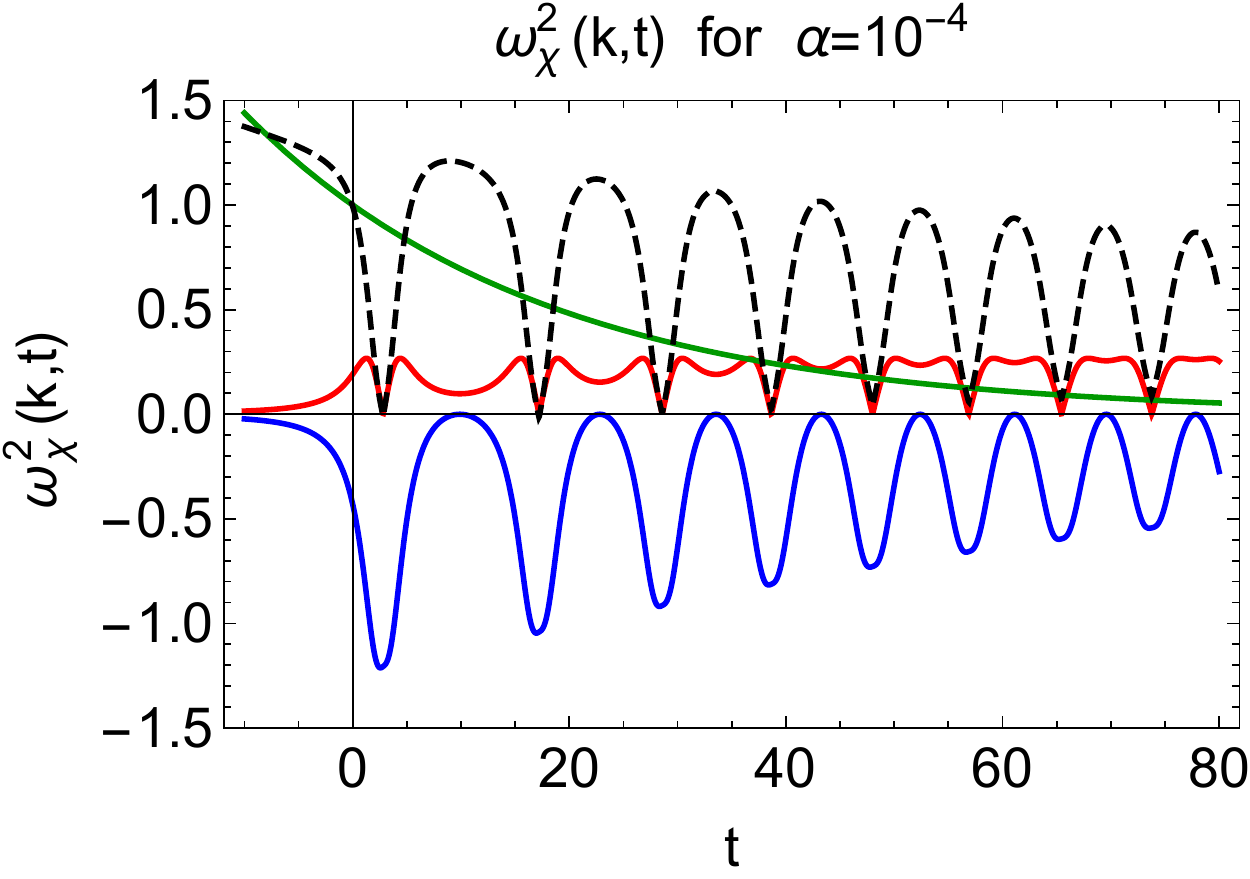}
\caption{The rescaled background field (top left) $\phi/\sqrt{\alpha}$ as a function of time for $\tilde\alpha=10^{-2},10^{-3},10^{-4}$ (blue, red-dashed and black-dotted respectively). The other three plots correspond to the isocurvature effective frequency-squared for the maximal marginally amplified wavenumber $k_{\rm max}$ (black-dotted), along with $(k/a)^2$ (green), the potential contribution (red) and the tachyonic Riemann term (blue).
 }
 \label{fig:omegachivstime}
\end{figure}

In the top left panel of Fig.~\ref{fig:omegachivstime} we  see the evolution of the background field $\phi(t)$, rescaled as $\delta(t)=\phi(t) / \sqrt{ \alpha}$ after the end of inflation and we take $t=0$ as the end of inflation. We see that inflation ends at $\phi(t) / \sqrt{\alpha} \simeq 3$ for all three cases considered here, consistent with Fig.~\ref{fig:phiendvsnvsalpha}. The main difference is both the frequency of oscillation and the decay of the amplitude of the background for different values of $\alpha$.

The maximum tachyonically excited wavenumber for the various cases under consideration is $k_{\rm max}\simeq 0.87 \mu$ for $\tilde\alpha=10^{-2}$, $k_{\rm max}\simeq 1.04 \mu$ for $\tilde \alpha=10^{-3}$ and $k_{\rm max}\simeq 1.11 \mu$ for $\tilde\alpha=10^{-4}$. So we can say\footnote{In the units of Appendix C and Ref. \cite{Krajewski:2018moi}, this corresponds to $k_{\rm max} \simeq {1\over \sqrt{\alpha}} M^2 /M_{\rm Pl}$, leading to $k_{\rm max} \simeq 33 M^2 /M_{\rm Pl}$ for $\alpha =10^{-3}$ and $k_{\rm max} \simeq 100 M^2 /M_{\rm Pl}$ for $\alpha =10^{-4}$. This is consistent with Figure 5 of Ref.~\cite{Krajewski:2018moi}.
} that $k_{\rm max} \simeq \mu$ for all values of $\tilde\alpha \ll 1$.
Furthermore, we see that background motion corresponding to larger values of $\tilde\alpha$ shows greater damping. This is consistent with the observation that $H_{\rm end} \sim \sqrt{\tilde \alpha}$, hence the Hubble damping term is smaller for highly curved field-space manifolds.

Examining the tachyonic contribution to $\omega^2_\chi(k)$,  a very simple scaling emerges
\beq
{1\over 2} {\cal R} \dot \phi^2 = -{1\over 2} \alpha \dot \delta^2 {4\over 3\alpha } = -{2\over 3}  \dot \delta^2 = {\cal O}(1) \,
\eeq
This is again consistent with Fig.~\ref{fig:omegachivstime}, especially as the value of $\tilde \alpha$ gets smaller.

\begin{figure}
\centering
 \includegraphics[width=0.45\textwidth]{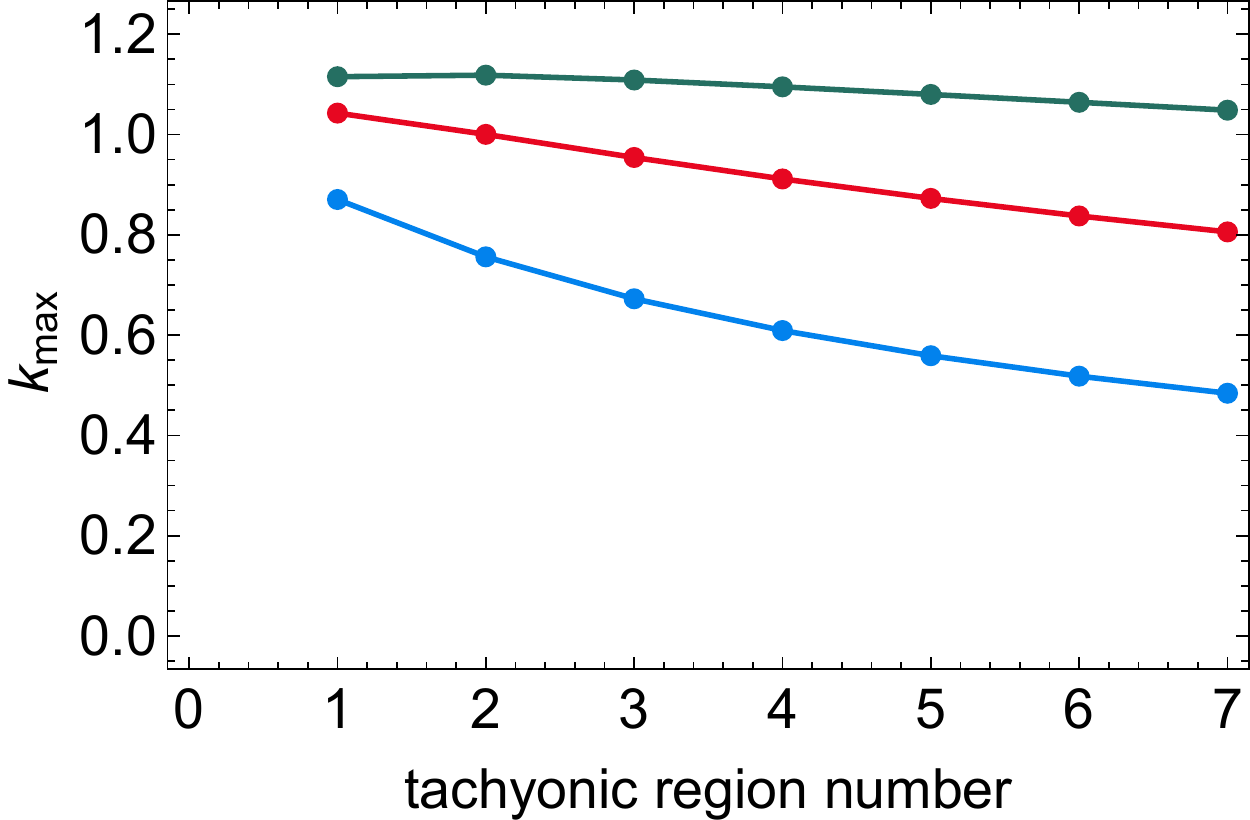}
\includegraphics[width=0.45\textwidth]{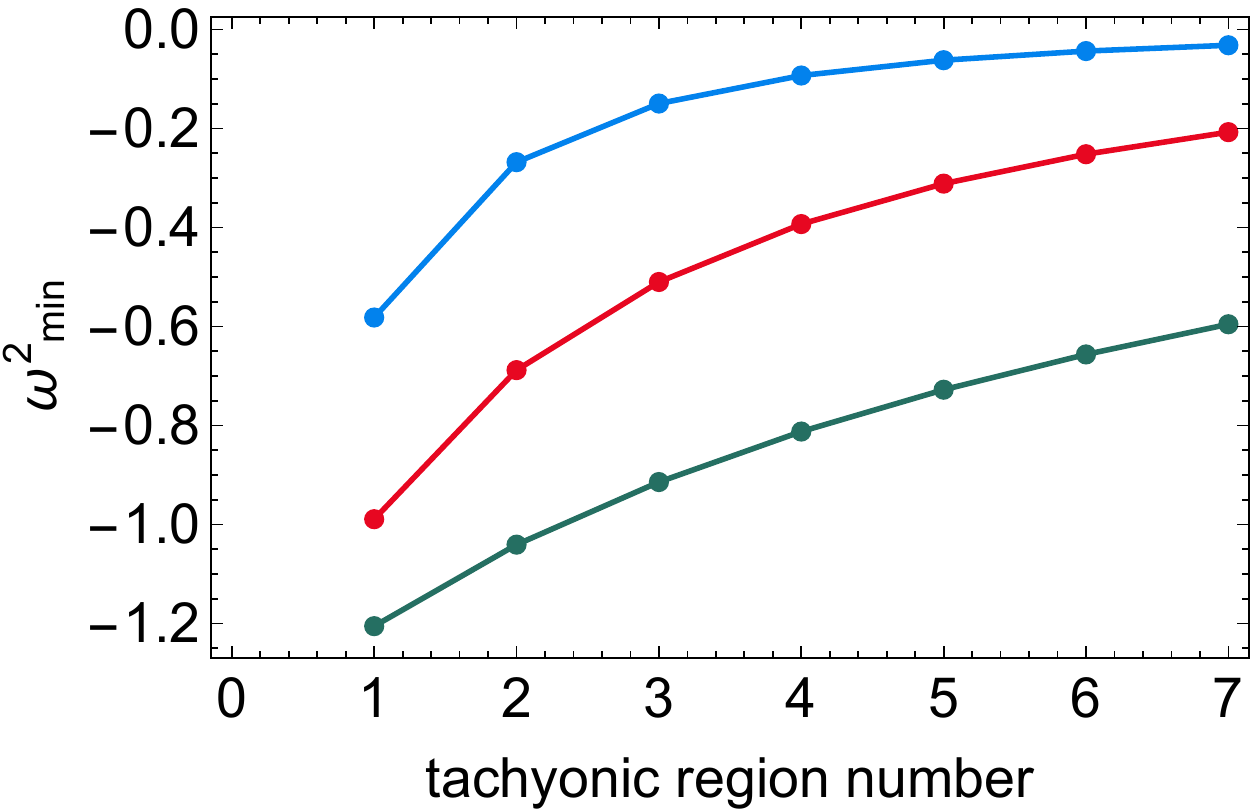}
\caption{
 {\it Left:} The dependence of  the maximum excited wavenumber $k_{\rm max}$ on the number of tachyonic regions for  $n=3/2$ and  $\tilde\alpha=10^{-2},10^{-3} ,10^{-4}$ (blue, red and green respectively).\\
 {\it Right:} The  minimum (maximally negative) value of the effective frequency-squared of $\chi$ fluctuations $\omega^2_{\rm min}$ as  a function of the number of tachyonic regions for the same parameters and color-coding.
 }
 \label{fig:kmaxOmegamin}
\end{figure}

Since the tachyonic contribution to the effective mass-squared is similar for models with different values of $\tilde \alpha$, the tachyonic amplification of the relevant mode-functions after each oscillation will be also similar. Fig. \ref{fig:kmaxOmegamin} shows the evolution of  $\omega^2_{\rm min}$ and $k_{\rm max}$ for each subsequent tachyonic region.
It is worth emphasizing that $\omega^2_{\rm min}$ is determined solely by the corresponding minimum (maximum negative) value of $m_{2,\chi}^2$. The maximum negative value of $m_{2,\chi}^2$ occurs when $|\dot \phi|$ is maximized, or equivalently when $\phi=0$. At this point, the potential can be Taylor-expanded as
\beq
V(\phi=0,\chi) \approx {M^4\over 4^n}  \beta^{2n} |\chi|^{2n} \, .
\label{eq:Vphieq0chi}
\eeq
For $n>1$ the effective mass component vanishes for $\chi=0$.
In particular for $n=3/2$ the effective mass component becomes $\partial^2_\chi V(\phi=0,\chi) \sim |\chi|$, as shown in Fig.~\ref{fig:omegachivstime}.
The case of $n=1$ is different and we consider it in Section~\ref{sec:potentials}.
 We see that both the maximum (negative) contribution of $m_{2,\chi}^2$, as well as the range of tachyonically excited wavenumbers decrease faster for larger values of $\tilde\alpha$, or less curved field-space manifolds. This can be traced back to the dependence of the Hubble scale on the field-space curvature, which scales as $H \propto \sqrt{\tilde\alpha}$. Hence in the first $e$-fold, or within the first Hubble-time after inflation, lower values of $\tilde \alpha$ will result in a larger number of tachyonic bursts and hence a larger overall amplification. Furthermore, a larger Hubble term for larger values of $\tilde \alpha$ will result in a faster red-shifting of the background field amplitude $\delta(t)$, resulting in a faster suppression of the parametric resonance, in line with Fig.~\ref{fig:omegachivstime}.

\subsection{WKB results }
\label{sec:WKBresults}

We  use the WKB analysis as described in Ref.~\cite{WKBtachy}, in order to make analytical progress in computing the amplification of the $\delta\chi$ modes during tachyonic preheating. In contrast to Refs.~\cite{Freese:2017ace, Adshead:2016iae, Adshead:2015pva}, where tachyonic preheating lasted for a few inflaton oscillations at most, in the present case, multiple inflaton oscillations might be required, in order to siphon enough energy from the inflaton into radiation modes. However, given the fact that the Hubble time is much larger than the period of oscillations, preheating will still be almost instantaneous in terms of  the number of $e$-folds. Based on $\omega / H_{\rm end} \sim 1/\sqrt{\tilde \alpha}$, we can estimate the number of background oscillations occurring during the first $e$-fold of preheating to be $N_{\rm osc.} \sim 0.2 / \sqrt{\tilde \alpha}$.

We neglect the effect of the expansion of the Universe, hence taking $H=0$. This is an increasingly good approximation for smaller values of $\tilde \alpha$, since $H_{\rm end} \sim \sqrt{\tilde \alpha}$. Furthermore, the static universe WKB analysis will provide a useful comparison to the Floquet analysis of Section \ref{sec:floquetcharts}. The equation of motion for the fluctuations in the $\chi$ field becomes\footnote{For the remainder of this work we denote the fluctuations of the $\chi$ field as $\chi_k$ rather than $\delta\chi_k$ for notational simplicity.}
\beq
\partial^2_t {\chi}_k  + \omega_\chi^2(k,t) \, \chi_k = 0 \, ,
\label{eq:wkbeom}
\eeq
where
\beq
\quad \omega_\chi(k,t)^2=k^2 + m_{\rm eff, \chi}^2=k^2 + m_{1, \chi}^2+m_{2, \chi}^2  \, ,
\eeq
where the components of the effective mass are given in Eq.~\eqref{miphi}.
Following Ref.~\cite{WKBtachy}, we write the WKB form of the mode-functions before, during and after a tachyonic transition  (regions I, II and III respectively).
\beq
 \begin{aligned}
\chi^I_k &=& {\alpha^n\over \sqrt {2\omega_k(t)}} e^{-i \int \omega_k(t)dt} + {\beta^n\over \sqrt {2\omega_k(t)}} e^{i \int \omega_k(t)dt}
\\
\chi_k^{II} &=& {a^{n}\over \sqrt {2\Omega_k(t)}} e^{- \int \Omega_k(t)dt} + {b^{n}\over \sqrt {2\Omega_k(t)}} e^{ \int \Omega_k(t)dt}
\\
\chi_k^{III} &=& {\alpha^{n+1}\over \sqrt {2\omega_k(t)}} e^{-i \int \omega_k(t)dt} + {\beta^{n+1}\over \sqrt {2k}} e^{ i \int \omega_k(t)dt}
\end{aligned}
\label{eq:WKBwavefunctions}
\eeq
where $\Omega_k^2(t) = -\omega_k^2(t)$.
The amplification factor after the first tachyonic region for each mode $k$ is
\beq
A_k = e^{ \int_{t_-}^{t_+} \Omega_k(t)dt}
\label{eq:AWKB}
\eeq
where $t_\pm$ are the points at which the effective frequency vanishes, $\omega_k^2(t_\pm) =\Omega_k^2(t_\pm) =0$.

\begin{figure}
\centering
\includegraphics[width=0.45\textwidth]{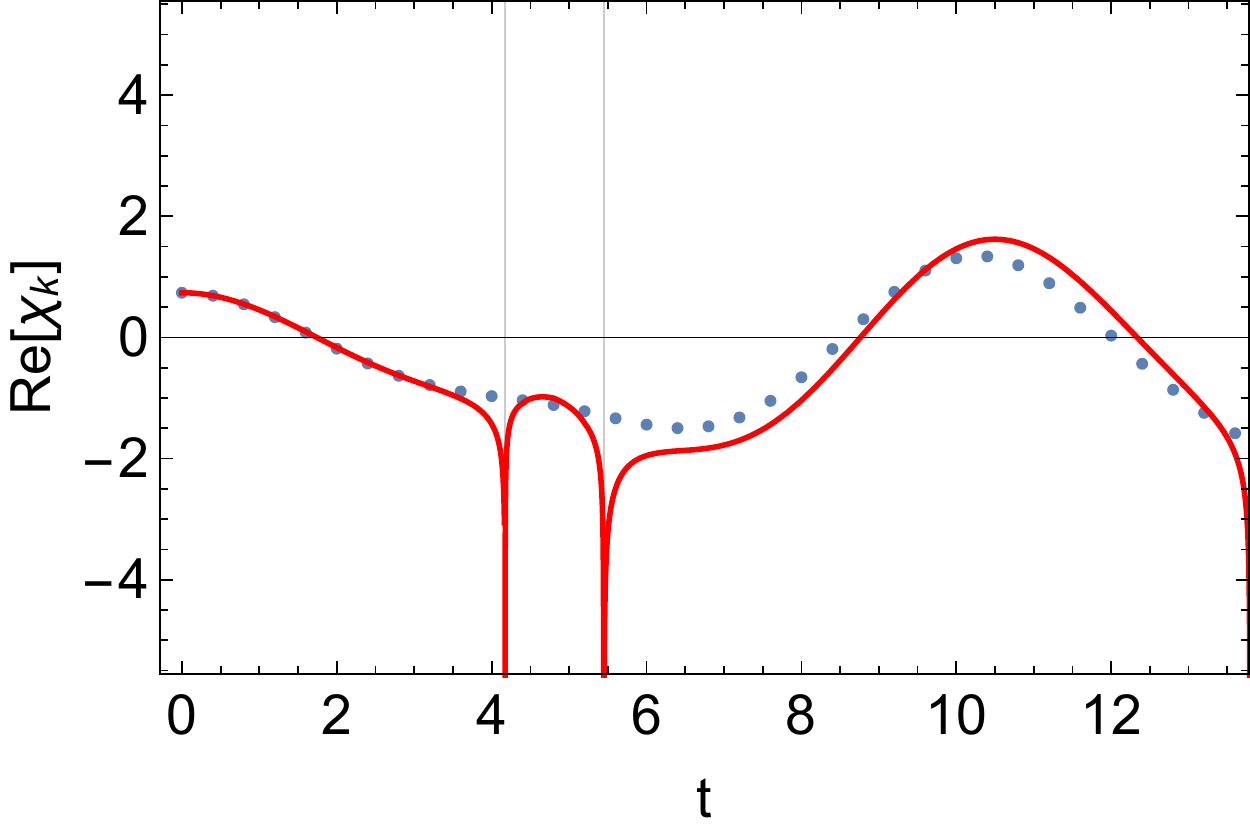}
 \includegraphics[width=0.45\textwidth]{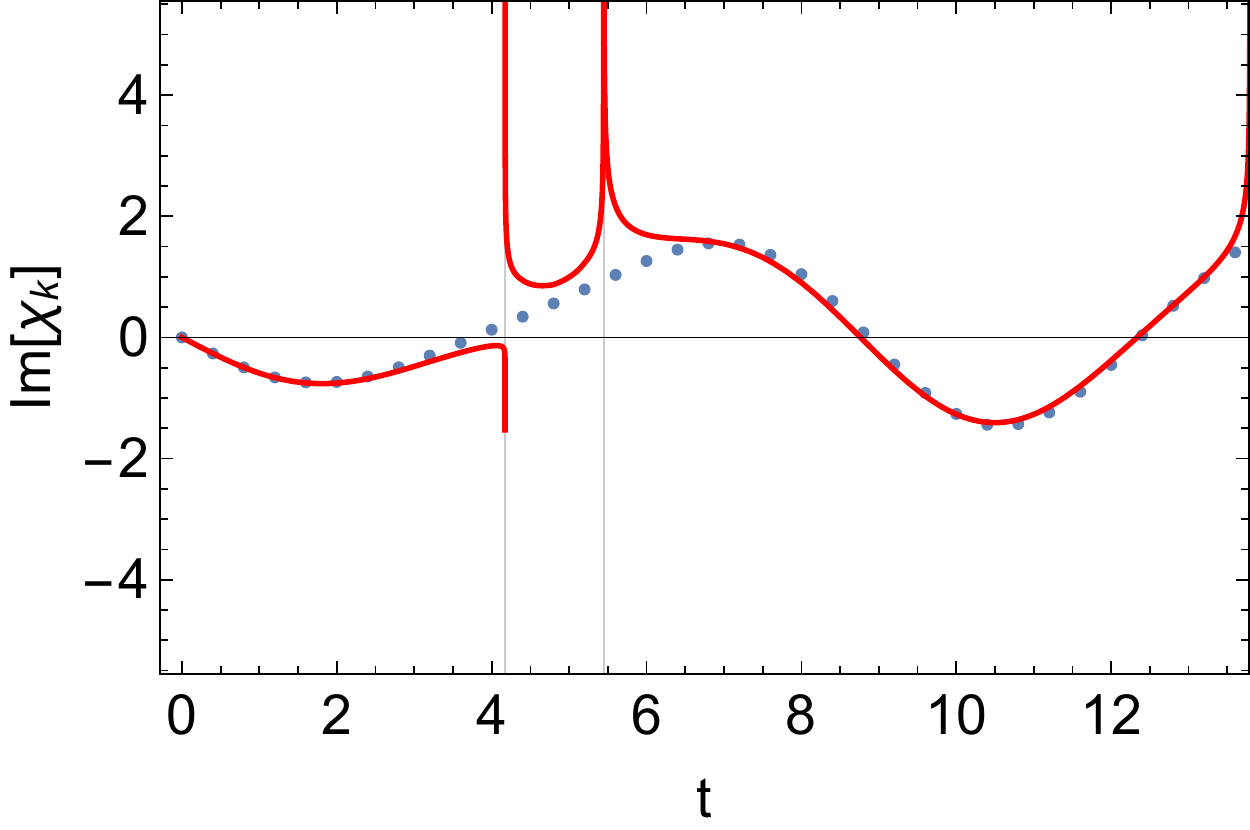}
\caption{
Comparison of the real (left) and imaginary (right) parts of the WKB solution (red line) and the numerical solution (blue dots) of the $\chi$ mode evolution in the static universe approximation
around the first tachyonic amplification burst
for $n=3/2$ , $\tilde\alpha=10^{-3}$ and $k=0.8\mu$. We see very good agreement, except in the vicinity of the points where $\omega^2=0$ and the WKB solution diverges.
 }
 \label{fig:wkbtest}
\end{figure}

Fig.~\ref{fig:wkbtest} shows the result of the numerical solution and the WKB result before, during and after the first tachyonic amplification phase. We see that the agreement is very good, hence we can use the expression of Eq.~\eqref{eq:AWKB} to estimate the growth rate of fluctuations.

\begin{figure}
\centering
\includegraphics[width=0.45\textwidth]{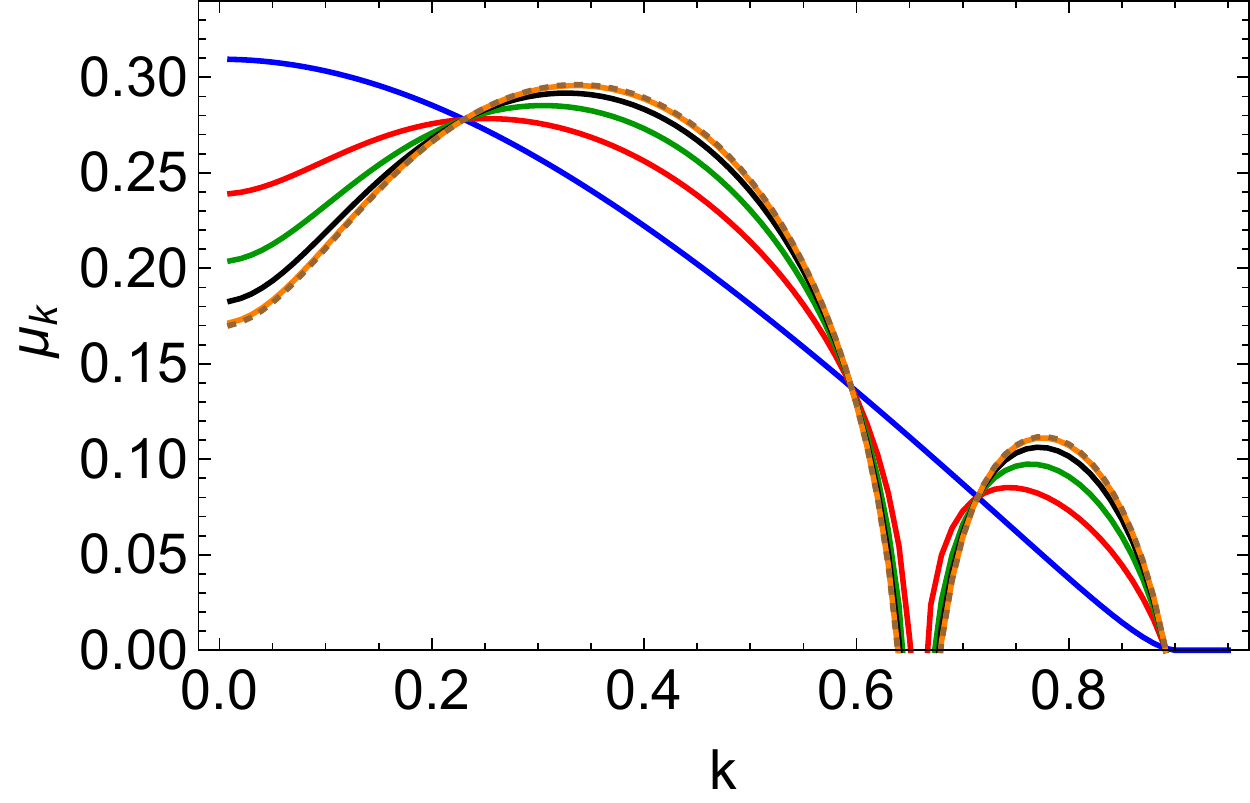}
\caption{
The Floquet exponent $\mu_k$ derived using the WKB approximation  in Eq.~\eqref{eq:mukWKB}. The Floquet exponent after $1, 2, 4, 10, 50, 100$ tachyonic regimes is shown (blue, red, green, black, orange and brown-dotted respectively).
 }
 \label{fig:wkbfloquet}
\end{figure}

As shown in Eq.~\eqref{eq:WKBwavefunctions}, following the first tachyonic burst all modes with wavenumbers $k\le k_{\rm max}$ will be amplified. Subsequent background oscillations will cause destructive or constructive interference, leading to the formation of stability and instability bands, the latter exhibiting no exponential growth. In Ref.~\cite{WKBtachy} it is shown that the amplitude of the wavefunction for a mode with wavenumber $k$ after the $j$'th tachyonic burst is
\beq
|\beta_k^j|^2 = e^{2 j A_k} (2\cos\Theta_k )^{2(j-1)} \, .
\label{eq:betak2}
\eeq
where $\Theta_k$ is the total phase accumulated between two consecutive tachyonic regimes.
We can define an averaged growth rate as
\beq
\chi_k(t) \sim e^{\mu_k t} P(t) \, ,
\eeq
where $P(t)$ is a bounded (periodic) function and $\mu_k$ is the Floquet exponent, as we discuss in detail in Section~\ref{sec:floquetcharts}. Since there are two tachyonic regimes for each background oscillation, the Floquet exponent $\mu_k$ is extracted from Eq.~\eqref{eq:betak2} as
\beq
\mu_k ={2\over T} {1\over 2j}  \log |\beta_k^j|^2 \, ,
\label{eq:mukWKB}
\eeq
where $T$ is the background period of oscillation.
As shown in Fig.~\ref{fig:wkbfloquet}, the Floquet exponent extracted from Eqs.~\eqref{eq:betak2} and \eqref{eq:mukWKB} depends on time, albeit mildly after the first few tachyonic bursts. However, there is a clear asymptotic regime that emerges after the background inflaton field has undergone multiple oscillations.
The asymptotic value should be compared to the ``true" Floquet exponent, which we compute in Section~\ref{sec:floquetcharts}.

\subsection{Floquet charts}
\label{sec:floquetcharts}

Floquet theory is a powerful tool for studying parametric resonance in the static universe approximation. The algorithm for computing Floquet charts can be found in the literature (see for example Ref.~\cite{AHKK}).

 We may further understand properties of the Floquet charts by examining the Fourier structure of certain field-space quantities.
In the rigid-spacetime limit, Eq.~\eqref{eq:wkbeom} for the isocurvature modes $\chi_k$ may be written in the suggestive form
\beq
{d \over dt} \begin{pmatrix}
\chi_k \\
\dot{\chi}_k
\end{pmatrix} =  \begin{pmatrix}
0 & 1 \\
-(k^2 + m_{{\rm eff},\chi}^2) & 0
\end{pmatrix} \begin{pmatrix}
\chi_k \\
\dot{\chi}_k
\end{pmatrix} ,
\label{eqn:floquetmatrix}
\eeq
again using $m_{\rm eff,\chi}^2 = m_{1,\chi}^2 + m_{2,\chi}^2$ in the rigid-spacetime limit. This equation is of the form
\beq
\dot{x}(t) = {\cal P}(t) \> x(t) \, ,
\label{eq:1stordereq}
\eeq
where $ {\cal P}(t)$ is a periodic matrix. The period of the background is $T$, but the period of $m^2_{{\rm eff},\chi}$ is $T/2$, since it depends quadratically on the background field $\phi(t+T)=\phi(t)$ and its derivative $\dot \phi(t+T)=\dot\phi(t)$.
In Ref.~\cite{DeCross:2016fdz} a semi-analytic method was described for computing the edges of the instability bands at arbitrary high accuracy, by reducing the system to an algebraic matrix equation. The truncation of the resulting matrices determines the number of Floquet bands that can be accurately computed. In the present work
we  determine the edges of the instability bands after
the computation of the full Floquet chart using {\it Mathematica}.

\begin{figure}
\centering
\includegraphics[width=0.45\textwidth]{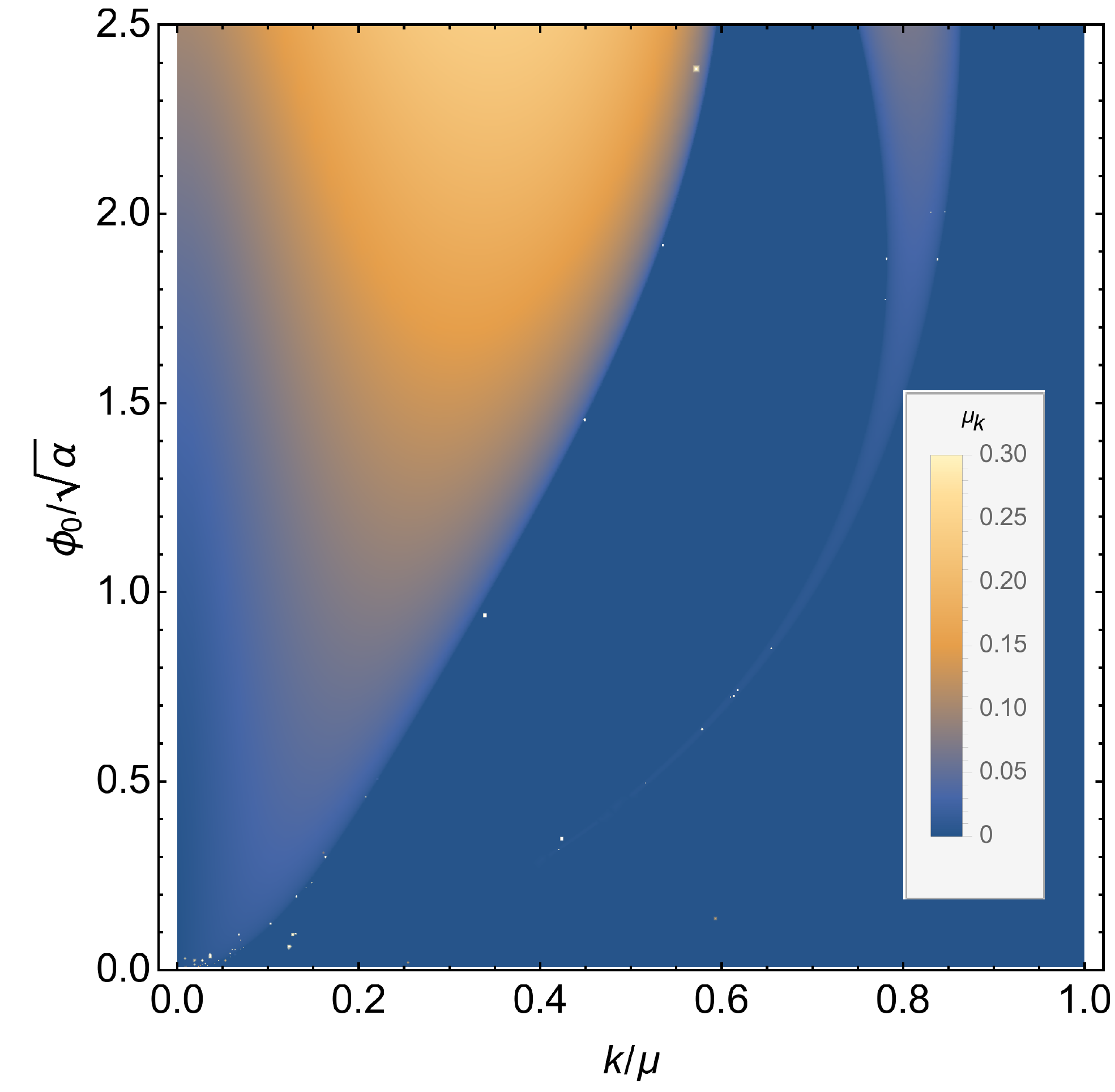}
 \includegraphics[width=0.45\textwidth]{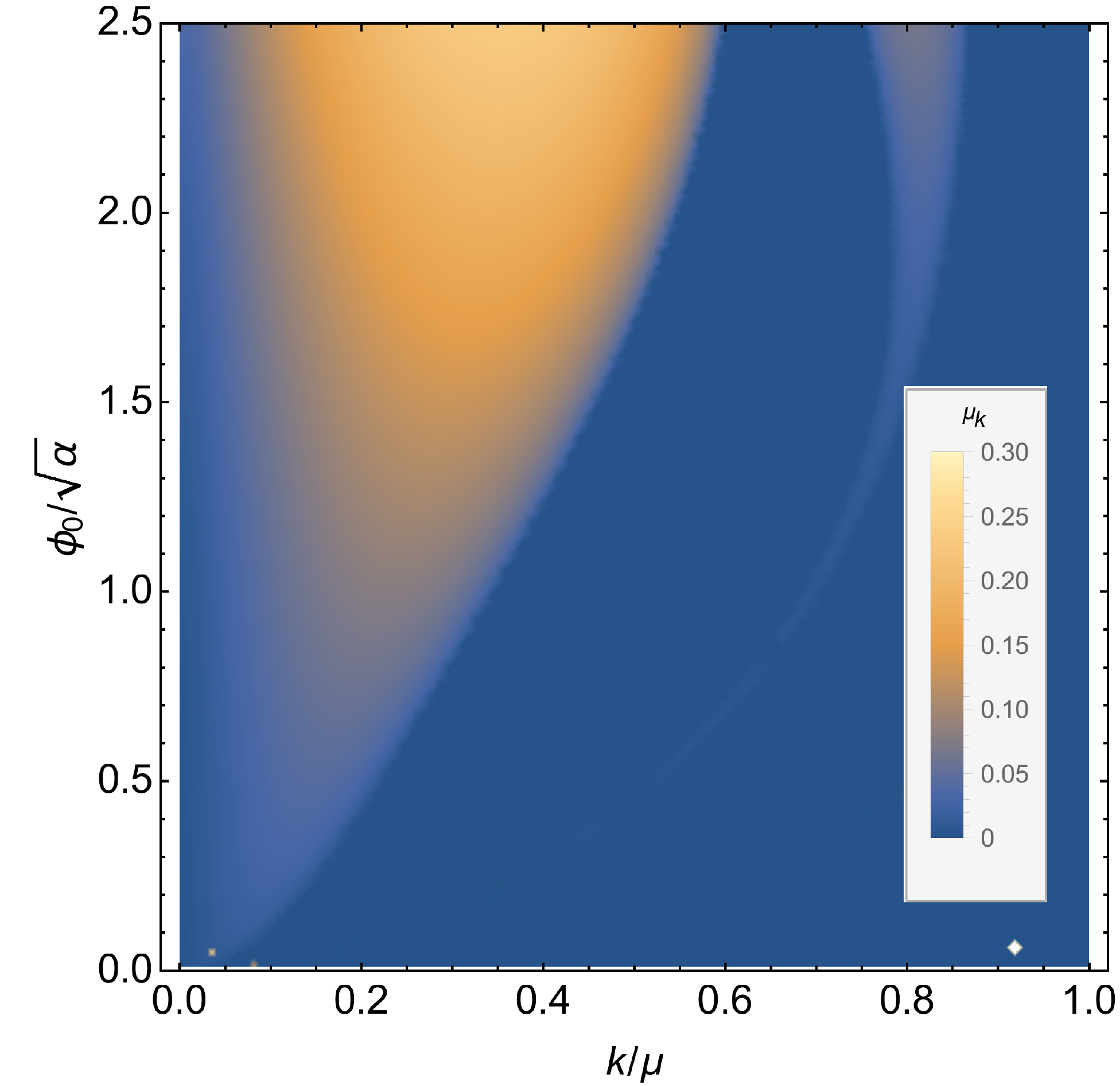}
 \\
 \includegraphics[width=0.45\textwidth]{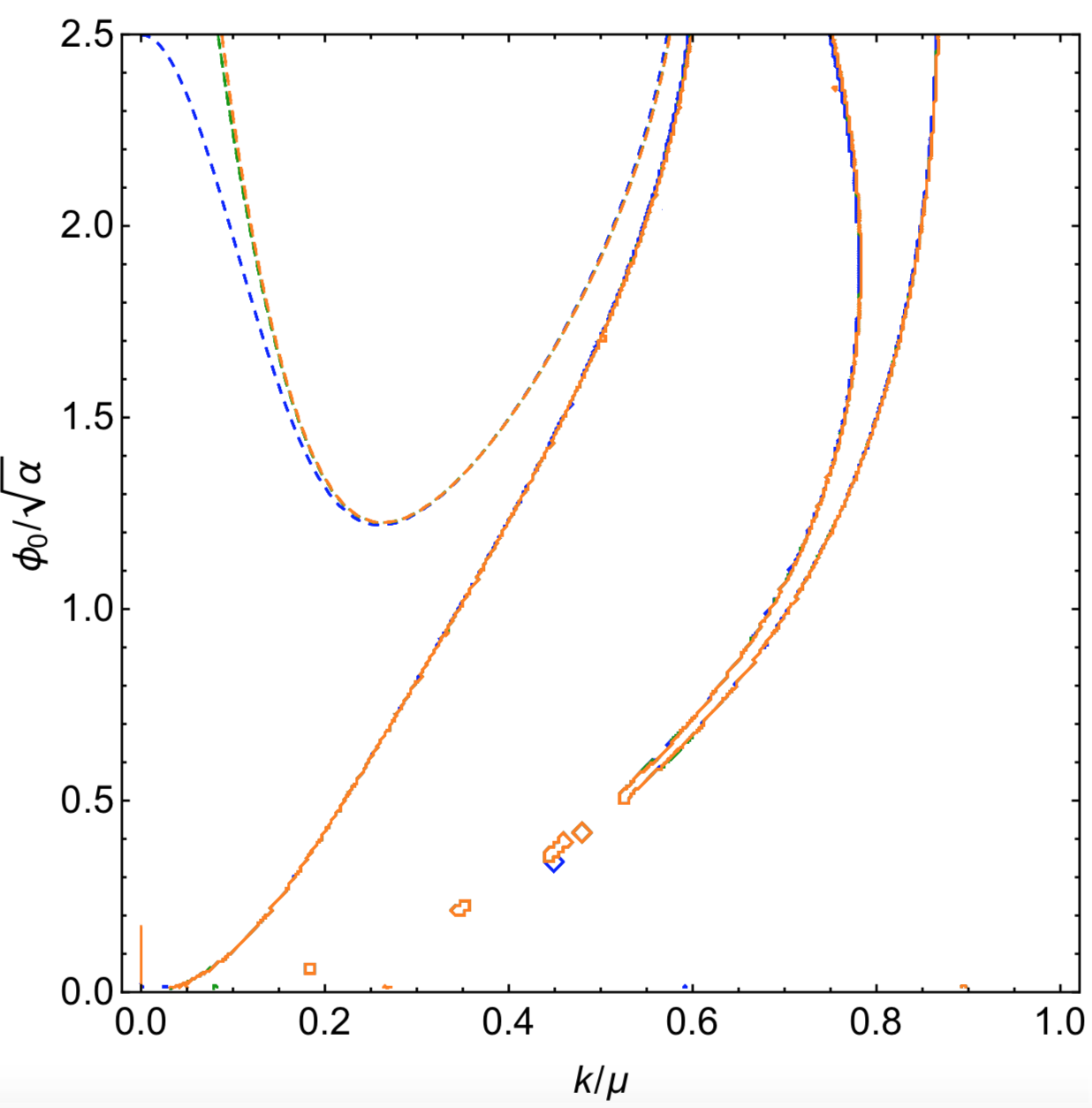}
 \includegraphics[width=0.45\textwidth]{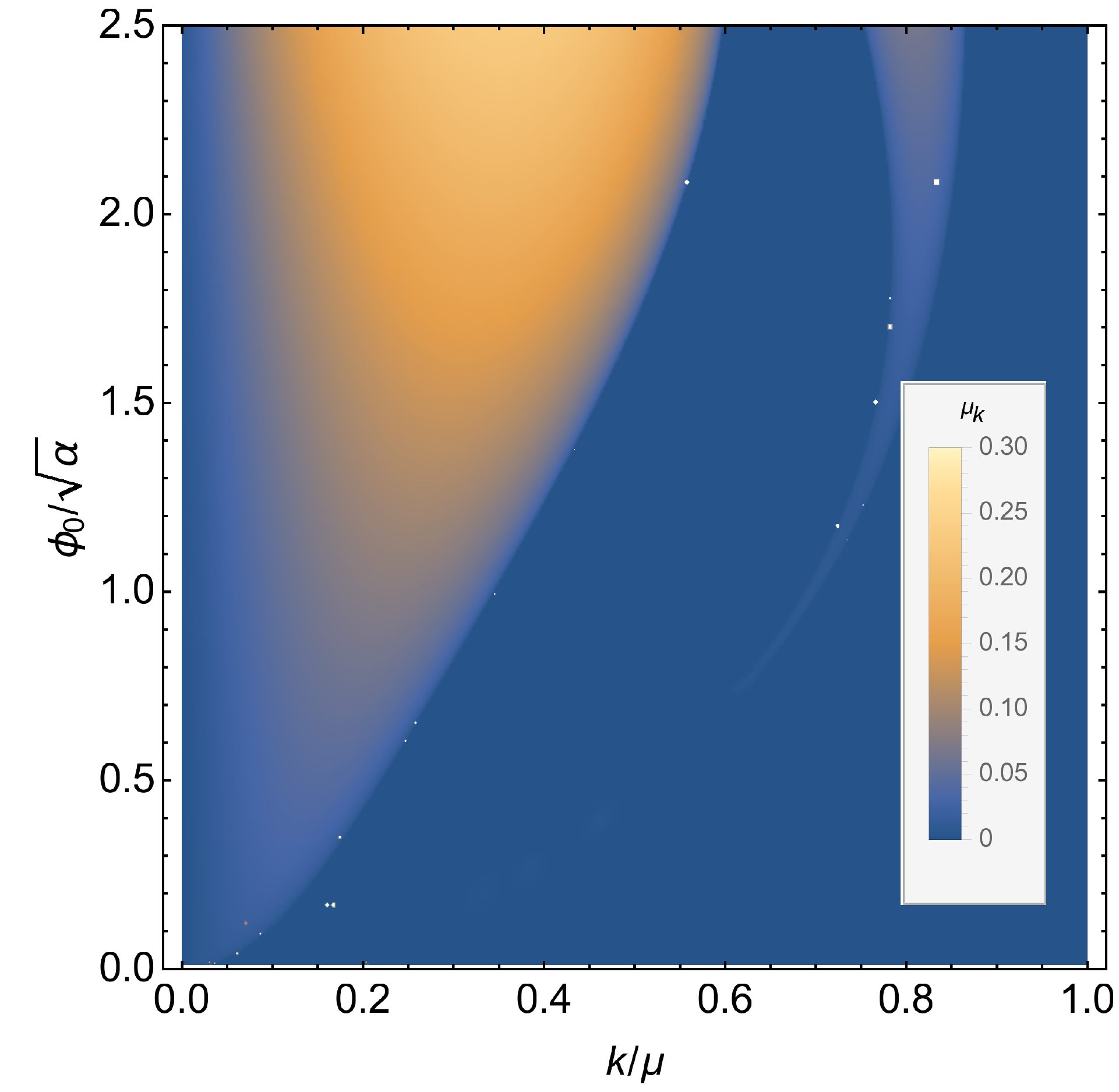}
\caption{
{\it Clockwise from the top:} The $3$-D Floquet charts for $n=3/2$ and $\tilde \alpha \equiv \alpha \, M_{\rm Pl}^{-2}=10^{-2}, 10^{-3}, 10^{-4}$.
{\it Bottom left panel:} The contour plots for $\mu_k=0$ (solid lines) and $\mu_k=0.1$ (dashed lines). The blue, green and orange curves are for $\tilde\alpha=10^{-2}, 10^{-3}, 10^{-4}$ respectively.
 }
 \label{fig:floquetn3over2}
\end{figure}

Fig.~\ref{fig:floquetn3over2} shows the Floquet charts for $n=3/2$ and $\tilde \alpha =10^{-2}, 10^{-3}, 10^{-4}$. We can see that, when normalized appropriately with $\tilde \alpha$, the Floquet charts look similar, especially when it comes to the first two instability bands, which essentially control the entirety of the parametric resonance. The relation between Floquet charts for different values of $\alpha$ becomes even more evident, when we show a few contours of the first instability bands on the same plot. It is then obvious that for $\tilde \alpha \lesssim 10^{-3}$ the parametric resonance in the static universe approximation is identical, regardless of the exact value of the field-space curvature\footnote{For $\tilde\alpha=10^{-2}$ the edges of the first two instability bands follow the ones exhibited by $\tilde\alpha \ll 1$, while the low-$k$ edge of the first band shows slightly larger Floquet exponents.}.
 This is no surprise, since the WKB analysis of Section \ref{sec:WKBresults} predicted the scaling behavior of the parametric resonance strength for low values of $\tilde \alpha$. The Floquet chart of {Fig.}~\ref{fig:floquetn3over2} can be considered a ``master diagram", from which the Floquet chart for arbitrary values of $\tilde\alpha \lesssim 0.01$ can be easily read-off by using the appropriate scaling with $\tilde \alpha$.

\begin{figure}
\centering
\includegraphics[width=0.6\textwidth]{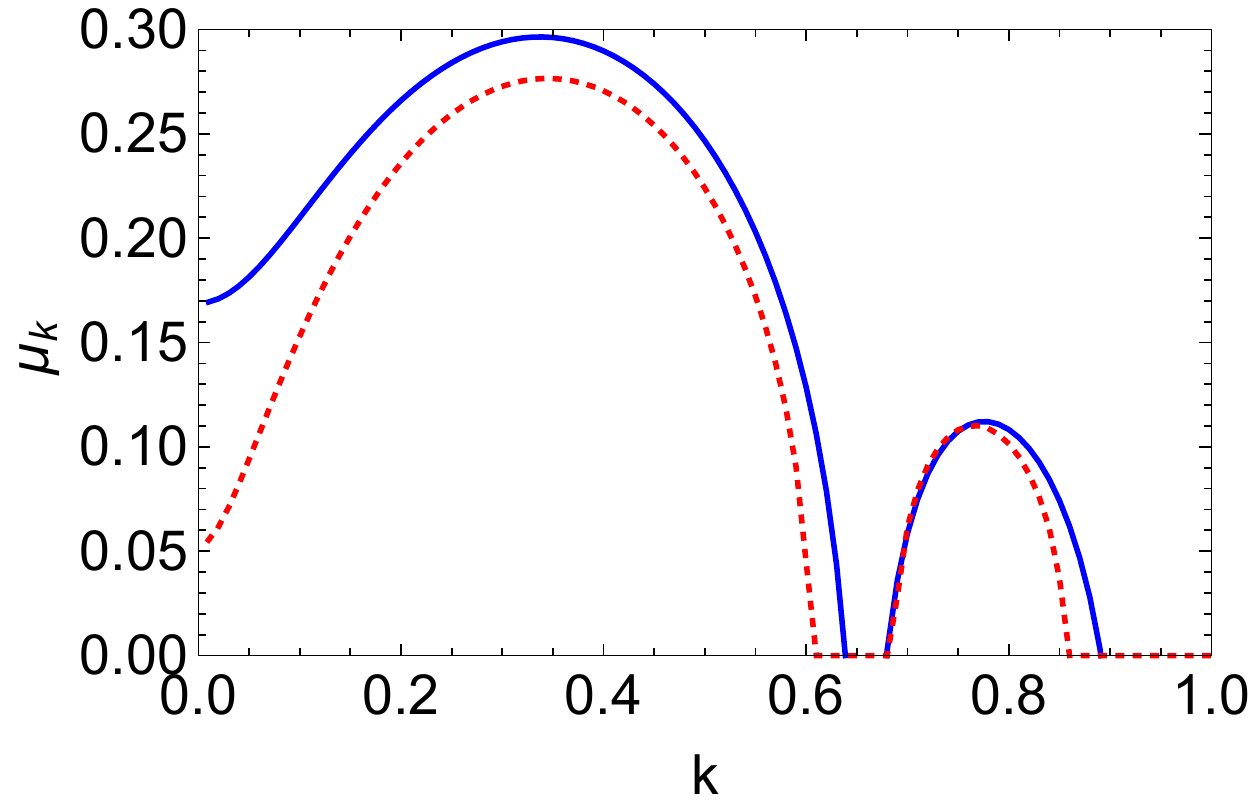}
 \caption{The (asymptotic) Floquet exponent computed using the WKB approximation (blue solid) and the Floquet exponent computed using the numerical algorithm described in Section~\ref{sec:floquetcharts} for $n=3/2$ and $\tilde \alpha=10^{-3}$. The agreement is very good, given the inherent limitations of the WKB approximation.
 }
 \label{fig:floquet0p001comp}
\end{figure}

Finally, Fig.~\ref{fig:floquet0p001comp} shows the comparison between the Floquet exponent computed using the algorithm described in Ref.~\cite{AHKK} and using the WKB analysis. We see that the WKB analysis is able to capture the existence of the first two instability bands, even though the shape does not perfectly match the fully numerical solution.

\subsection{Expanding Universe}

There are two complications introduced by studying preheating in an expanding universe:
the  (slow) decay of the amplitude of the background oscillations due to the non-zero Hubble drag and the red-shifting of the physical wavenumber $k_{\rm phys}  =k_{\rm comoving}/a$ due to the increasing scale-factor $a(t)$.
Both effects are comparable, so they must be studied together. While a WKB analysis can be performed in an expanding universe \cite{WKBtachy}, it must take into account the evolution of both $k_{\rm phys}$ and $\phi(t)$ numerically.  Since we believe that it will not add significantly to building intuition on the model at hand, we will not pursue it here. Instead we numerically solve the equations of motion for the $\chi$ fluctuations, {working in the linear regime as follows:
The evolution for the background inflaton field and the Hubble rate are solved numerically using Eqs.~\eqref{eq:backgroundfield} and \eqref{eq:backgroundHubble}.
We subsequently compute the produced $\chi$ fluctuations driven by the background inflaton field.
 The back-reaction of the produced $\chi$ fluctuations on the inflaton field or the Hubble rate is ignored. This is a valid approximation until the energy density of the $\chi$ fluctuations becomes comparable to the background inflaton energy density. We briefly discuss back-reaction effects in Appendix~D.}
We start {our  computations} several $e$-folds before the end of inflation, in order for  the effective mass to be positive for all modes, according to Fig.~\ref{fig:meffGD} and so that the WKB solutions of Eq.~\eqref{eq:WKBz} provide accurate initial conditions for our code.

\begin{figure}
\centering
\includegraphics[width=0.5\textwidth]{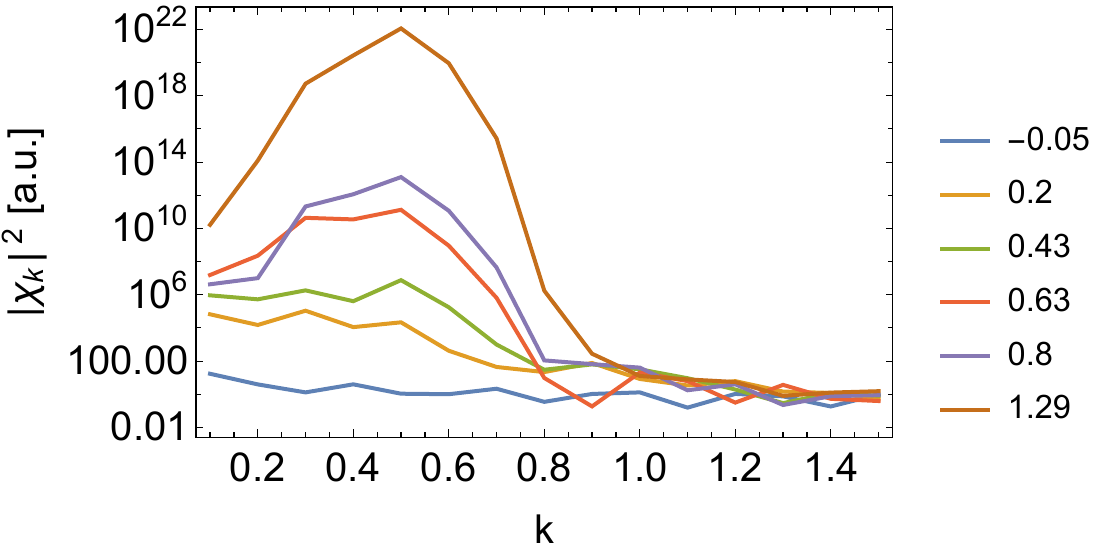}\quad \includegraphics[width=0.47\textwidth]{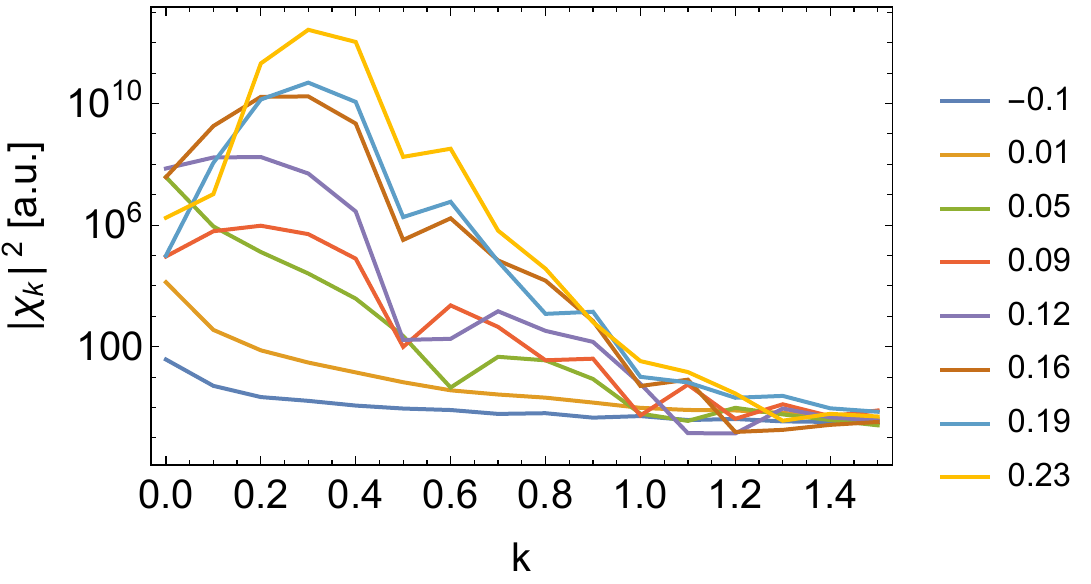}
\caption{
The spectra of the fluctuations in the $\chi$ field $|\chi_k|^2$ (in arbitrary units) as a function of the wavenumber $k$ (in units of $\mu$) at different times for $n=3/2$ and $\tilde \alpha=10^{-3}$ (left) and $\tilde \alpha=10^{-4}$ (right). The comparison with Fig.~5 of Ref.~\cite{Krajewski:2018moi} shows  agreement in the initial stages, when the linear analysis is valid.  {{The comparison is most easily done by considering the amplification occurring between the various time-points shown in the figures, both here and in Ref.~\cite{Krajewski:2018moi}.}} Note that Ref.~\cite{Krajewski:2018moi} uses a different normalization for $k$, as discussed in Appendix C.
 The linear analysis presented here cannot capture the re-scattering effects leading to the broadening of the $\chi$ spectrum at late times that was observed in Ref.~\cite{Krajewski:2018moi}.
 {The times corresponding to the various curves are shown in the legend of each panel, measured in $e$-folds after the end of inflation (negative values correspond to spectra during the last stages of inflation).
 }
 }
 \label{fig:chispectra}
\end{figure}

Fig.~\ref{fig:chispectra} shows the spectra of the fluctuations in the $\chi$ field at different times. We see that the band structure of the static universe Floquet charts of Section~\ref{sec:floquetcharts} has disappeared, essentially leaving behind a region of excited modes with comoving wavenumbers that satisfy $k\le k_{\rm max} \approx \mu$. This occurs because each mode with a specific wavenumber $k$ redshifts through the bands of Fig.~\ref{fig:floquetn3over2}, hence a mode with $k\le k_{\rm max}$ will eventually redshift into the main instability band. Even though the exact band structure is erased, the WKB analysis can still capture very well the behavior after the first tachyonic burst. We see that the amplification factor computed in Eq.~\eqref{eq:AWKB} matches very well with the actual amplification. For small values of $\tilde\alpha$, where the Hubble scale is much smaller than the frequency of background oscillations,  Eq.~\eqref{eq:AWKB} can provide useful intuition for the behavior of the $\chi$ fluctuations during the  first few $\phi$ oscillations.
Using Eq.~\eqref{eq:Hendslowroll} the maximum excited wavenumber can be immediately compared to the Hubble scale at the end of inflation to give
\beq
{k_{\rm max} \over H_{\rm end} } \simeq {2 \over  \tilde \alpha } \gg 1
\eeq
Hence tachyonic amplification occurs  predominately for sub-horizon modes, meaning that they will behave like radiation after the end of preheating.

\begin{figure}
\centering
\includegraphics[width=0.45\textwidth]{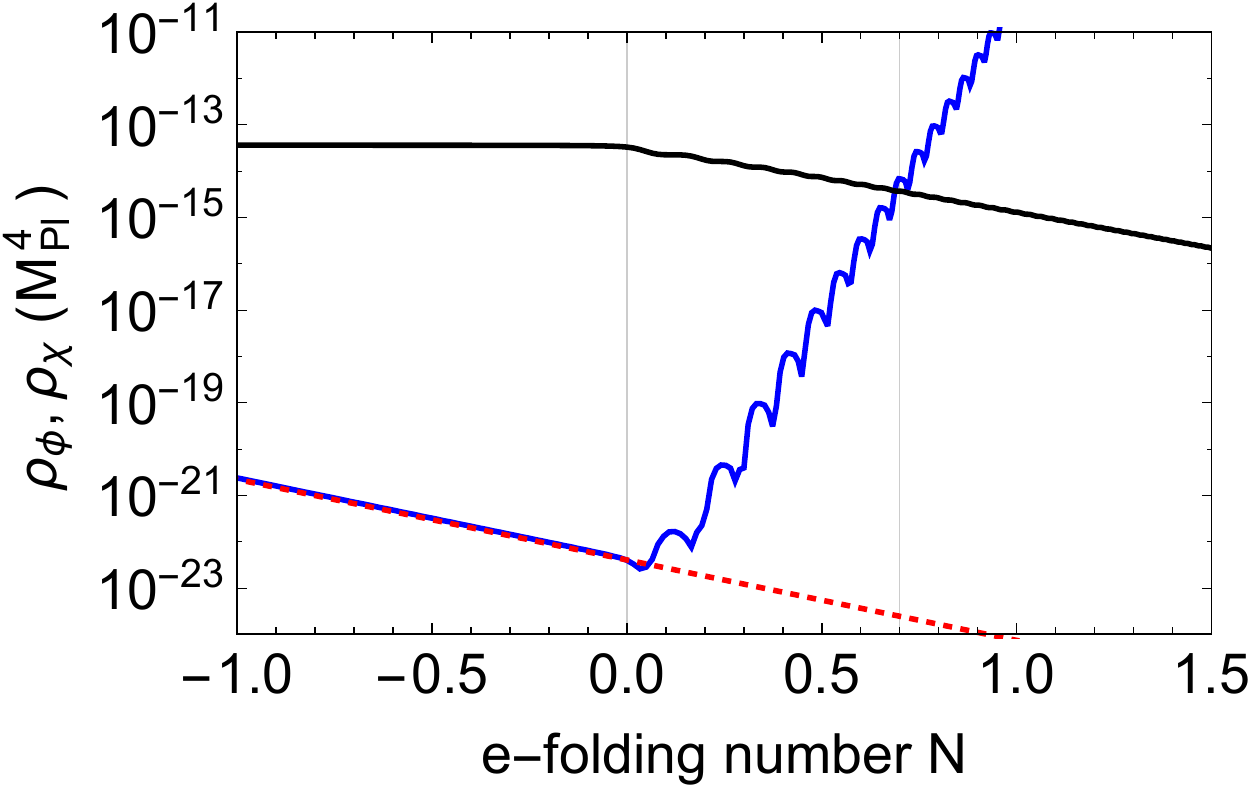}\quad \includegraphics[width=0.45\textwidth]{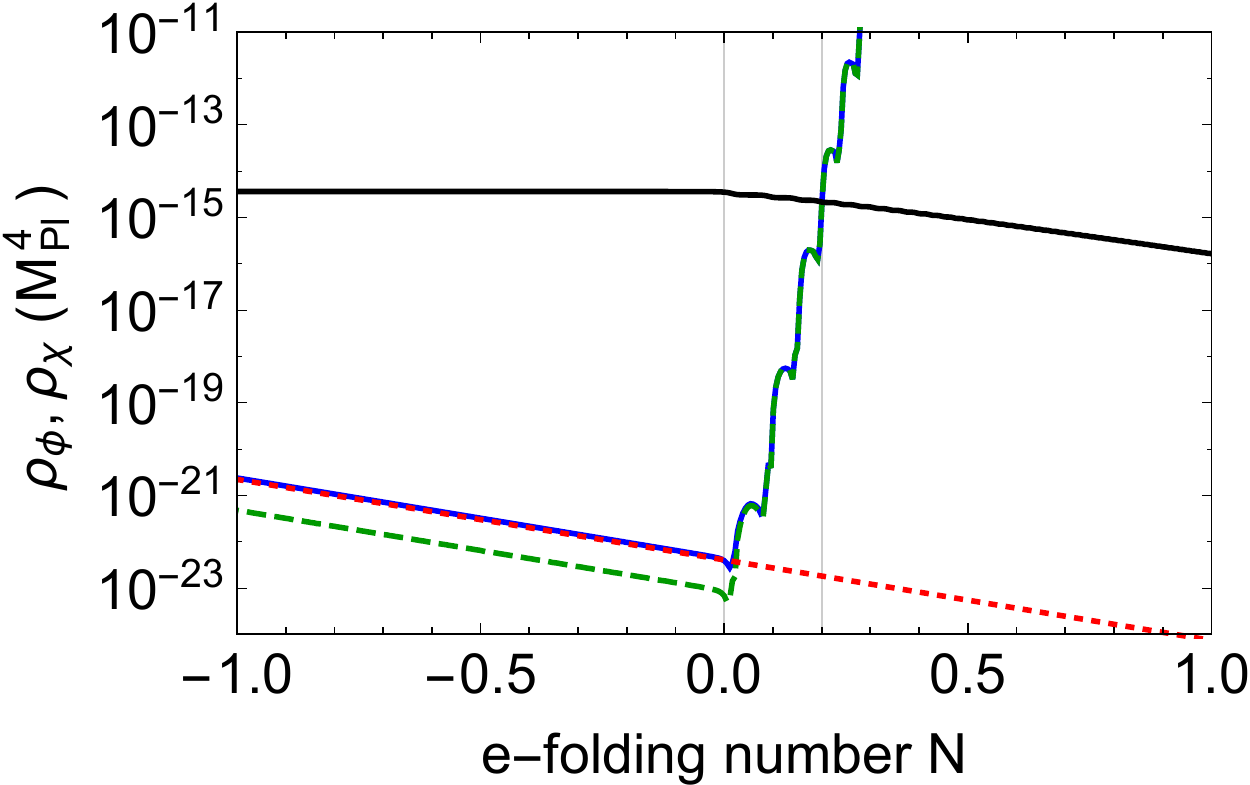}
\caption{
The energy density in the background inflaton field (black) and the $\chi$ fluctuations (blue) for $n=3/2$ and $\tilde \alpha=10^{-3}$ (left)  and $\tilde \alpha=10^{-4}$ (right). The red-dashed lines show the scaling $a^{-4}$, which is observed by the fluctuations before the onset of the tachyonic preheating regime. The green-dashed line on the right panel shows a calculation using a different range of $UV$ modes, as explained further in the main text. $N=0$ marks the end of inflation and we see that preheating concludes within a fraction of an $e$-fold in both cases.
 }
 \label{fig:rhoevol}
\end{figure}

 Fig.~\ref{fig:rhoevol} shows the evolution of the energy density in the background inflaton field $\phi$ and the fluctuations of the $\chi$ field.
Considering a finite amount of wavenumbers $k<k_{\rm UV}$ initialized at the Bunch Davies vacuum, we can compute their energy density at the end of inflation to be
\beq
\rho_{\chi}  = \int {d^3 k\over (2\pi)^3} k^2 {1\over 2k} = {1\over (2\pi)^2} {k_{UV}^4  \over 4}
\eeq
This corresponds to the red-dashed line of Fig.~\ref{fig:rhoevol}, {where we took $k_{\rm UV}=1.5\mu$}. This is not a physical energy density, since these are vacuum modes. It is however useful as a check of our numerical calculation. Using different values of $k_{\rm UV}$ leads to different early time behavior, as shown from the green-dashed line in Fig.~\ref{fig:rhoevol}. As long as $k_{\rm UV} \ge k_{\rm max}$, the exact choice of $k_{\rm UV}$ becomes irrelevant once tachyonic resonance begins and all modes within $k<k_{\rm max}$ become exponentially amplified. Hence the blue and green-dashed curves of Fig.~\ref{fig:rhoevol}, {corresponding to $k_{\rm UV}=1.5\mu$ and $k_{\rm UV}=\mu$ respectively},  become indistinguishable shortly after the end of inflation. In is interesting to note that we find for $n=3/2$ and $\alpha=10^{-4}$ that preheating will conclude at $N_{\rm reh}=0.2$, where the energy density in $\chi$ fluctuations equals the energy density in the background field. This result agrees well with the findings of Fig.~4 of Ref.~\cite{Krajewski:2018moi}, where the results of a full lattice code are shown for the same model parameters.

\section{Potential dependence}
\label{sec:potentials}

So far we have used the T-model potential of Eqs.~\eqref{eq:KH} and \eqref{eq:WH} with $n=3/2$ as a concrete example to study in detail both analytically and numerically. This potential has the added benefit of allowing for an easy comparison with the full lattice simulations presented in Ref.~\cite{Krajewski:2018moi}\footnote{{After submission of the present manuscript, an updated version of Ref.~\cite{Krajewski:2018moi} appeared. This includes results for two potential types, corresponding to $n=3/2$ and $n=1$, as well as two values of the field-space curvature parameter $\alpha$. These match our results, as we  describe in Section~\ref{sec:summary}. }}. We now extend the analysis to arbitrary values of $n$, hence to the whole family of the generalized T-model  potentials. The background dynamics is summarized in Figs.~\ref{fig:phiendvsnvsalpha} and \ref{fig:periodend} through the dependence of $H_{\rm end}$, $\phi_{\rm end}$ and the period of oscillation $T$ on $\tilde \alpha$ and $n$.

\begin{figure}
\centering
\includegraphics[width=0.45\textwidth]{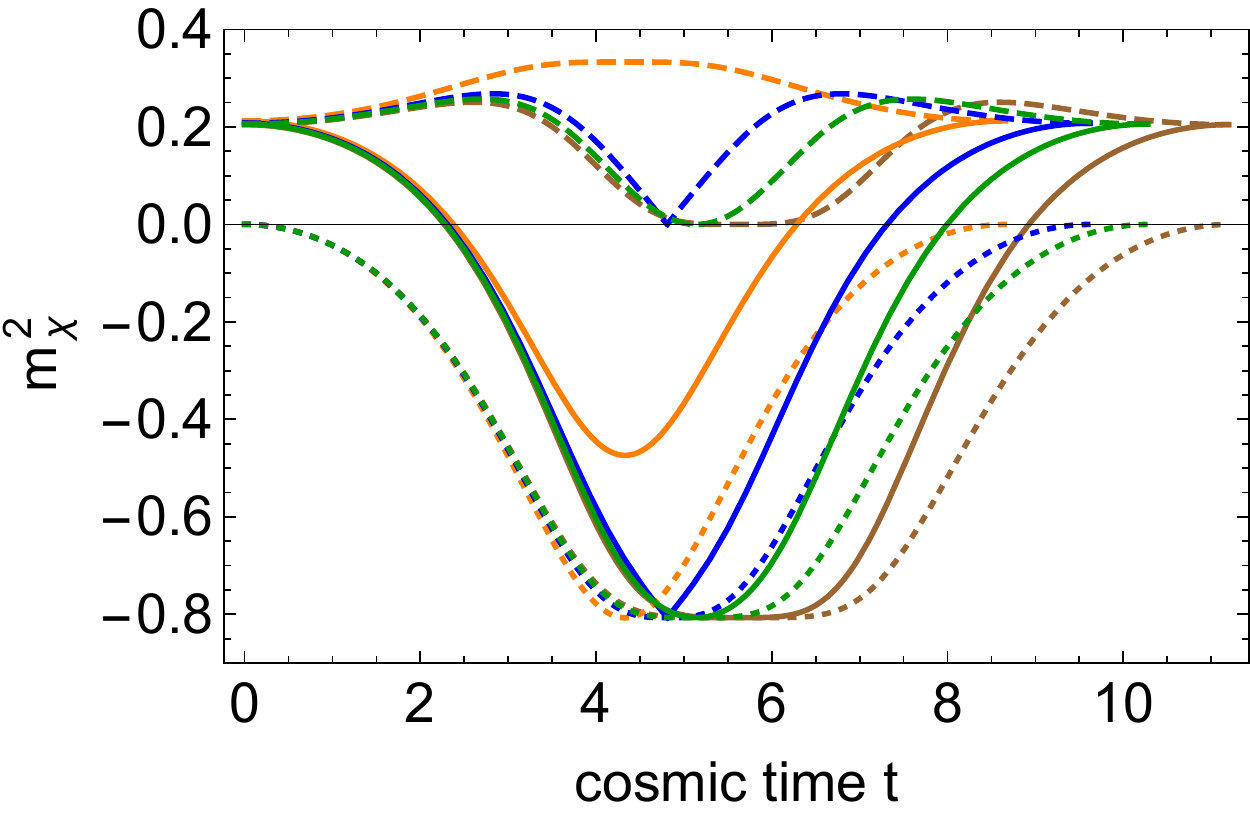}
\quad
\includegraphics[width=0.45\textwidth]{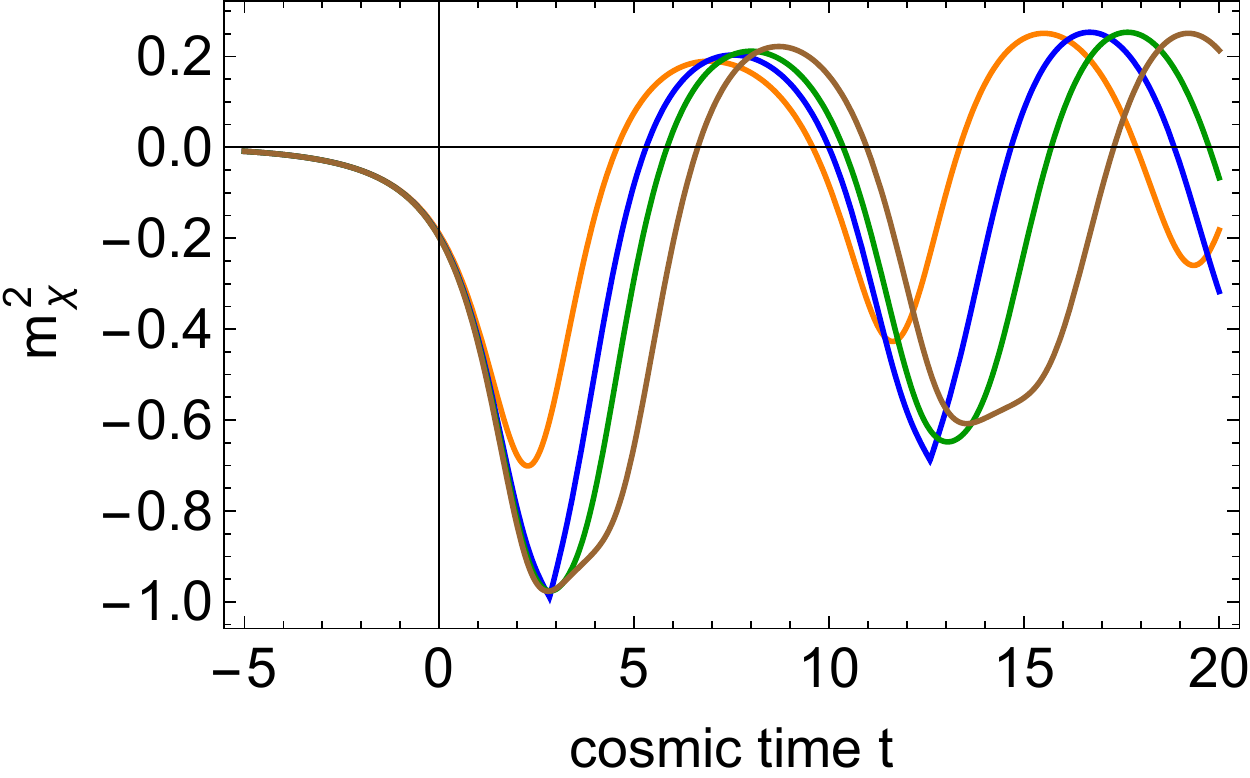}
\caption{
{\it Left:} The two main components of the effective mass squared for $\chi$ fluctuations: the potential contribution (dashed) and the field-space Ricci contribution (dotted), along with the sum (solid) for $\tilde \alpha=10^{-3}$ and $n=1,3/2,2,3$ (orange, blue, green and brown respectively). The plot shows one period in the static universe approximation with $\phi_{\rm max} = \phi_{\rm end}$.
\\
{\it Right:}  The sum $m_{1,\chi}^2  +m_{2,\chi}^2  $ using the full expanding universe solution for the background field $\phi(t)$. Inflation is taken to end at $t=0$.
 }
 \label{fig:meffovern}
\end{figure}

Fig.~\ref{fig:meffovern} shows the effective mass-squared for $\tilde\alpha=10^{-3}$ and varying $n$ as a function of time, both in the static universe approximation and using the full expanding universe background solution. The former will be used for computing the resonance structure.
It is worth noting that the maximally negative value of $m_{{\rm eff},\chi}^2$ is larger in the expanding universe case, compared to the static universe one. This is due to the fact that we consider the initial conditions $\{ \phi_0, \dot\phi_0\} = \{\phi_{\rm end},0\}$. In reality, the inflaton velocity is not zero at the end of inflation, hence the Ricci-driven component of the effective mass, which is proportional to $|\dot\phi|^2$ is underestimated in our static universe calculations.
One important difference between the various values of $n$ shown in Fig.~\ref{fig:meffovern} can be traced back to Eq.~\eqref{eq:Vphieq0chi}, which defines the potential contribution of the effective mass near the point $\phi(t)=0$ or equivalently $\left |\dot \phi (t) \right |= {\rm max}$, where the Riemann contribution $m_{2,\chi}^2$ is maximized. For $n=1$ the potential is locally quadratic, hence describing massive fields\footnote{{A  locally quadratic potential that becomes less steep at larger field values can also support oscillons. It was shown in Ref.~\cite{MustafaDuration} that oscillons can emerge during preheating in a single-field T-model for $n=1$. Since oscillons are massive objects, a period of oscillon domination causes the universe to acquire an equation of state of $w=0$, identical to that of a matter dominated era. }}. This leads to a non-zero positive contribution to the effective mass-squared for all values of  time and wavenumber, thus reducing the overall efficiency of tachyonic resonance,  through reducing both $A_k$ of Eq.~\eqref{eq:AWKB} and $k_{\rm max}$. For $n\ge 3/2$, the potential contribution vanishes for $\phi(t)=0$, hence the Riemann term completely determines the maximally negative value of $m_{{\rm eff},\chi}^2$. Furthermore, $\left |\dot \phi (t) \right |_{\rm max}$ is found to be almost identical for all values of $n$. The main difference for increasing the value of $n$ is the increased  duration of the regime where $m_{1,\phi}^2 \approx 0$. Overall, for $n\ge 3/2$ the maximum excited wavenumber $k_{\rm max}$ is the same, while the amplification factor $A_k$ grows, because each tachyonic burst lasts longer.
\begin{figure}
\centering
\includegraphics[width=0.45\textwidth]{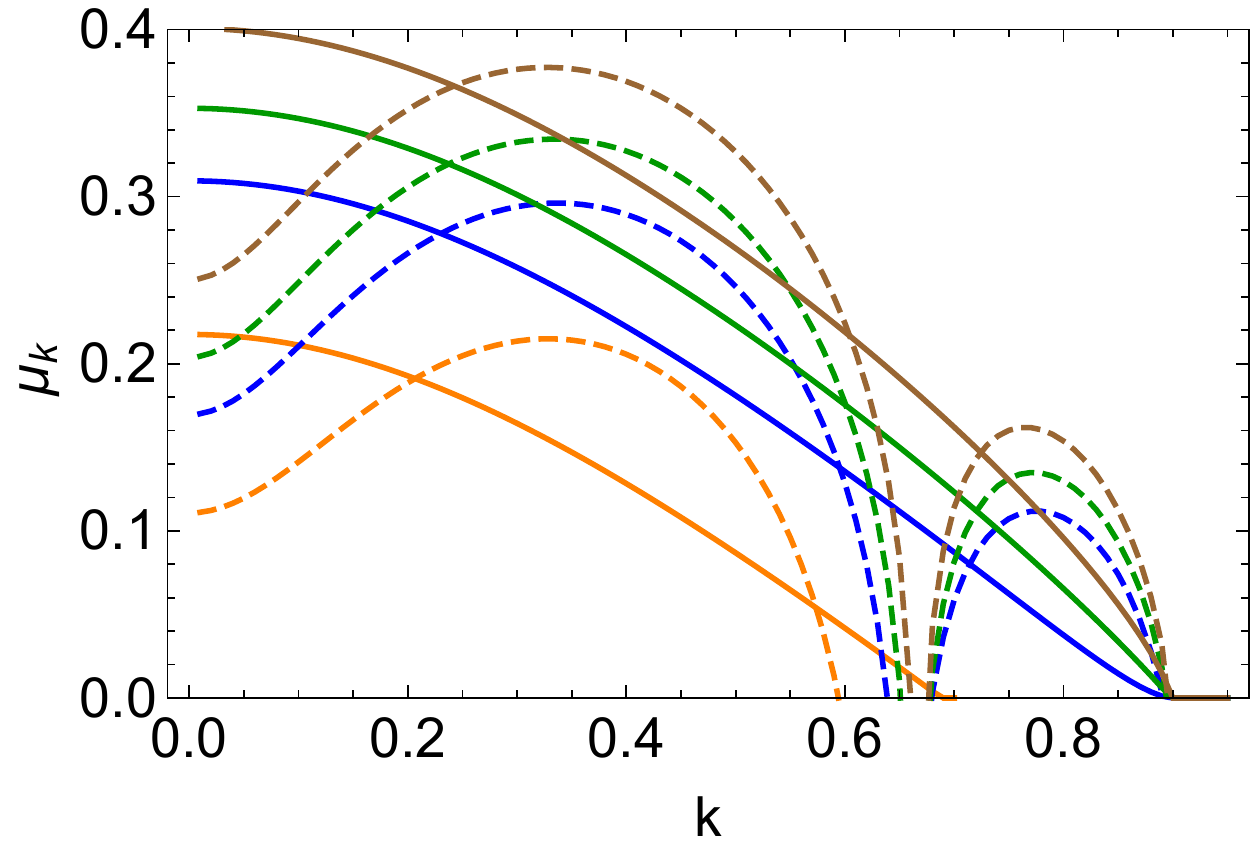}
\quad
\includegraphics[width=0.45\textwidth]{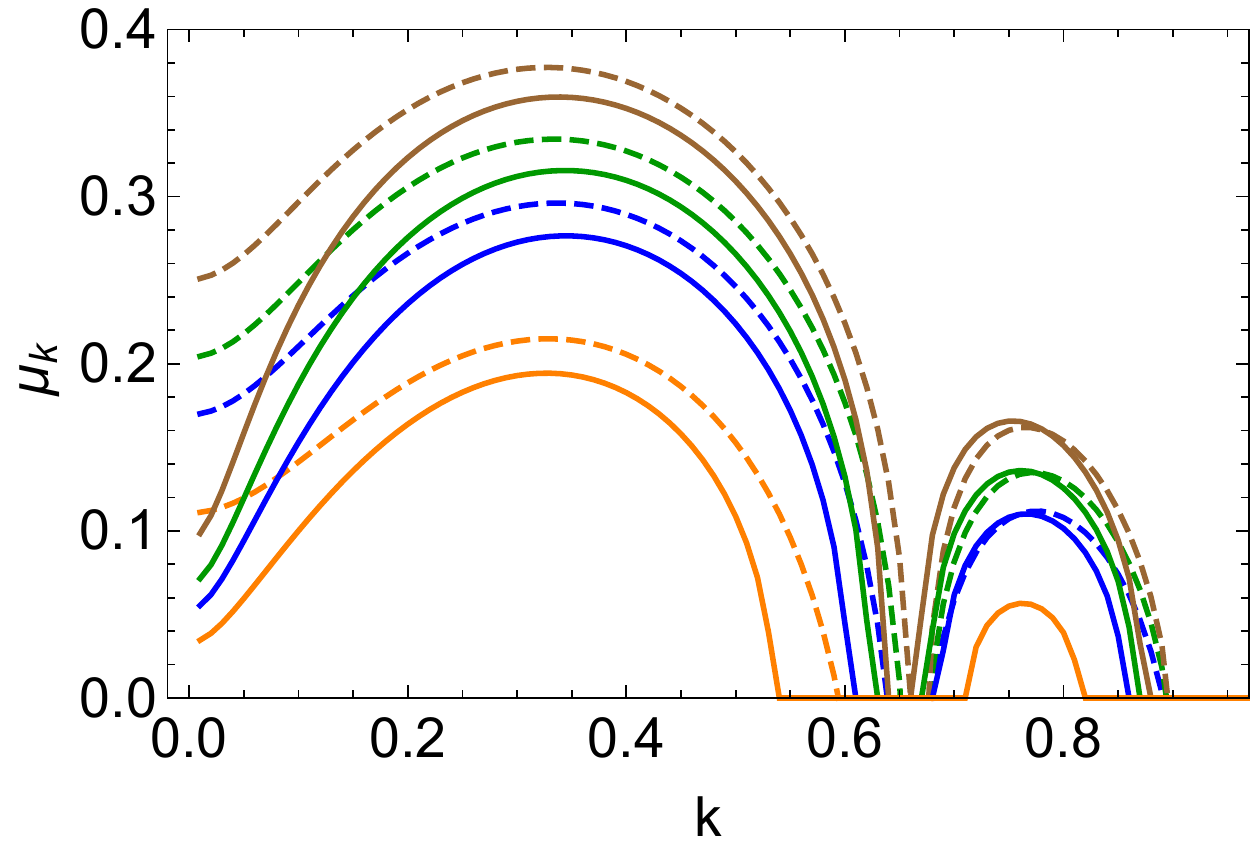}
\caption{
{\it Left:} The asymptotic Floquet exponent (dashed) and the Floquet exponent after the first tachyonic burst (solid) using the WKB approximation for $n=1,3/2,2,3$ (orange, blue, green and brown respectively).
{\it Right:} The asymptotic Floquet exponent using the WKB method (dashed) and using the algorithm of Section~\ref{sec:floquetcharts} (solid). The agreement is remarkable
given the limitations of the WKB approximation.  }
 \label{fig:mukovern}
\end{figure}
This is shown in Fig.~\ref{fig:mukovern} using both the WKB approximation, as well as by computing the Floquet exponent  numerically following Section~\ref{sec:floquetcharts}. We see that for $n=1$ the WKB approximation captures only the first instability band, while for $n\ge 3/2$ the first two instability bands are well described.

\begin{figure}
\centering
\includegraphics[width=0.45\textwidth]{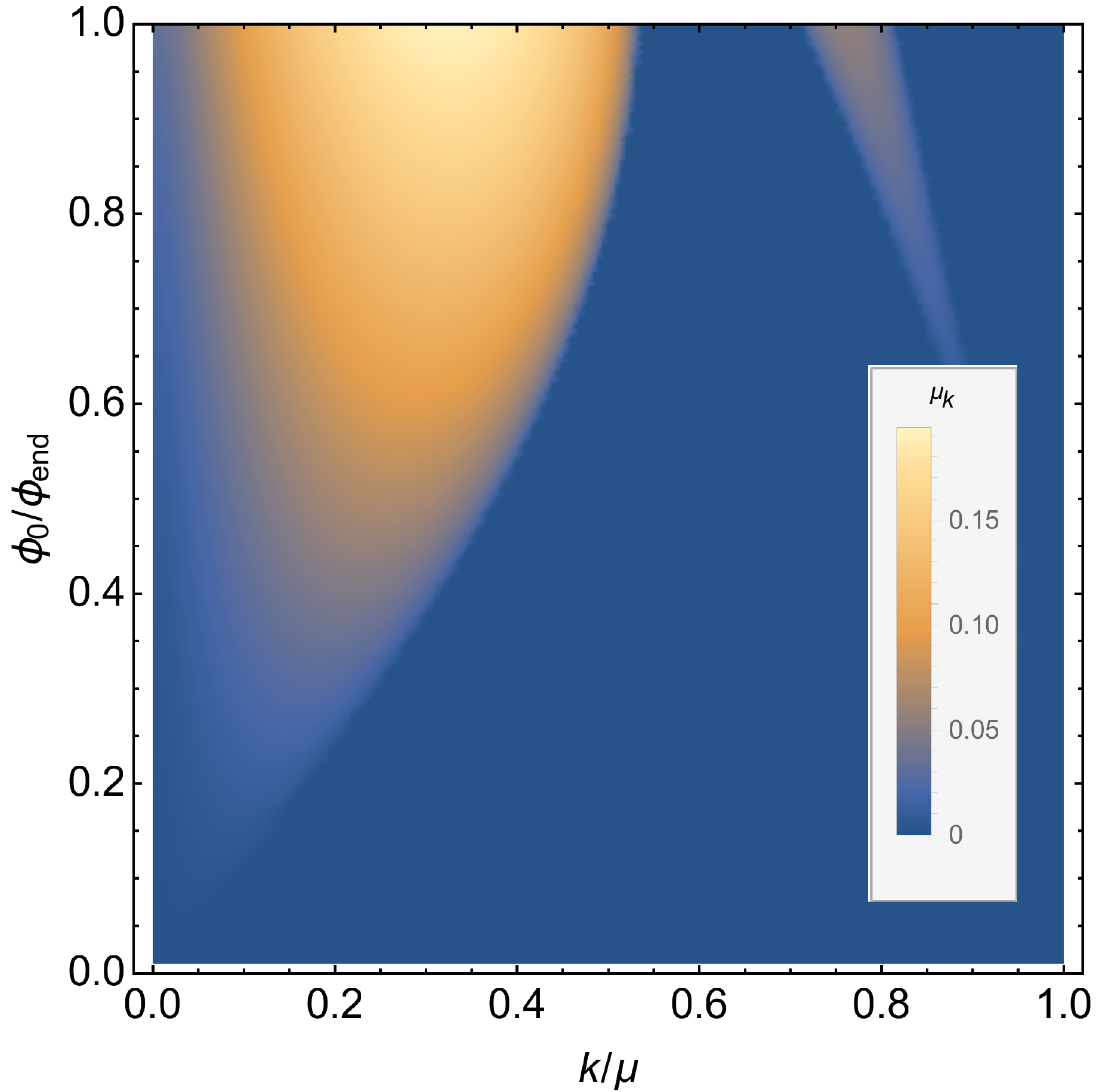}\quad
\includegraphics[width=0.45\textwidth]{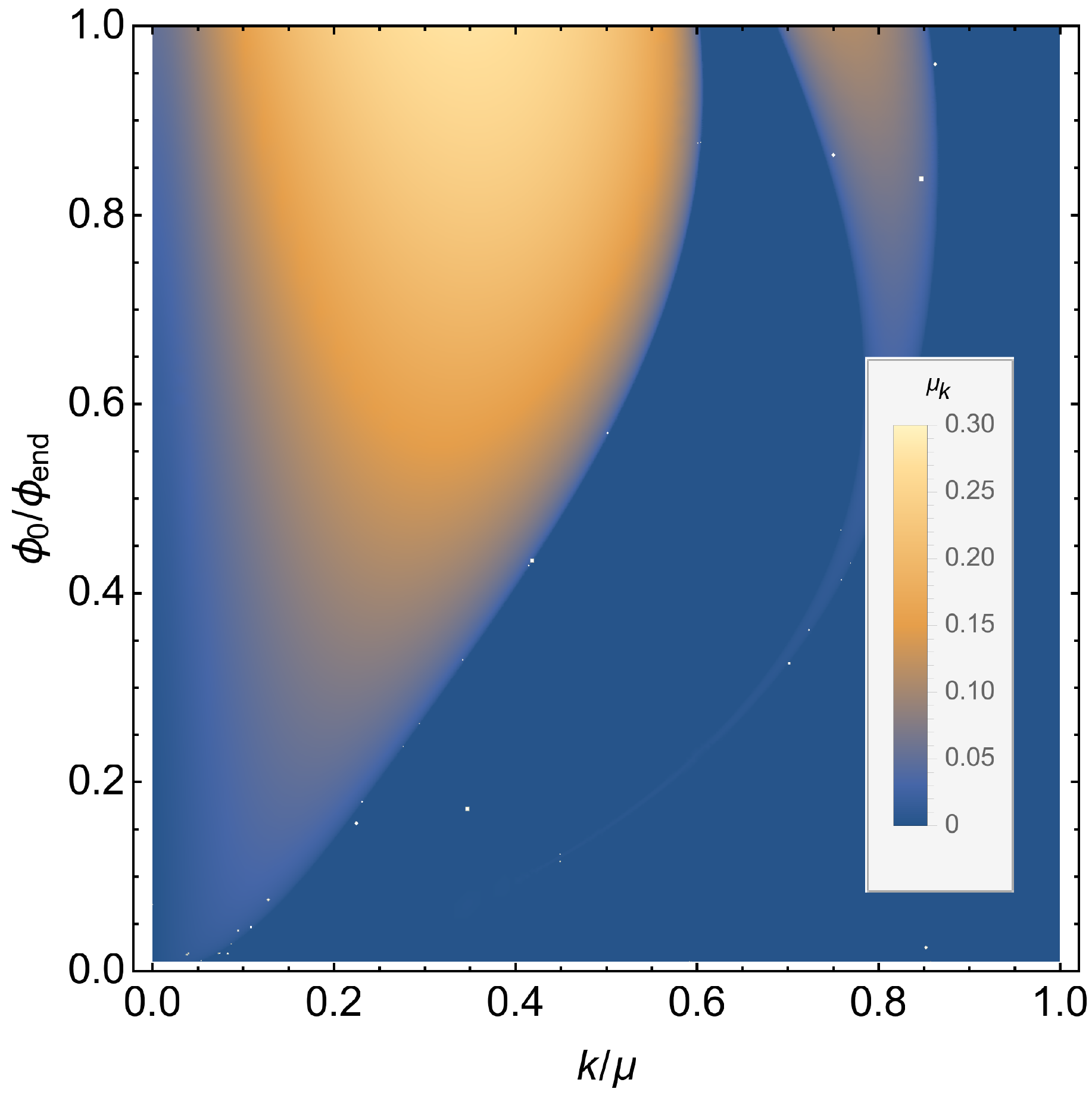}
\\
\includegraphics[width=0.45\textwidth]{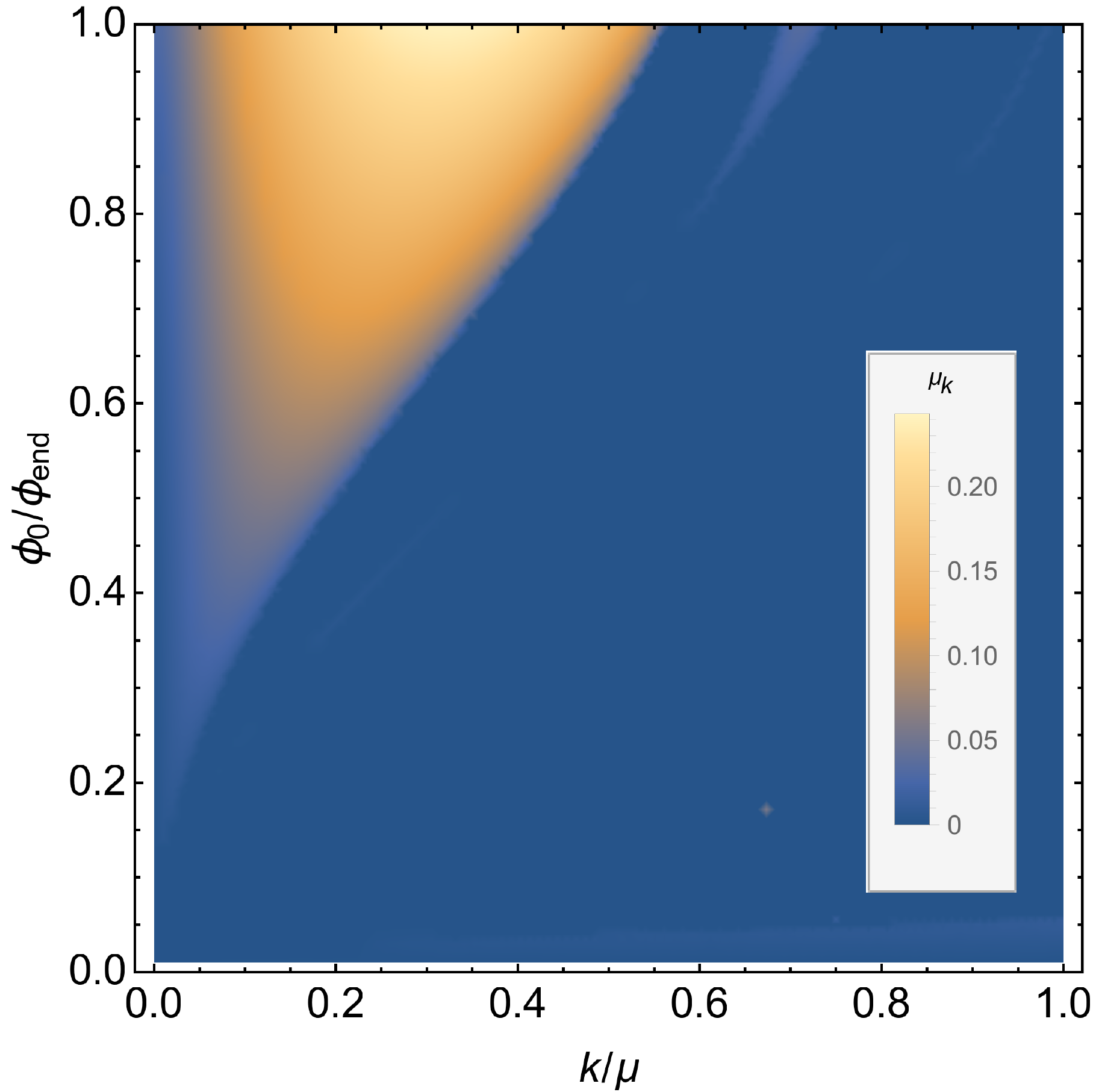}\quad
\includegraphics[width=0.45\textwidth]{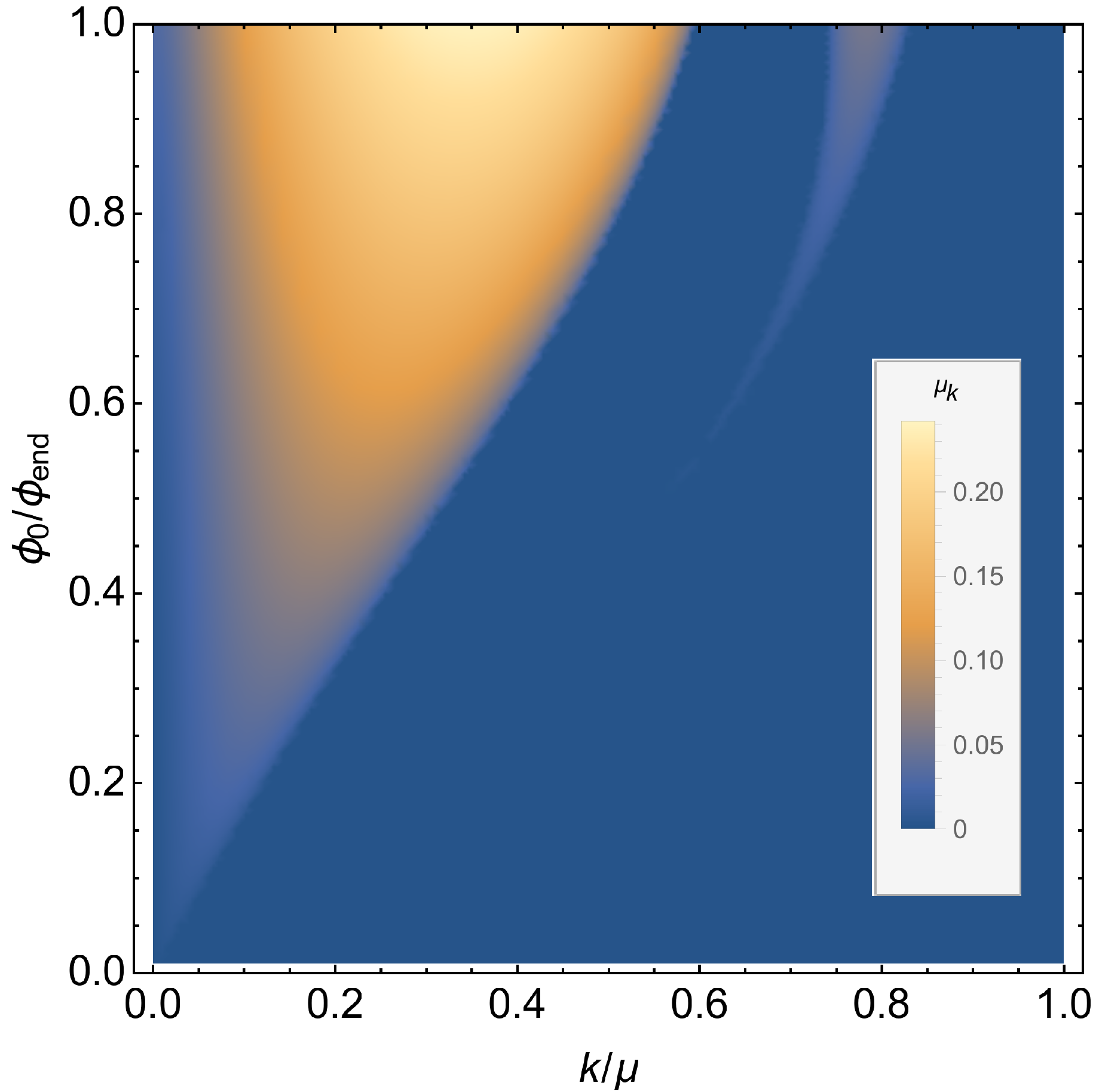}
\caption{
Clockwise from top left: The Floquet charts for $\alpha=10^{-3}$ and $n=1,3/2,2,3$ }
 \label{fig:3Dfloquetchartsovern}
\end{figure}
\begin{figure}
\centering
\includegraphics[width=.6\textwidth]{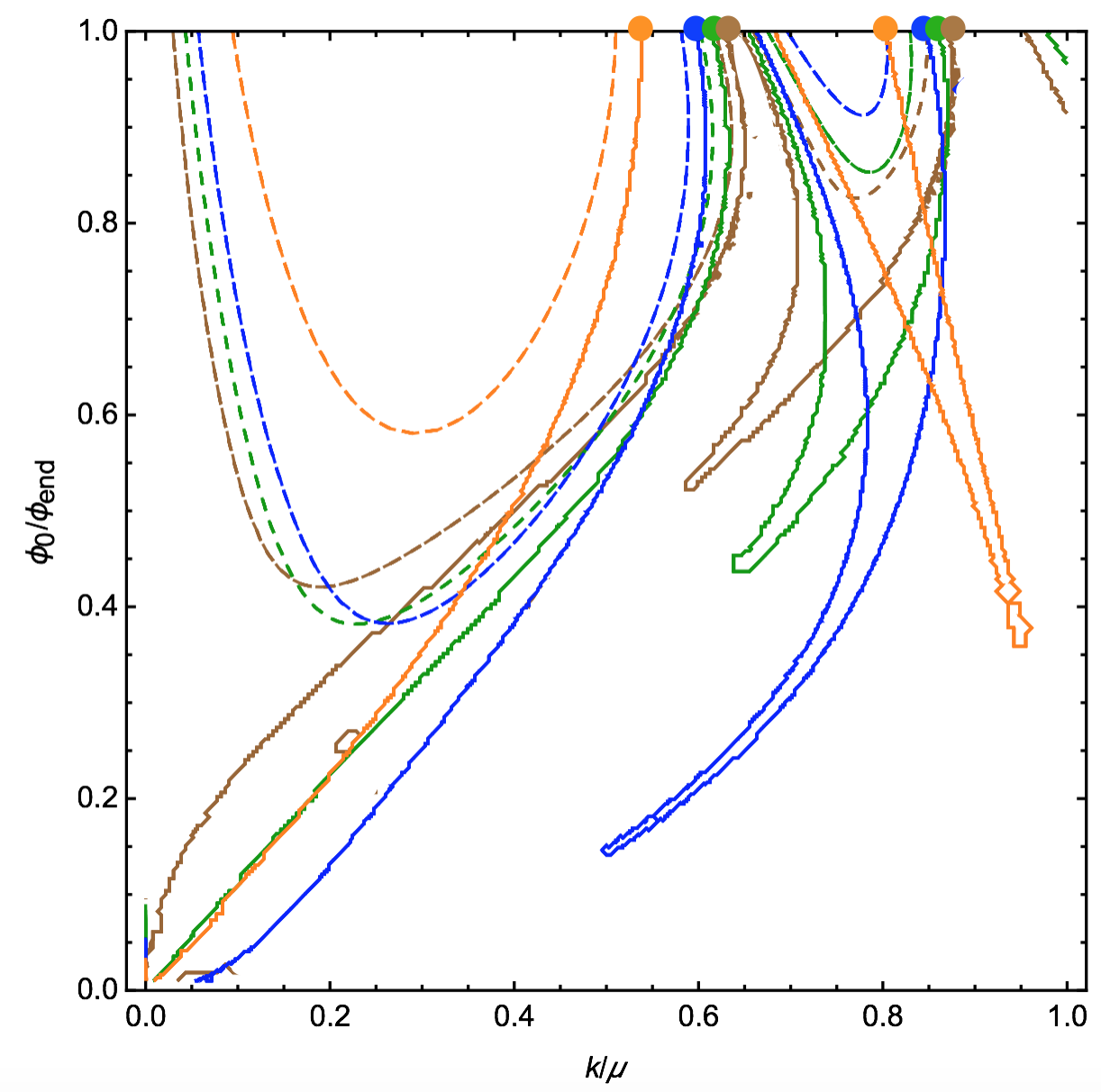}
\caption{
The contour plots for $\mu_k=0$ (solid lines) and $\mu_k=0.1$ (dashed lines) for  $\tilde \alpha=10^{-3}$ and $n=1,3/2,2,3$ (orange, blue, green and brown respectively). The colored dots on the top denote the right edges of the first and second instability bands. We can see that the edges of the bands for $n\ge 3/2$ are almost overlapping, while the range of excited wavenumbers for $n=1$ is significantly smaller.
}
 \label{fig:floquetchartsovern}
\end{figure}

If one tries to plot the three dimensional Floquet diagrams using the field rescaling $\phi_0/\sqrt{\alpha}$, which was used in Fig.~\ref{fig:floquetn3over2}, no unifying pattern emerges. The proper scaling however is $\phi_0/\phi_{\rm end}$, since the comparison must begin at the background field value present at the end of inflation. Using this field rescaling, we can see in Fig.~\ref{fig:3Dfloquetchartsovern} and more clearly in Fig.~\ref{fig:floquetchartsovern} that the edges of the instability bands for $\phi_0=\phi_{\rm end}$ are almost identical for $n\ge 3/2$ and significantly higher than the case of $n=1$. Also, the overall Floquet exponents exhibited are larger for larger values of $n$, as expected from the behavior of the effective frequency-squared shown in Fig.~\ref{fig:meffovern}.

\begin{figure}
\centering
\includegraphics[width=0.7\textwidth]{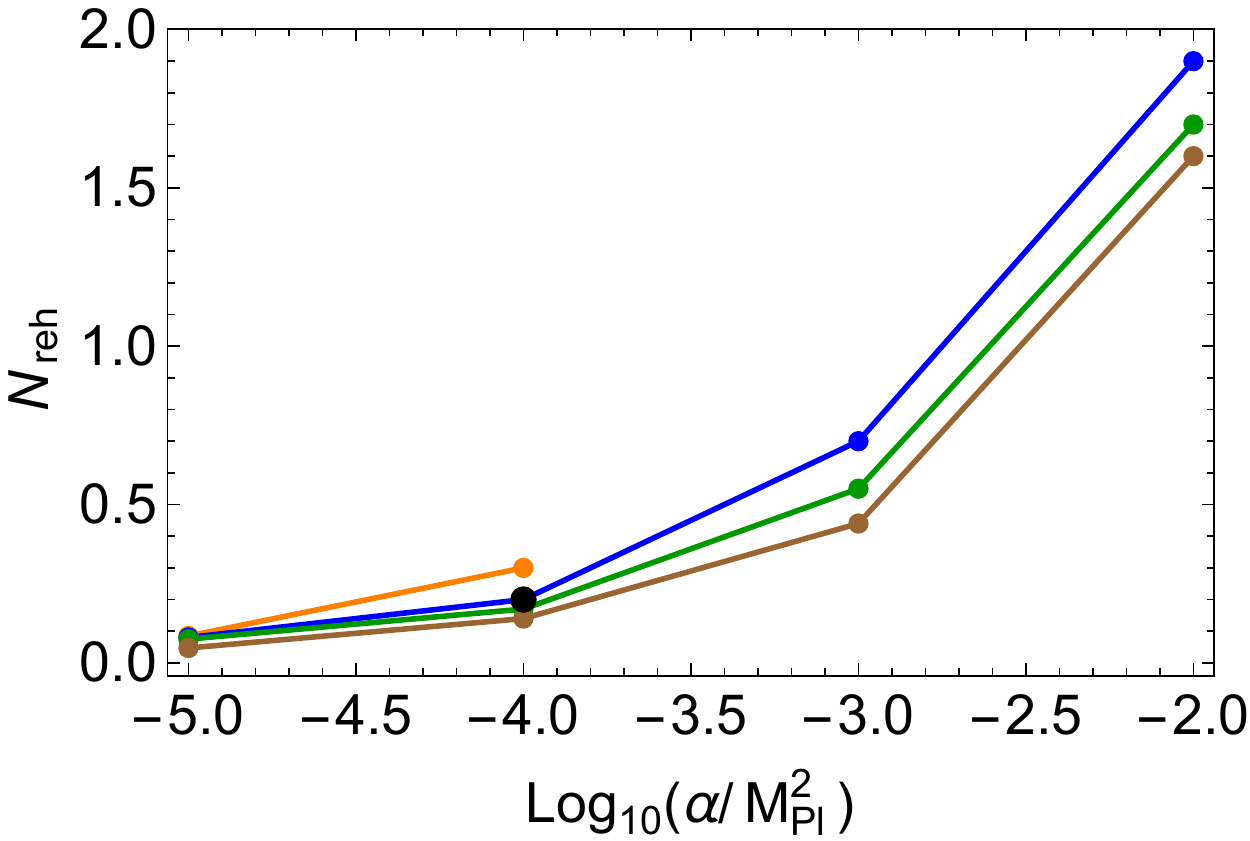}
\caption{
The time required (in $e$-folds) for the transfer of the entire inflaton energy density into modes of the $\chi$  field as a function of the field-space curvature parameter $\alpha$ for $n=1, \, 3/2, \, 2, \, 3$ (orange, blue, green and brown respectively). The black point shows the parameters used in Ref.~\cite{Krajewski:2018moi}. The linear no-backreaction approximation is used. We see that preheating is essentially instantaneous for $ \alpha \lesssim 10^{-4} M_{\rm Pl}^2$.
 }
 \label{fig:Nreh}
\end{figure}

Starting from Bunch-Davies initial conditions during inflation, specifically initializing our computations at $N_{\rm init} \simeq -4$, we evolved the fluctuations in the $\chi$ field on the single-field $\phi$ background, taking into account the expansion of the universe and working in the linear regime, hence neglecting any mode-mode coupling and back-reaction effects. Fig.~\ref{fig:Nreh} shows the time needed for the complete transfer of energy from the $\chi$ background field to $\chi$ radiation modes\footnote{{An updated version of Ref.~\cite{Krajewski:2018moi} includes simulations for $\{\tilde \alpha, n  \} =
\{10^{-3}, 3/2  \},\{10^{-4}, 3/2  \}, \{10^{-4}, 1  \}  $ exhibiting complete preheating at $N_{\rm reh}\approx 0.7, 0.15, 0.2 $ respectively, which match the values shown in Fig.~\ref{fig:Nreh} for these parameter values. }}. For $n=1$ and $ \alpha \gtrsim 10^{-3}M_{\rm Pl}^2$ preheating did not complete through this channel. Overall we see faster preheating for larger values of $n$, hence steeper potentials. However the differences are diminishing for highly curved field-space manifolds,  practically disappearing for $ \alpha \lesssim 10^{-4}M_{\rm Pl}^2$, where preheating occurs almost instantaneously.

\section{Summary and Discussion}
\label{sec:summary}

In the present work we studied preheating in a two-field generalization of the T-model, which is part of the larger family of $\alpha$-attractors, characterized by a field-space manifold of constant negative curvature. We focused on the production of non-inflaton particles, since inflaton self-resonance in the single-field T-model has been extensively studied (e.g. Ref.~\cite{MustafaDuration}), finding reheating to complete within a few $e$-folds for $n\ne1$ and oscillon formation leading to a prolonged matter-dominated phase for $n=1$.

We examined the possibility of multi-field effects arising during inflation and found a strong single-field attractor along a straight background trajectory $\chi=0$. In order for multi-field effects to produce observable signatures, like ``ringing" patterns on the CMB, the initial conditions have to be extremely fine-tuned, which makes such an event unlikely.
{The strong single-field inflationary attractor ensures that preheating will also occur around a single-field background, at least during the initial stage, when back-reaction effects can be safely ignored. Different multi-field potentials on hyperbolic manifolds might support genuinely multi-field background trajectories, leading to significantly different preheating dynamics. This remains an intriguing possibility worth further study.}

We found that most key preheating quantities rely crucially on the field-space curvature parameter $\alpha$, in fact exhibiting simple scaling behaviors. The Hubble scale at the end of inflation scales as $H_{\rm end} \sim \sqrt \alpha$ and is largely independent of the potential steepness, a characteristic trait of $\alpha$-attractors. However the period of background oscillations does not involve $\alpha$, meaning that more background oscillations ``fit" in the first $e$-fold after inflation for higher values of the field-space curvature (low values of $\alpha$). The maximum amplified wavenumber is roughly constant for all values of $\alpha$ and potential steepness parameter $n$, with the exception of $n=1$, where $k_{\rm max}$ is smaller by about $25\%$.

Since the frequency of background oscillations is much larger than the Hubble scale at the end of inflation, the static universe is an increasingly good approximation for larger values of the field-space curvature. This makes Floquet theory a useful tool for understanding preheating in the two-field $T$-model. We found that when plotting the Floquet charts for a specific value of the potential steepness parameter $n$ using the wavenumber and the background field amplitude rescaled by $\sqrt{\alpha}$, all Floquet charts collapse into a single ``master diagram" for small values of $\alpha$.

This scaling behavior of the Floquet charts persists even for different potentials within the T-model. In the case of varying $n$ the background field must be normalized by the field value at the end of inflation $\phi_{\rm end}$ in order for the Floquet chart scaling behavior to appear. As expected, the scaling between Floquet charts of different potentials is not exact, but similarities are enough to explain the similar preheating behavior shown in Fig.~\ref{fig:Nreh}. There we see that preheating lasts longer for larger values of $\alpha$ and smaller values of $n$, while recovering the results of Ref.~\cite{Krajewski:2018moi} for $n=3/2$ and $\tilde \alpha =10^{-4} M_{\rm Pl}^2$, {$n=3/2$ and $\tilde \alpha =10^{-3} M_{\rm Pl}^2$, as well as for $n=1$ and $\tilde \alpha =10^{-4} M_{\rm Pl}^2$.}

While observing reheating is difficult due to the inherently small length scales involved, knowledge of the duration of reheating is essential to correctly match the CMB modes to the exact point during inflation when they left the horizon \cite{Liddle:2003as}. Expanding on the lattice simulations of Ref.~\cite{Krajewski:2018moi} we showed that preheating in the two-field T-model is essentially instantaneous for highly curved field-space manifolds, regardless of the exact form of the potential. This reduces the uncertainty of the predictions of this class of models for the scalar spectral index $n_s$. Unfortunately the low values of $\alpha$ required for the onset of instantaneous preheating makes the observation of tensor modes in these models unlikely even with the CMB Stage 4 experiments, since the resulting tensor-to-scalar ratio is too small $r<10^{-4}$.

The scaling behavior found  in $T$-model preheating does not guarantee that similar effects will arise in other $\alpha$-attractor models. Our results can be applied to study preheating in broader classes of multi-field inflationary models with hyperbolic field-space manifolds. We leave an exhaustive analysis for future work.

\acknowledgements{
It is a pleasure to thank K.~Turzynski and Y.~Welling for helpful discussions.
This work was supported by the Netherlands' Organisation for Scientific Research (NWO) and the Netherlands' Organization for Fundamental Research in Matter (FOM).
The work of AA is partially supported by the Basque Government (IT-979-16) and by the Spanish Ministry MINECO (FPA2015-64041-C2-1P).
}


\section*{Appendix A: Generalization of the T-model}

A simple generalization of the $T$-model \cite{Tmodel, Krajewski:2018moi} is given by the super-potential
\beq
W_H = \sqrt{\alpha}\mu \, S\, F(Z)
\eeq
and K\"ahler potential
\beq
K_H = {-3\alpha \over 2} \log \left [{ (1-Z\bar Z)^2 \over (1-Z^2)(1-\bar Z^2)}\right ]+S \bar S \, .
\eeq
Using the relation between the K\"ahler potential and the superpotential
\beq
Z = {T-1\over T+1}
\eeq
and choosing
\beq
F(Z)=Z^n
\eeq
we get
\beq
K_H = {-3\alpha \over 2} \log \left [{      (T+\bar T)^2
 \over 4 T \bar T
 }\right ]+S \bar S
 \label{eq:KH}
\eeq
and
\beq
W_H = \sqrt{\alpha}\mu S \left ( {T-1\over T+1}  \right )^n \, .
 \label{eq:WH}
\eeq
as in Ref.~\cite{Tmodel}.
The potential follows to be of the form
\beq
V=\alpha \mu^2 \left(Z \bar Z\right)^n \left ({ (1-Z^2)(1-\bar Z^2) \over (1-Z\bar Z)^2}\right )^{3\alpha /2}\, .
\eeq
One can use multiple field-space bases to describe these models.
The choice
\beq
Z=\tanh\left ( {\phi + i\theta \over \sqrt {6\alpha}}\right )
\label{eq:Tmodeloriginal}
\eeq
was used in Ref.~\cite{Tmodel}, leading to the kinetic term
\beq
{\cal L}_{kin} = {1\over 2} {\cal G}_{\phi\phi} \,\partial_\mu \phi\, \partial^\mu \phi
+
{1\over 2} {\cal G}_{\theta\theta}\, \partial_\mu \theta\, \partial^\mu \theta
\,
\eeq
with
\beq
{\cal G}_{\phi\phi} = {\cal G}_{\theta\theta}  ={1\over \cos^2 \left( \sqrt{2\over 3\alpha} \theta \right ) }
\eeq
and the two-field potential
\beq
V(\phi,\theta) = \alpha\mu^2 \left( {\cosh \left ( \sqrt{2\over 3\alpha} \phi \right ) - \cos \left ( \sqrt{2\over 3\alpha} \theta \right )\over \cosh \left ( \sqrt{2\over 3\alpha} \phi \right ) + \cos \left ( \sqrt{2\over 3\alpha} \theta \right ) } \right)^n\left (
\cos \left ( \sqrt{2\over 3\alpha} \theta \right )
\right )^{-3\alpha} \, .
\eeq
 We instead choose the basis used in Ref.~\cite{Krajewski:2018moi},
 which can be derived from Eq.~\eqref{eq:Tmodeloriginal} by performing the transformation
 \beq
 \cos \left ( \sqrt{2\over 3\alpha} \theta \right ) = {1\over \cosh \left ( \sqrt{2\over 3\alpha} \chi \right ) }\, .
 \eeq
 This leads to the kinetic term  ~\eqref{eq:Lphichi}
 \beq
{\cal L}_{kin} = {1\over 2}  \partial_\mu \chi \partial^\mu \chi +{1\over 2} \cosh^2 \left ( \sqrt{2\over 3\alpha} \chi \right ) \partial_\mu \phi \partial^\mu \phi   \, ,
\eeq
 and potential ~\eqref{eq:Vphichi}
 \beq
V(\phi,\chi) =  \alpha \mu^2 \left ( {\cosh(\beta \phi)\cosh(\beta \chi)-1 \over \cosh(\beta \phi)\cosh(\beta \chi)+1}\right)^n \left( \cosh(\beta \chi) \right )^{2/\beta^2} \, ,
\eeq
 where $\beta = \sqrt{2/3\alpha}$.

This choice of the field-space basis allows an easier comparison between our work and Ref.~\cite{Krajewski:2018moi} and simple equations of motion, both for the background as well as for the fluctuations. This comes at a price, namely the illusion that the two field-space directions are inherently different, one of them even being canonically normalized. However, as can be seen in Appendix B, this basis describes a field-space with a constant curvature at every point.

\section*{Appendix B: Field-Space quantities for hyperbolic space}

The kinetic term for the two-field model at hand is written as
\beq
{\cal L} = {1\over 2} {\cal G}_{IJ} \partial_\mu \phi^I \partial^\mu \phi^J \, ,
\eeq
where $\{\phi^1,\phi^2\} \equiv \{\phi,\chi\}$.
In the basis used the non-zero field-space quantities are
\begin{itemize}
\item The metric
\beqn
{\cal G}_{\phi\phi} = e^{2b(\chi)} = e^{2 \log(cosh(\beta \chi))} =  cosh^2(\beta \chi)
\, , \quad
{\cal G}_{\chi\chi}=1
\eeqn
\item The inverse metric
\beqn
{\cal G}^{\phi\phi} =e^{-2b(\chi)} = e^{-2 \log(cosh(\beta \chi))} =  \text{sech}^2(\beta  \chi )
\, , \quad
{\cal G}^{\chi\chi}=1
\eeqn
\item The Christoffel symbols
\beqn
\Gamma^\phi_{\chi\phi} =\beta  \tanh (\beta  \chi )
\, , \quad
\Gamma^\chi_{\phi\phi} =-\frac{1}{2} \beta  \sinh (2 \beta  \chi )
\eeqn
\item The Riemann tensor
\begin{equation}
\begin{aligned}
{\cal R}^\phi_{\chi\phi\chi} = -\beta^2 \, ,\quad
{\cal R}^\phi_{\chi\chi\phi} = \beta^2
\, , \quad
{\cal R}^\chi_{\phi\phi\chi} =\beta ^2 \cosh ^2(\beta  \chi )\, , \quad
{\cal R}^\chi_{\phi\chi\phi} =  -\beta ^2 \cosh ^2(\beta  \chi )
\end{aligned}
\end{equation}
\item The Ricci tensor
\beqn
{\cal R}_{\phi\phi} =-\beta ^2 \cosh ^2(\beta  \chi )
\, , \quad
{\cal R}_{\chi\chi} =  -\beta^2
\eeqn
\item Finally, the Ricci scalar
\beqn
{\cal R}=-2\beta^2 =- {4\over 3\alpha} \, ,
\eeqn
\end{itemize}
where we used
\beq
\beta = \sqrt{2\over 3\alpha} \, .
\eeq

\section*{Appendix C: Alternative time parametrization}

For completeness and ease of comparison with Ref.~\cite{Krajewski:2018moi} we present a different rescaling prescription. Specifically in Ref.~\cite{Krajewski:2018moi} the field-space curvature is rescaled using the reduced Planck mass as $ \alpha = M_{\rm Pl}^2 \tilde \alpha$ and the equation of motion for the background field becomes
\beq
\ddot {\tilde \phi} + 3 H \dot {\tilde  \phi }+   \left (  {M^2\over M_{\rm Pl}}  \right )^2  { \sqrt{6}  n\over \sqrt{\tilde\alpha}    }  \text{csch}\left(\frac{\sqrt{\frac{3}{2}} \tilde\phi  }{\sqrt{\alpha }}\right) \tanh ^{2n}\left(\frac{\sqrt{\frac{3}{2}} |\tilde\phi|  }{2 \sqrt{\alpha }}\right)=0 \, .
\eeq
Time is then rescaled by $m\equiv M^2/M_{\rm Pl}$, leading to the equation
\beq
{{d^2 {\tilde \phi} \over d(mt)^2}+ 3 \tilde H {d {\tilde  \phi } \over d(mt)}+   { \sqrt{6}  n\over \sqrt{\tilde\alpha}    }  \text{csch}\left(\frac{\sqrt{\frac{3}{2}} \tilde\phi  }{\sqrt{\alpha }}\right) \tanh ^{2n}\left(\frac{\sqrt{\frac{3}{2}}| \tilde\phi|  }{2 \sqrt{\alpha }}\right)=0}  \, ,
\eeq
where $\tilde H = H/m$.
The relevant plots, Floquet exponents and  comoving wavenumbers in Ref.~\cite{Krajewski:2018moi} are presented and measured in units of $M^2/M_{\rm Pl}$.

The Hubble scale is
\beqn
\nonumber
\tilde H^2 = {1\over 3 } \left [ {1\over 2}\left ({d {\tilde \phi}  \over d(mt)} \right )^2 + \tanh^{2n} \left ( {|\tilde\phi | \over \sqrt{6\tilde\alpha}} \right ) \right]    \, .
\eeqn

The fluctuation equations with this definition of time become
\beqn
{d^2 Q_\phi\over d(m t)^2} + 3 {H\over m} {d Q_\phi\over d(mt)} + \left [ {(k/m)^2\over a^2} + {V_{\phi\phi}\over m^2} \right ]Q_\phi =0   \, ,
\\
{d^2 Q_\phi\over d(m t)^2} + 3 {H\over m} {d Q_\phi\over d(mt)} + \left [ {(k/m)^2\over a^2} + {V_{\phi\phi}\over m^2} \right ]Q_\phi =0    \,
\eeqn
with \beq
{V_{\phi\phi}\over m^2} = {d^2\over d\tilde\phi^2} \left [  \tanh^{2n} \left (|\tilde\phi |\over \sqrt{6 \tilde \alpha} \right ) \right ]  \, .
\eeq

The ratio of the two mass-scales that can be used to normalize time and wave-numbers is
\beq
 {  { {m\over \mu } = \sqrt{\tilde \alpha}}}  \, ,
\eeq
making the comparison of our linear results with the full lattice simulations of Ref.~\cite{Krajewski:2018moi} straightforward.


\section*{Appendix D: Back-reaction}
\begin{figure}
\centering
\includegraphics[width=0.7\textwidth]{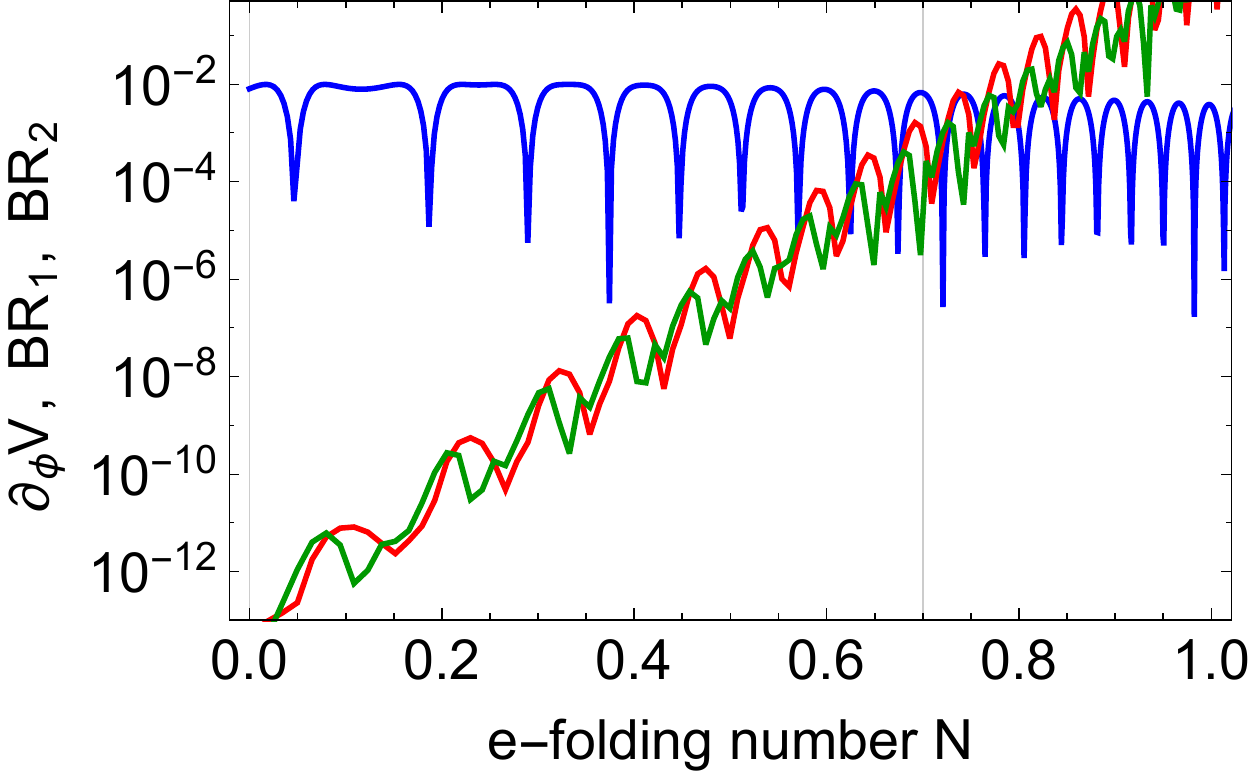}
\caption{
{
The magnitude of the inflaton potential term $|V_{,\phi}|$ (blue) and the two back-reaction terms $BR_1\equiv |\beta^2 \langle \chi\dot\chi \rangle \dot \phi|$ (green) and
$BR_2\equiv  \Delta V \langle \chi^2 \rangle |V_{,\phi}|$ (red) for $\tilde\alpha=0.001$ and $n=3/2$. The vertical line at $N=0.7$ corresponds to the time of complete preheating, according to Fig.~\ref{fig:rhoevol}. We see that  back-reaction effects only become important close to the point of complete preheating and they do not affect the early time dynamics, as expected.
 }}
 \label{fig:BRplot}
\end{figure}

{
Since the present work is focused on extracting semi-analytical arguments, based on the WKB approximation, it is worth examining some back-reaction effects more closely. There are several sources of back-reaction and the only way to accurately describe their combined effects is through lattice simulations, as done for the system under study in Ref.~\cite{Krajewski:2018moi}. On a qualitative level, we can distinguish various back-reaction effects:
\begin{itemize}
\item Mode-mode mixing: This refers to non-linear mixing between the modes $\delta \chi_k$ and usually leads to a power cascade towards the UV. Mode-mode mixing is required for thermalization and is outside of the scope of linear theory. Even in lattice simulations, proper study of thermalization processes usually requires even more UV modes than are usually available.
\item Induced $\delta\phi$ fluctuations due to $\delta \chi$ modes scattering off the inflaton condensate $\phi$.
\item Siphoning energy off the inflaton condensate and acting as a extra drag term for the inflaton motion $\phi(t)$, thus suppressing  background oscillations.
\end{itemize}
We will focus on estimating the last term, as it is the one that can damp the background motion and thus suppress tachyonic preheating\footnote{{Thermalization can affect Bose enhancement by altering the produced $\delta\chi_k$ spectrum, but it typically operates close to or after the point of complete preheating. Since we only intend to estimate back-reaction effects, we will not discuss it further.
  }}.
\newline
The full equation of motion for the $\phi$ field is
\beq
\ddot \phi + \Gamma^\phi_{\chi\phi} \dot \chi\dot\phi + 3H\dot\phi + {\cal G}^{\phi\phi}V_{,\phi} =0
\eeq
In order to estimate the terms arising from the back-reaction of the produced $\chi$ particles, we Taylor expand all terms involving $\chi$ and use a Hartree-type approximation to substitute all quadratic quantities with their average value
\beqn
\chi^2 &\to& \langle \chi^2 \rangle = \int {d^3k\over (2\pi)^3} |\delta\chi_k|^2
\\
\chi \dot \chi &\to& \langle \chi \dot \chi \rangle = \int {d^3k\over (2\pi)^3} \delta\chi_k \cdot \dot{\delta\chi_k^* } \, .
\eeqn
The background equation of motion for the inflaton $\phi$ thus becomes
\beq
\ddot \phi + 3H\dot\phi + {\cal G}^{\phi\phi}V_{,\phi} = - \beta^2  \langle \chi \dot \chi \rangle \dot \phi + \Delta V
   \langle \chi^2 \rangle V_{,\phi}
\eeq
where $\Delta V$ arises from expanding ${\cal G}^{\phi\phi}$ and  $V_{,\phi}$ around $\chi=0$.
The term in the equation of motion involving $\langle \chi \dot\chi\rangle$ arises from the Christoffel symbol and acts as an extra drag term, whereas $\Delta V$ can be thought of as an extra force. Fig.~\ref{fig:BRplot} shows the potential term ${\cal G}^{\phi\phi} V_{,\phi}$ along with the back-reaction contributions to the equations of motion for the case of $\tilde\alpha=0.001$ and $n=3/2$. We see that the back-reaction terms only become important close to the point of complete preheating, defined as $\rho_\phi = \rho_{\delta\chi}$. This means that  during the last inflaton oscillation(s) before complete preheating is achieved, the background inflaton motion will be suppressed due to the produced modes. This has the potential of quenching the resonance and causing the stop of $\chi$ particle production. However tachyonic resonance is usually very robust, since --as we described using the WKB analysis-- as long as the inflaton velocity is non-zero, the hyperbolic metric will lead to a tachyonic instability of $\delta\chi_k$. A careful numerical investigation of tachyonic resonance, albeit in another context, can be found in Ref.~\cite{Adshead:2016iae}, where lattice results were compared to linear calculations, like the ones presented here. It was shown that for the case where the linear calculations pointed to complete tachyonic preheating after a few inflaton oscillations, lattice simulations led to very similar results. The lattice simulations of Ref.~\cite{Krajewski:2018moi} indeed point to a decay of the inflaton condensate and complete preheating, but an evolution of $\phi(t)$ identical to the back-reaction-free case up until very close to that point. Hence linear analysis can successfully capture the initial growth of $\delta\chi$ fluctuations and provide strong indications for parameter choices that allow for complete preheating.
}

\end{document}